\documentclass[a4paper,11pt]{article}
\usepackage{jheppubMod}
\usepackage[utf8]{inputenc}
\usepackage{graphicx}
\usepackage{svg}
\usepackage{xcolor}
\usepackage{wrapfig}

\usepackage{amsmath}
\usepackage{amssymb}
\usepackage{amsfonts}
\usepackage{bm}
\usepackage{physics}
\usepackage{bbold}
\usepackage{mathtools}
\usepackage{slashed}
\usepackage{tensor}
\usepackage{stmaryrd}
\usepackage{arydshln}

\usepackage{tikz} 
\usetikzlibrary{patterns}

\renewcommand{\cos}[1]{\mathrm{cos}\left(#1\right)}

\renewcommand{\exp}[1]{\mathrm{exp}\left(#1\right)}

\renewcommand{\ln}[1]{\mathrm{ln}\left(#1\right)}

\renewcommand{\dim}{\mathrm{dim}}    
\newcommand{\img}{\mathrm{Img}\;}    
\renewcommand{\ker}{\mathrm{Ker}\;}  

\newcommand{\unity}{\mathbb{1}}         
\newcommand{\pepsilon}{{\bm{\epsilon}}}   
\newcommand{\Cmat}{C}                   
\newcommand{\EFmat}{\omega}              
\newcommand{\ECmat}{S}                  
\newcommand{\ESmat}{E}                  
\newcommand{\sigmab}{\overline{\sigma}} 

\newcommand{\dr}{\mathrm{d}_\R}  
\newcommand{\R}{\mathcal{R}} 
\newcommand{\F}{\mathcal{F}} 

\renewcommand{\tr}[1]{\mathrm{Tr}\left\{#1\right\}}

\renewcommand{\det}[1]{\mathrm{det}\left(#1\right)}

\renewcommand{\d}{\mathrm{d}} 
\newcommand{\D}[2]{\mathrm{D}_\mu^{#1} \left[#2\right]} 
\newcommand{\Pd}{\mathcal{D}} 

\newcommand{\combr}[1]{\left[ #1 \right]}     
\newcommand{\acombr}[1]{\left\{ #1 \right\}}  

\newcommand{\Exv}[1]{\bra{0}#1\ket{0}}        

\newcommand{\hg}{\bm{\pi}} 

\newcommand{\pic}{\Tilde{\pi}}       
\newcommand{\pid}{\pi}       
\newcommand{\Ac}{A}                  
\newcommand{\Af}{B}                  
\newcommand{\Afh}{\hat{B}}                  
\newcommand{\Fc}{A}                  
\newcommand{\Ff}{B}                  
\newcommand{\Av}{V}                  
\newcommand{\Fv}{V}                  
\newcommand{\qm}{q_M}                
\newcommand{\zp}{Z^{\prime}}         
\newcommand{\etap}{{\eta^\prime}}    
\newcommand{\ngb}{\xi}    
\newcommand{\fs}{X}                  
\newcommand{\mcf}{\Omega}            
\newcommand{\mcfh}{\hat{\Omega}}     

\newcommand{\qconm}{\chi}
\newcommand{\qcon}{\qconm_c}          

\newcommand{\nc}{{N_C}}                
\newcommand{\nf}{{N_F}}                
\newcommand{\mpi}{m_\pid}             
\newcommand{\mv}{m_\Av}              
\newcommand{\meta}{m_{{\etap}}}      
\newcommand{\fpi}{f_\pi}             
\newcommand{\fetap}{f_\etap}         
\newcommand{\mzp}{m_{{\zp}}}         
\newcommand{\ed}{e_D}                
\newcommand{\gd}{g_D}                
\newcommand{\gv}{g_\Av}     
\newcommand{\Q}{\mathcal{Q}}         
\newcommand{\Qtopo}{\mathcal{Q}_\mathrm{Topo}}         
\newcommand{\cmix}{\varepsilon}     
\newcommand{\clhs}{C_\mathrm{HLS}\,}   

\newcommand{\CC}{\mathcal{C}}    
\newcommand{\Par}{\bm{P}}        
\newcommand{\Nar}{\hat{\sigma}}  

\newcommand{\liea}[1]{\mathfrak{#1}}    
\newcommand{\ks}{\textbf{k}}            

\newcommand{\needCite}[1]{{\color{blue}[??]}}
\newcommand{\needEqref}[1]{{\color{blue}(??)}}
\newcommand{\needRef}[1]{{\color{blue}??}}
\newcommand{\jpc}[1]{\textcolor{blue}{#1}}

\newcommand{\OA}[0]{\quad \quad \text{or} \quad \quad }

\renewcommand{\i}{i}

\title{Low energy effective theories of composite dark matter with real representations}
\abstract{}
\author[a,b]{Joachim Pomper}
\author[a]{, Suchita Kulkarni}
\affiliation[a]{Institute of Physics, NAWI Graz, University of Graz, Universit\"atsplatz 5, A-8010 Graz, Austria}
\affiliation[b]{Pisa University, Physics Department, 
Largo Bruno Pontecorvo 3, 56127 Pisa, Italy}
\emailAdd{suchita.kulkarni@uni-graz.at}
\emailAdd{joachim.pomper@phd.unipi.it}

\abstract{
We consider pseudo Nambu-Goldstone bosons arising from Dirac fermions transforming in real representations of a confining gauge group as dark matter candidates. We consider a special case of two Dirac fermions and couple the resulting dark sector to the Standard Model using a vector mediator. Within this construction, we develop a consistent low energy effective theory, with special attention to Wess-Zumino-Witten term given the topologically non-trivial coset space. We furthermore include the heavier spin-0 flavour singlet state and the spin-1 vector meson multiplet, by using the Hidden Local Symmetry Lagrangian for the latter. Although we concentrate on special case of two flavours, our results are generic and can be applied to a wider variety of theories featuring real representations. We apply our formalism and comment on the effect of the flavour singlet for dark matter phenomenology. Finally, we also comment on generalisation of our formalism for higher representations and provide potential consequences of discrete symmetry breaking. 
}

\begin{document}

\maketitle

\section{Introduction} 
    \label{sec:introduction}
    A class of particle physics models dubbed Strongly-Interacting Massive Particles (SIMP) \cite{Hochberg:2014kqa} reconciling correct relic density together with large self interaction consistent with current limits from astrophysics realized in QCD-like models have gathered a lot of attention in recent years. These are models of fermions, transforming under a non-trivial representation of a non-Abelian gauge group in the ultra-violet (UV) and resulting in pseudo Nambu-Goldstone bosons (pNGBs) due to spontaneously broken (approximate) symmetry in the infra-red (IR). These particles are dubbed ``dark pions'' ($\pi$), in analogy to QCD. An additional mediator is introduced in order to maintain kinetic equilibrium between the new non-Abelian sector and the SM. Dark pions are stabilised against decays through mediator via careful charge assignments. Such models feature a $3\pi\rightarrow2\pi$ cannibalization process resulting due to Wess-Zumino-Witten (WZW) term \cite{Wess:1971yu, Witten:1983tw}, that may be used to set the relic density via a freeze-out process and a $2\pi\to2\pi$ self scattering mechanism for generating large enough dark matter self-interactions.

While these models seem very tempting, the sheer complexity of such a dark sector should not be underestimated. The amount of physical bound states can be numerous and dependent on the details of the theory. States other than the dark pions may become relevant for DM physics \cite{Bernreuther:2023kcg, Bernreuther:2019pfb, Berlin:2018tvf, Choi:2018iit}. Most investigations so far use effective field theory approaches such as chiral perturbation theory to describe the dynamics of the relevant parts of the particle spectrum.  However, it is hard to say in general which states will be relevant, if we do not know the exact mass spectrum, which depends on the details of the UV model. There have been novel approaches \cite{Kulkarni:2022bvh, Appelquist:2015zfa, Cline:2022leq} in combining effective field theories and lattice field theory methods in the context of DM, in order to constrain or calculate the mass spectrum and low energy effective constants (LEC) for an effective DM description. 

In this work we will focus on Dirac fermions transforming under a finite dimensional, unitary, real representation of a gauge group. The defining feature of such representation is that it is unitary-equivalent to its complex conjugate representation. Thus, there is no way to distinguish particles and anti-particles with respect to this gauge group on physical grounds. The prototypical theory is an $SO(\nc)$ gauge theory, with fermions transforming under the so-called vector representation of $SO(\nc)$. These theories have been studied very little in the context of DM~\cite{Hochberg:2014kqa,Kamada:2017tsq, Davighi:2024zip,Beauchesne:2019ato}. They are also studied in the context of composite Higgs dynamics \cite{Cacciapaglia:2019ixa, Agashe:2004rs, Agashe:2005dk, Contino:2006qr, Agugliaro:2019wtf,Ferretti:2013kya, Cacciapaglia:2019bqz, Ferretti:2016upr}. The meson spectrum resulting from real representations is also studied on lattice. Investigations for the $SO(4)$ gauge group with two Dirac fermions are available in \cite{Hietanen:2012sz}. In \cite{Ayyar:2018zuk, Ayyar:2017qdf} lattice simulations for $SU(4)$ gauge theory with fermions simultaneously transforming fundamental and  two-index antisymmetric (sextet) representation were performed, while results for $Sp(4)$ gauge group with dynamical fermions simultaneously in fundamental and antisymmetric representation are available \cite{Bennett:2022yfa}. Lattice simulations for fermions in several representations of $Sp(4)$ gauge group in quenched limit are also available in \cite{Bennett:2023qwx}. The formalism we derive in this work can readily utilise results from these lattice works.

We focus on the scenario with $\nf=2$ Dirac flavors as a minimal candidate theory containing a WZW term, $\nf=1$ contains no WZW interactions. We examine in-depth the UV and IR behaviour of this theory with a detailed analysis of associated symmetries, construct the low energy chiral Lagrangian including the vector mesons and the pseudo-scalar singlet $\etap$. We point out that in this case a topological obstruction renders the standard construction and classification of WZW terms \cite{Witten:1983tw, DHoker:1994rdl, DHoker:1995mfi} inconclusive. However, since these terms are essential for the SIMP model, we exploit a different approach, first explored in \cite{Chu:1996fr} to construct the WZW term, even if the standard approach seems to be not available. 

Finally, we investigate the effect of the light $\etap$ on DM freeze-out due to an anomalous decay channel, once we couple the dark sector to the SM via a dark photon. We therefore derive a representation theoretic criterion that characterizes for which theories the physics of the $\etap$ meson becomes important for DM. To the best of our knowledge, the role of this particle for DM physics was not investigated within the SIMP model so far, mostly because its QCD analog is rather heavy. However, no statements exist for general theories.  With this setup we also lay the foundation for lattice studies of these strong dark sectors by offering classifications and construction recipes for interpolating operators of all the relevant particle states. Further, we provide some technical details on the structure of continuous and discrete symmetries of the underlying UV theory. 

The structure of the paper is as follows. In section~\ref{sec:short_range_description}, we introduce the UV Lagrangian for the dark matter model based on an $SO(\nc)$ gauge theory with mass degenerate fermions and identify the symmetries. In section~\ref{sec:long_range_description}, we derive the associated chiral Lagrangian including non-anomalous and anomalous (WZW) terms and include the $\etap$. We use this formalism and develop dark matter phenomenology in section~\ref{sec:first_phenomenological_applications}, establishing the interplay of $2\to2$ and $3\to2$ annihilation processes and comment on the viable regions of parameter space compatible also with the pion self-scattering cross section. In section \ref{sec:generalizations} we discuss generalizations to other gauge groups and higher order representations. Finally we conclude in section~\ref{sec:summary_and_conclusion}. 

\subsection*{How to read this paper?}

A big part of the paper is an in-depth discussion of the construction and properties of QCD-like theories with fermions in real representations. Given the familiarity with $SU(N_C)$ gauge groups, a large part of SIMP literature is focused on it. This article is aimed at closing the gap in the literature by providing a cohesive formalism while being as self contained as possible. The price one pays for providing such a framework is the length of the paper. For an efficient first read, especially from the point of view of DM phenomenologists, we point towards a couple of relevant results, beyond the brief explicit phenomenological applications in section \ref{sec:first_phenomenological_applications}. We note here that our construction of low energy chiral Lagrangian is generic and can be applied to a wide variety of theories featuring real representations.
\begin{itemize}
    \item Figure \ref{fig:comparison_of_symmetries_in_fundamental_two_flavor_theories} summarizes the global symmetry structure of the theories. 
    \item The criterion \eqref{eq:large_nc_criterion} can be used to estimate if a light $\etap$ particle can be expected in a given theory. 
    \item Equation \eqref{eq:HLS_lagrangian_1_expanded}-\eqref{eq:HLS_lagrangian_6_expanded} states the lowest order Lagrangian for massless dark pions, massless dark photon and vector mesons. It demonstrates modification of pion self-interactions for mass degenerate theories as explained in \eqref{eq:interactions_pions}. The mass term for the dark photon is given in \eqref{eq:massterm_dark_photon}. 
    \item Modifications to the pion Lagrangian, when including the $\etap$ state can be found in \eqref{eq:lagrangian_eta_pi_interactions}.
    \item  The WZW term expanded to lowest order without vector mesons is given in \eqref{eq:wzw_action_expanded_1}-\eqref{eq:wzw_action_expanded_3} and with vector mesons in \eqref{eq:generalized_wzw_action_expanded_1}-\eqref{eq:generalized_wzw_action_expanded_6}. Inclusion of vector mesons results in four additional low-energy effective constants $\clhs$, $C^\mathrm{anom.}_{1,3,4}$. The values of the relevant low-energy constants may be estimated by assuming vector meson dominance, which allows to develop some phenomenological intuition. We discuss the potential values using eqn. \eqref{eq: clhs_fixing} and \eqref{eq:generalized_wzw_action_expanded_1}-\eqref{eq:generalized_wzw_action_expanded_6}.
    \item In section \ref{sec:generalizations} we discuss a potential source of gravitational waves from domain wall collapse due to the $U(1)_A$ axial symmetry. This would be complementary to first order transition signals and unique to sectors with fermions in non-fundamental representations. If such signals can be observed remains an open question.  
\end{itemize}

\section{Short range description}
    \label{sec:short_range_description}
    Successful construction of a low energy effective theory starts by investigation of the symmetries of the underlying microscopic theory in the ultraviolet (UV). The dark sector model we want to investigate comprises a new strong dark force, that mimics features of QCD, and an abelian sector that acts as a mediator between the dark sector and the SM. Within our setup the dark sector is QCD-like, in other words it features a chirally broken phase in the IR and the coupling behaves asymptotically free. We describe the IR properties via chiral perturbation theory methods. The coupling of the abelian sector shows the opposite behaviour in the IR. It thus is a fair assumption to treat it as a small perturbation to the strong sector. Accordingly, our discussion will treat these sectors separately. 

\subsection{The isolated strong dark sector}
    The strong dark sector consists of $\nf = 2$ Dirac fermions $q^{(k)}$ transforming under the non-abelian gauge group $G_C = SO(\nc)$ in the vector representation $\R$ of dimension $\dr = \nc$. We call the Dirac fermions dark quarks, in analogy to QCD. The dynamics of the dark gluons $\Ac_\mu^\alpha$ is described by a Yang-Mills Lagrangian
    \begin{equation} \label{eq:dark_strong_yang_mills_lagrangian}
        \mathcal{L}_\mathrm{YM}^\mathrm{UV} = -\frac{1}{4} \Fc_{\mu\nu}^\alpha\Fc^{\mu\nu}_\alpha 
    \end{equation}
    with $\Fc_{\mu\nu}^\alpha = \partial_\mu\Ac_\nu^\alpha - \partial_\mu\Ac_\nu^\alpha + \gd C^{\alpha}_{\;\;\beta\gamma}\Ac_\mu^\beta\Ac_\nu^\gamma$ the field strength tensor of the dark gluons and $\gd$ the gauge coupling of the strong dark force. The dark quarks are coupled to the dark gluons by virtue of the gauge principle
    \begin{equation} \label{eq:dark_strong_quark_lagrangian}
        \mathcal{L}_\mathrm{q}^\mathrm{UV} = \sum_{j=1}^\nf \left(\overline{q}^{(j)}i\gamma^\mu\D{\R}{\Ac}q^{(j)} - m\, \overline{q}^{(j)}q^{(j)}\right)
    \end{equation}
    with $\overline{q}^{(j)}$ the adjoint Dirac spinor and the covariant derivative given by
    \begin{equation} \label{eq:covariant_derivative_dark_quarks}
        \D{\R}{\Ac}q := \partial_\mu q - i\gd \Ac_\mu^\alpha T_\alpha^\R q, 
    \end{equation}
    where $T_\alpha^\R$ denotes the generators in representation $\R$. The vector representation $\R$ is a real representation. On physical grounds this means that fermions and anti-fermions are indistinguishable with respect to the strong gauge group $G_C$. Mathematically, this can be formulated via existence of a unitary matrix $\ECmat$ that maps the representation $\R$ equivariantly onto its complex conjugate representation i.e. 
    \begin{equation} \label{eq:equivariance_condition}
        \ECmat U^\R \ECmat^{-1} = U^{\R*} 
        \OA
        \ECmat T^\R_\alpha \ECmat^{-1} = -\left(T^{\R}_\alpha\right)^\top.
    \end{equation}
    Here $^*$ denotes complex conjugation of a matrix. For a real representation, $S$ is symmetric and $S^* = S^{-1}$ \cite{Georgi:2000vve}.  Due to the reality of the theory, the fundamental degrees of freedom are not $\nf$ Dirac fermions $q^{(j)}$ but $2\nf$ Majorana fermions $\qm^{(n)}$, with respect to an augmented charge conjugation operator
    \begin{equation} \label{eq:augmented_charge_conjugation}
        \CC : q \xmapsto{} q_\CC =  \Cmat \ECmat q^*, 
    \end{equation}
    where $\Cmat = -i\gamma_2$ is the charge conjugation matrix as defined in \eqref{eq:charge_conjugation_and_gamma_five_matrix} and the matrix $\ECmat$ makes the equivalence between $\R$ and its conjugate representation explicit. Since each Majorana fermion satisfies $\CC \qm^{(n)} = \qm^{(n)}$, every Dirac fermion may be decomposed according to $q^{(j)} = \qm^{(j)} + i \qm^{(j+\nf)}$. 
    
    Rewriting the dark quark Lagrangian in terms of these Majorana fermions makes the chiral symmetry of the Lagrangian explicit and would result in the Lagrangian stated in \cite{Hochberg:2014kqa, Hochberg:2015vrg} for the $SO(\nc)$ case. Instead we would like to employ the Nambu-Gorkov formalism \cite{Nambu:1960tm}, since it pronounces the flavour structure and makes it easier to compare features with symplectic gauge theories. For this we fix a chiral basis of the $\gamma$-matrices \eqref{eq:gamma_matrices} and decompose the $\nf$ Dirac spinor $q^{(j)}$ into $2\nf$ left-handed Weyl {(anti-)spinors}
    \begin{equation} \label{eq:dark_strong_quark_lagrangian_nambu_gorkov}
        q^{(j)} = 
        \begin{pmatrix}
            \psi^{(j)} \\
            \ESmat \ECmat \psi^{(j+\nf)*}
        \end{pmatrix}.
    \end{equation}
    Here\footnote{For the conventions on $\gamma$-matrices, Pauli-matrices and charge conjugation see appendix \ref{sec:conventions}.} $\ESmat = \i\sigmab^2$ is a non-zero off-diagonal block of the charge conjugation matrix $\Cmat$. We can rewrite the Lagrangian using $ E^{-1}\,\overline{\sigma}^\mu\,E = g^{\mu\mu}(\overline{\sigma}^\mu)^\top = (\sigma^\mu)^\top$, eqn. \eqref{eq:equivariance_condition}, anti-commutativity of fermions and partial integration\footnote{We assume appropriate boundary conditions.}    
    \begin{align} \label{eq:quark_}
        \mathcal{L}_\mathrm{q}^\mathrm{UV} 
        &= \sum_{n=1}^{2\nf} \quad \psi^{(n)\dagger}i\sigmab^\mu\D{R}{\Ac}\psi^{(n)} - \frac{1}{2}m\,\EFmat_{mk} \left(\psi^{(m)\dagger}\ESmat\ECmat\psi^{(k)*} - \psi^{(m)\top}\ESmat^*\ECmat^*\psi^{(k)}\right) \\
        &=i\Psi^\dagger \sigmab^\mu\D{R}{\Ac}\Psi - \frac{m}{2}\left(\Psi^\dagger \ESmat\ECmat\EFmat^* \Psi^* - \Psi^\top \ESmat^*\ECmat^*\EFmat \Psi\right).
    \end{align}
    In the second line we collected all Weyl spinors in $\Psi^\top = \left(\psi^{(1)\top}, \dots, \psi^{(2\nf)\top}\right)$. The symmetric tensor $\EFmat_{ij}$, defining the structure of the mass term, may be represented by the following matrix
    \begin{equation} \label{eq:invaraint_flavor_tensor}
        \EFmat = 
        \begin{pmatrix}
            0 & \unity_\nf \\
            \unity_\nf & 0\\
        \end{pmatrix}.
    \end{equation}
    To investigate non-degenerate masses one can replace $m\,\EFmat_{ij} = M_{ij}$ with a generic symmetric rank 2 mass tensor. 
    \subsubsection{Anomalous symmetry breaking}
    In the chiral limit $m \xrightarrow{} 0$, the Lagrangian \eqref{eq:dark_strong_quark_lagrangian_nambu_gorkov} in the Nambu-Gorkov formulation demonstrates that the action is invariant under complex rotations of the $2\nf$ Weyl fermions, which substitutes a global $U(2\nf)$ symmetry on the classical level. The associated currents are given by
    \begin{equation} \label{eq:flavor_currents}
        j_N^\mu = \Psi^\dagger \sigmab^\mu T_N^\F \Psi,
    \end{equation}
    with $T_N^\F$ the generators of $U(2\nf)$ in the fundamental representation. On quantum level, only a subgroup of the global symmetry may be an actual symmetry due to the potential non-invariance of the fermionic path integral measure \cite{Fujikawa:1979ay}. This is similar to the anomalous breaking of the $U(1)_A$ in QCD, resolving the so-called ``$U(1)$-problem'' \cite{Weinberg:1975ui}. The derivation works analogously to that of standard QCD \cite{Witten:1979vv}\cite[Chpt. 22]{Weinberg:1996kr}. 
    \begin{figure}
        \centering
        \resizebox{0.70\textwidth}{!}{\tikzset{every picture/.style={line width=0.75pt}} 

\begin{tikzpicture}[x=0.75pt,y=0.75pt,yscale=-1,xscale=1]

\draw    (188,70) -- (248,110) ;
\draw [shift={(222.16,92.77)}, rotate = 213.69] [fill={rgb, 255:red, 0; green, 0; blue, 0 }  ][line width=0.08]  [draw opacity=0] (8.93,-4.29) -- (0,0) -- (8.93,4.29) -- cycle    ;
\draw    (248,40) -- (188,70) ;
\draw [shift={(213.53,57.24)}, rotate = 333.43] [fill={rgb, 255:red, 0; green, 0; blue, 0 }  ][line width=0.08]  [draw opacity=0] (8.93,-4.29) -- (0,0) -- (8.93,4.29) -- cycle    ;
\draw    (248,110) -- (248,40) ;
\draw [shift={(248,70)}, rotate = 90] [fill={rgb, 255:red, 0; green, 0; blue, 0 }  ][line width=0.08]  [draw opacity=0] (8.93,-4.29) -- (0,0) -- (8.93,4.29) -- cycle    ;
\draw    (188,70) .. controls (186.33,71.67) and (184.67,71.67) .. (183,70) .. controls (181.33,68.33) and (179.67,68.33) .. (178,70) .. controls (176.33,71.67) and (174.67,71.67) .. (173,70) .. controls (171.33,68.33) and (169.67,68.33) .. (168,70) .. controls (166.33,71.67) and (164.67,71.67) .. (163,70) .. controls (161.33,68.33) and (159.67,68.33) .. (158,70) -- (158,70) ;
\draw    (258,20) .. controls (258.75,22.24) and (258,23.73) .. (255.76,24.47) .. controls (253.53,25.22) and (252.78,26.71) .. (253.53,28.94) .. controls (254.28,31.18) and (253.53,32.67) .. (251.29,33.42) .. controls (249.06,34.17) and (248.31,35.66) .. (249.06,37.89) -- (248,40) -- (248,40) ;
\draw    (258,130) .. controls (255.76,129.26) and (255.01,127.77) .. (255.76,125.53) .. controls (256.51,123.3) and (255.76,121.81) .. (253.53,121.06) .. controls (251.29,120.31) and (250.54,118.82) .. (251.29,116.58) .. controls (252.04,114.35) and (251.29,112.86) .. (249.06,112.11) -- (248,110) -- (248,110) ;
\draw    (411,70) -- (471,110) ;
\draw [shift={(445.16,92.77)}, rotate = 213.69] [fill={rgb, 255:red, 0; green, 0; blue, 0 }  ][line width=0.08]  [draw opacity=0] (8.93,-4.29) -- (0,0) -- (8.93,4.29) -- cycle    ;
\draw    (471,40) -- (411,70) ;
\draw [shift={(436.53,57.24)}, rotate = 333.43] [fill={rgb, 255:red, 0; green, 0; blue, 0 }  ][line width=0.08]  [draw opacity=0] (8.93,-4.29) -- (0,0) -- (8.93,4.29) -- cycle    ;
\draw    (471,110) -- (471,40) ;
\draw [shift={(471,70)}, rotate = 90] [fill={rgb, 255:red, 0; green, 0; blue, 0 }  ][line width=0.08]  [draw opacity=0] (8.93,-4.29) -- (0,0) -- (8.93,4.29) -- cycle    ;
\draw    (411,70) .. controls (409.33,71.67) and (407.67,71.67) .. (406,70) .. controls (404.33,68.33) and (402.67,68.33) .. (401,70) .. controls (399.33,71.67) and (397.67,71.67) .. (396,70) .. controls (394.33,68.33) and (392.67,68.33) .. (391,70) .. controls (389.33,71.67) and (387.67,71.67) .. (386,70) .. controls (384.33,68.33) and (382.67,68.33) .. (381,70) -- (381,70) ;
\draw    (481,20) .. controls (481.75,22.24) and (481,23.73) .. (478.76,24.47) .. controls (476.53,25.22) and (475.78,26.71) .. (476.53,28.94) .. controls (477.28,31.18) and (476.53,32.67) .. (474.29,33.42) .. controls (472.06,34.17) and (471.31,35.66) .. (472.06,37.89) -- (471,40) -- (471,40) ;
\draw    (481,130) .. controls (478.76,129.26) and (478.01,127.77) .. (478.76,125.53) .. controls (479.51,123.3) and (478.76,121.81) .. (476.53,121.06) .. controls (474.29,120.31) and (473.54,118.82) .. (474.29,116.58) .. controls (475.04,114.35) and (474.29,112.86) .. (472.06,112.11) -- (471,110) -- (471,110) ;

\draw (141,42.4) node [anchor=north west][inner sep=0.75pt]    {$j_{N }^{\mu }(x)$};
\draw (269,7.4) node [anchor=north west][inner sep=0.75pt]    {$J_{\alpha }^{\nu }( y)$};
\draw (269,116.4) node [anchor=north west][inner sep=0.75pt]    {$J_{\beta }^{\sigma }(z)$};
\draw (364,42.4) node [anchor=north west][inner sep=0.75pt]    {$j_{N }^{\mu }(x)$};
\draw (492,117.4) node [anchor=north west][inner sep=0.75pt]    {$J_{\alpha }^{\nu }( y)$};
\draw (492,6.4) node [anchor=north west][inner sep=0.75pt]    {$J_{\beta }^{\sigma }(z)$};

\end{tikzpicture}}
        \caption{Triangle diagram contributing to the axial anomaly. The axial anomaly leads to non-conservation of the singlet flavour current $j_0^\mu$, sourced by the dark gluons.}
        \label{fig: axial_anomaly}
    \end{figure}
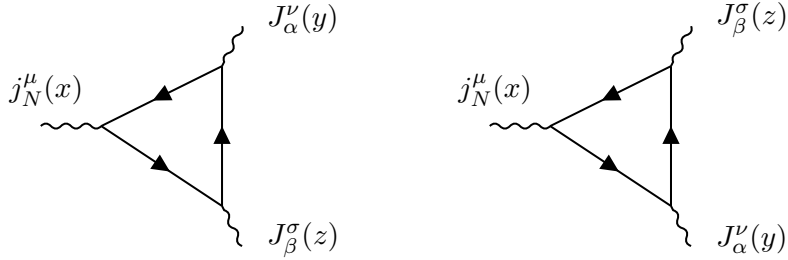
    Under a global transformation $U^\F = \exp{-\pepsilon^\F}$, with $\pepsilon^\F = -i\epsilon^N T_N^\F$, the path integral measure shifts the phase according to
    \begin{equation} \label{eq:anomalous_phaseshift}
        \Pd\overline{\Psi}\Pd\Psi \xrightarrow{}  e^{i\mathcal{A}[\pepsilon^\F, \Ac]}\Pd\overline{\Psi}\Pd\Psi\,,   
    \end{equation}
     which is determined by the anomaly functional $\mathcal{A}[\pepsilon, \Ac] = \int \d^4x\,\epsilon^N \mathcal{A}_N[\Ac]$. The anomaly functional for these global symmetries calculated by a perturbative one-loop calculation \cite[Chpt. 22.3]{Weinberg:1996kr}, involving the triangle diagrams in figure \ref{fig: axial_anomaly} is given by
    \begin{align}
        \mathcal{A}[\pepsilon, \Ac] 
        &= 2i \,T_\R \tr{\pepsilon^\F} \frac{\gd^2 \epsilon^{\mu\nu\rho\sigma}\delta_{\alpha\beta}}{64\pi^2} \int \d^4x\,\Fc_{\mu\nu}^\alpha(x) \Fc_{\rho\sigma}^\beta(x) \\
        \label{eq:anomaly_functional_axial_anomaly}
        & = 2i \,T_\R \,\tr{\pepsilon^\F}\,\Qtopo[\Ac].
    \end{align}
    Here $T_\R$ is the Dynkin\footnote{In principle the value of $T_\R$ is defined up to a multiplicative constant that can be absorbed in the running-coupling. In appendix \ref{sec:topological_charge} we explain why $T_\R = 1$, which is related to the definition of topological charge of the gluon field configuration.} index of the representation $\R$. $T_\R = 1$ for the vector representation of $SO(\nc)$. The topological charge operator $\Qtopo[\Ac]$ takes on only integer values in a dark gluon background. The existence of topologically non-trivial gauge field configuration has first been proven in \cite{Belavin:1975fg} for $SU(2)$ and later for all simple Lie-groups \cite{Atiyah:1978ri, Dorey:2002ik}. Since $\tr{\pepsilon^\F} = 0$ implies vanishing anomaly, global symmetries within the $SU(2\nf)$ subgroup are non-anomalous. Moreover, there exists a non-anomalous set of discrete symmetries for which $\tr{\pepsilon^\F} \neq 0$, discussed below in section \ref{sec:charge_conjugation}.

\subsubsection{Explicit symmetry breaking}
    Like in QCD, the mass-term introduces a source of explicit symmetry breaking. We restrict the generic mass matrix $M$ to be real in order to avoid explicit $CP$-violating terms.  In a tensorial notation the mass matrix is a symmetric rank 2 tensor under the flavour symmetry group $G_F = SU(2\nf)$. The isotropy condition for the unbroken flavour group $H_F$ is given by 
    \begin{equation} \label{eq:invariance_condition_mass_term}
        \left\{U^\F_h\right\}^{k}_{l} M_{km} \left\{U^\F_h\right\}^{m}_{n} = M_{ln}  
        \OA
        U^{\F\top}_h M U^\F_h = M,
    \end{equation}
    where $U^\F_h \in H_F$. Taking the determinant of this equation one arrives at the constraint $\mathrm{det}\left(U^\F_h\right)^2 = 1$. In the mass degenerate case i.e. $M= m \EFmat$, the unbroken subgroup is spanned by the generators of $\mathfrak{so}(2\nf)$. The isotropy condition \eqref{eq:invariance_condition_mass_term} can be translated to the level of Lie-Algebras 
    \begin{align} \label{eq: broken_generators_condition}
        \text{\scriptsize Broken $U(4)$ generators}
        \;\,\quad\quad\quad\quad
        T^{\F\top}_{a}\omega - \omega T^\F_{a} &= 0 
        \quad\quad\quad\quad
        a = (0),1,\dots, 9 \\
        \label{eq: unbroken_generators_condition}
        \text{\scriptsize Unbroken $U(4)$ generators}
        \quad\quad\quad\quad 
        T^{\F\top}_{A}\omega + \omega T^\F_{A} &= 0 
        \quad\quad\quad\quad
        A = 10,\dots,15.
    \end{align} 
    Here $A$ denotes the index of the unbroken and $a$ that of broken generators of the flavour algebra $\mathfrak{g}_F$. The zeroth index always refers to the generator defined by $\sqrt{4\nf} T_0^\F = \unity$, which generates the anomalously broken $U(1)_A$ component in $U(2\nf)$.  We may introduce the gauge invariant operators 
    \begin{equation}\label{eq:pseudo_scalar_bilinear}
        \mathcal{O}^{\mathrm{PS}}_a := \Psi^\top \ESmat^*  \ECmat^* \EFmat T_a^\F \Psi +  \Psi^\dagger \ESmat \ECmat \EFmat^* T_a^{\F*} \Psi^*,
    \end{equation}
    which help express (partial) conservation laws of the (broken) currents of the global flavour symmetries (PCBC-Relations). These are the analog of the PCAC-relations \cite{Scherer:2012xha} in real world QCD. 
    \begin{align}
    \partial_\mu j^\mu_A & = 0 \\
    \partial_\mu j^\mu_a & = - i\, m\, \mathcal{O}^{\mathrm{PS}}_a \\
    \partial_\mu j^\mu_0 & =  -i\,m\, \mathcal{O}^{\mathrm{PS}}_0 - \gd^2 T_\R \frac{\epsilon^{\mu\nu\rho\sigma}\delta_{\alpha\beta}}{16 \sqrt{2}\pi^2} \Fc_{\mu\nu}^\alpha \Fc_{\rho\sigma}^\beta. \quad\quad \label{eq:pcbc_eta}
    \end{align}
    Let us note that for non-degenerate fermions masses, the symmetry breaking pattern may be investigated in exactly the same way. The flavour symmetry is then generated from the algebra $\mathfrak{so}(2) \oplus \mathfrak{so}(2)$. The PCBC relations must be modified accordingly. 

\subsubsection{Spontaneous symmetry breaking}
    The order parameter may be defined via a quark condensate
    \begin{equation} \label{eq:definition_quark_condensate}
        \qcon := \Exv{\Psi^\dagger \ESmat\ECmat\EFmat\Psi^*} -  \Exv{\Psi^\top\ECmat\ESmat\EFmat\Psi} = 2\delta_{ij}\Exv{\overline{q}^{(i)}q^{(j)}}\,,
    \end{equation}
    who's isotropy group is the same as the degenerate mass term \eqref{eq:invariance_condition_mass_term}. Hence, the unbroken symmetries are exact symmetries of the quantum theory and the mass term, acting as a perturbation to the system in the chiral limit, allows to argue why we expect to see this specific breaking pattern. The Nambo-Goldstone theorem then tells us that we expect
    \begin{equation}
        \text{\#NGb's} = \mathrm{dim}\,\mathfrak{g}_F - \mathrm{dim}\,\mathfrak{h}_F \overset{\footnotesize \nf = 2}{=} 9
    \end{equation}
    pNGb's states in the theory, which are the lightest states in the theory if we are reasonably close to the chiral limit.
    
\subsubsection{Spatial parity}
    For Dirac fermions the spatial parity transformation may be represented by $\Par: q(t,\Vec{x}) \xmapsto{} \eta_P \gamma_0 q(t, -\Vec{x})$, with $\eta_P$ an arbitrary complex phase \cite{Coleman:2018mew}. It is possible to adapt a choice of $\eta_P = -i$ such that $\Par$ commutes with the flavour symmetries. This can be seen explicitly by expressing the action of parity in the Nambu-Gorkov basis
    \begin{equation} \label{eq:spatial_parity_nambu_gorkov_basis}
        \Par: \Psi(t, \Vec{x}) \xmapsto{} i \EFmat \ECmat \ESmat\, \Psi^*(t, -\Vec{x}).
    \end{equation}
    This also demonstrates a connection between spatial parity and the properties of so-called Riemann symmetric spaces, which will be very convenient later in the description of the low energy effective theory. A coset space $G_F/H_F$ is said to be symmetric if it is connected, compact and if the Lie-algebra $\mathfrak{g}_F$ decomposes according to $\mathfrak{g}_F = \mathfrak{h}_F \oplus \ks$, with $\ks$ being spanned by the broken generators, such that 
    \begin{equation}\label{eq:symmetric_lie_algebra_splitting}
        \combr{\mathfrak{h}_F,\mathfrak{h}_F} \subset \mathfrak{h}_F 
        \quad \quad
        \combr{\mathfrak{h}_F, \ks} \subset \ks 
        \quad \quad  
        \combr{\ks,\ks} \subset \mathfrak{h}_F.
    \end{equation}
    Due to this decomposition, such a space allows for an involutive Lie-algebra automorphism $\Nar: \mathfrak{g}_F \rightarrow \mathfrak{g}_F$ with positive eigenspace $\mathfrak{h}_F$ and negative eigenspace $\ks$. In case of $G_F = SU(2\nf)$ and $H_F = SO(2\nf)$, this automorphism is given explicitly via 
    \begin{equation} \label{eq:lie_algebra_involutive_automorphism}
        \forall \Af \in \mathfrak{su}(2\nf): \Nar(\Af) := -\EFmat^{-1} \Af^\top \EFmat  
    \end{equation}
    and will be dubbed ``naive parity''. To highlight the relation to spatial parity consider for example the flavour current composite field given in \eqref{eq:flavor_currents}.
    Using \eqref{eq:spatial_parity_nambu_gorkov_basis} and $\ESmat^\dagger \overline{\sigma}^\mu \ESmat = g^{\mu\mu}\overline{\sigma}^{\mu\top}$ we obtain
    \begin{equation*} \label{eq:parity_and_currents}
        \Psi^\dagger \overline{\sigma}^\mu T_N^F \Psi \xmapsto{\Par} 
        - \Psi^\dagger \left(\ESmat^\dagger \overline{\sigma}^\mu \ESmat\right)^\top \,\left(\EFmat^\dagger T_N^\F \EFmat\right)^\top\,\Psi = g^{\mu\mu}\Psi^\dagger \overline{\sigma}^\mu \Nar(T_N^F) \Psi
    \end{equation*}
    The result depends only on whether the index $N$ refers to an element of $\mathfrak{h}_F$ or $\ks$. 
    For (axial)vectors fields we can express  $\Af_\mu^N(t,\vec{x}) \xmapsto{\Par} (-)+g_{\mu\mu}\Af_\mu^N(t,-\vec{x})$. This can be more conveniently formulated by defining the connection 1-form $\Af = -i \Af_\mu^N T_N^F \mathrm{d}x^\mu$. The correct parity transformation depends on the index $N$ and is given by
    \begin{equation}
       \Par( \Af(t,\Vec{x})) = \Nar(\Af(t, -\Vec{x})).
    \end{equation}
    Extracting the coordinates again gives the correct transformation behaviour, where $\Nar$ determines the transformation of the flavour algebra index $N$ and $\mathrm{d}x^\mu \vert_{(t, -\Vec{x})} = g^{\mu\mu} \mathrm{d}x^\mu \vert_{(t, \Vec{x})}$ supplements the correct factors from changing the spatial argument. We can use this to define spatial (and naive) parity for any kind of $\liea{g}_F$-valued field e.g. the dark pions. Since the Lagrangian \eqref{eq:dark_strong_quark_lagrangian} and \eqref{eq:dark_strong_yang_mills_lagrangian} are invariant under spatial parity and, by virtue of the Vafa-Witten theorem \cite{Vafa:1984xg}, spatial parity can not be broken by quantum effects, parity is a good symmetry of our quantum theory. More importantly, since it commutes with the global symmetry $G_F$ due to our choice of $\eta_P$, we can classify physical states by their parity and flavour quantum numbers. 
    
\subsubsection{Charge conjugation} \label{sec:charge_conjugation}
    In the Nambu-Gorkov formulation, charge conjugation manifests as flavour symmetry
    \begin{equation}
        \CC: \Psi(t,\Vec{x}) \xmapsto{} \EFmat \Psi(t,\Vec{x}).
    \end{equation}
    This reflects the fact that dark quarks cannot be physically distinguished from dark anti-quarks in this theory. Since charge conjugation respects the isotropy condition specified in \eqref{eq:invariance_condition_mass_term}, leaves the Lagrangian invariant and 
    $\det \EFmat = (-1)^{\nf} = 1$  for $\nf = 2$, it is a good symmetry of quantum theory. However, since it manifests as a flavour symmetry it does not give us any new information.
    One might consider what happens if only a single Dirac fermion is charge conjugated. In principle, this should also be a symmetry of quantum theory, since dark quarks and anti-quarks are indistinguishable. If we agree to only charge conjugate $q^{(1)}$, the transformation manifest as a left-multiplication of $\Psi$ with the matrix 
    \begin{equation} \label{eq:cu_mat}
        C_u =  \left(\begin{matrix}0 & 0 & 1 & 0\\0 & 1 & 0 & 0\\1 & 0 & 0 & 0\\0 & 0 & 0 & 1\end{matrix}\right).
    \end{equation}
    Again, this matrix respects the isotropy condition \eqref{eq:invariance_condition_mass_term} and is a symmetry of the Lagrangian. However, it has negative determinant i.e. $\det{C_u} = -1$. If we assume that $\pepsilon^\F = -\ln{C_u}$ we obtain from \eqref{eq:anomaly_functional_axial_anomaly} that in instanton backgrounds with $\Qtopo[\Ac] = 1$  the following condition must hold in order for the transformation to be non-anomalous
    \begin{equation} \label{eq:set_of_non_anomalous_determinants}
        \det{C_u} \in \left\{e^{-i k \pi/T_\R} \;\Big\vert \; k = 0,1,\dots, 2 T_\R-1 \right\} \;\overset{T_\R = 1}{=}\; \left\{1,-1\right\}.
    \end{equation}
    This shows that this symmetry is not anomalous. Further, one observes that this action of charge conjugation does not commute with the rest of the flavours symmetries. Hence it does not seem to be useful to classify states via their charge conjugation quantum numbers. However, the symmetry enlarges the physically realised $SU(2\nf)$ chiral symmetry to $ \mathbb{Z}_{2}\ltimes SU(2\nf)$ and the unbroken flavour symmetry to $\mathbb{Z}_2 \ltimes SO(2\nf) \cong O(2\nf)$. The semi-direct product reflects the fact that the discrete symmetry does not commute with the rest of the flavour symmetries. While the pNGb states remain completely ignorant of this enlargement, the discrete transformations relate elements of the self-dual and anti-self-dual antisymmetric 2 index representation of $SO(4)$, causing the lightest vector mesons states to be mass-degenerate. 
    
\begin{figure}[]
        \centering
        \resizebox{\textwidth}{!}{\input{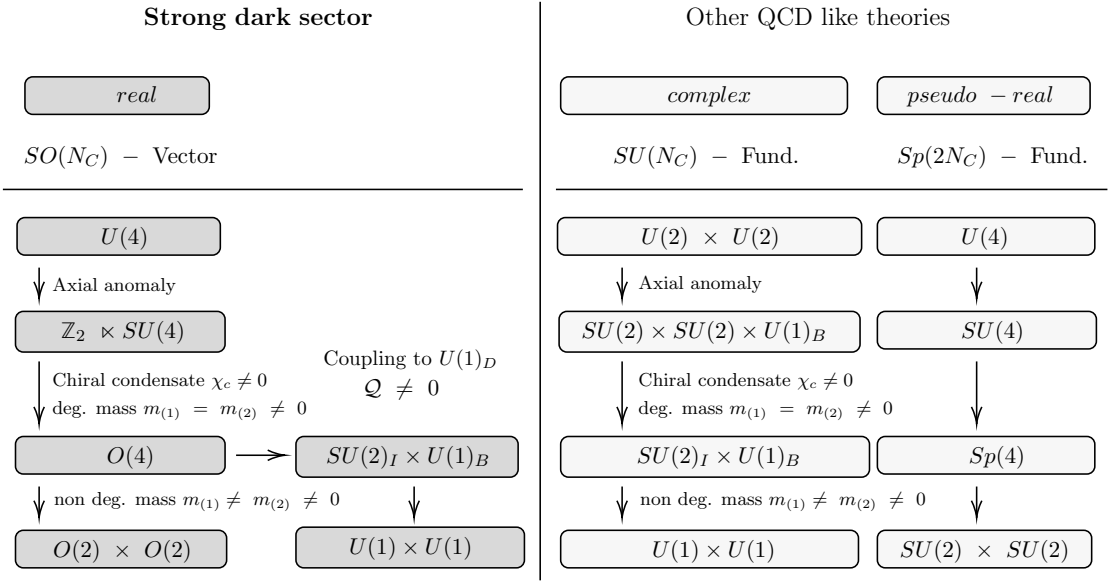}}
        \caption{Comparison of symmetry breaking patterns in QCD-like theories with two Dirac fermions i.e. $N_F = 2$. The main features of the patterns are determined by the gauge group representation being real, pseudo-real or complex. On the left: The breaking pattern for the dark sector considered here for Dirac fermions gauged under $SO(\nc)$-vector representation. On the right: 2-flavour QCD and a dark $Sp(2\nc)$ theory with two fundamental Dirac fermions discussed in \cite{Kulkarni:2022bvh}. The explicit breaking via charge assignments $\Q$ is discussed in section \ref{sec:charge_assignments}. }
        \label{fig:comparison_of_symmetries_in_fundamental_two_flavor_theories}
    \end{figure}

    In principle $T_\R > 1$ can hold for higher tensor representation, leading to the appearance of larger discrete symmetries. These nevertheless are dynamically broken by the chiral condensate. Their precise structure and potential phenomenological consequences will be discussed in section \ref{sec:generalizations}. A summary of the symmetries of the strong dark sector in isolation can be found in figure \ref{fig:comparison_of_symmetries_in_fundamental_two_flavor_theories}. 

\subsection{The dark photon} \label{sec:dark_photon_in_uv}
    As a mediator between the strong dark sector and the Standard Model (SM) we consider a massive dark photon \cite{Fabbrichesi:2020wbt, Holdom:1985ag}, which is implemented by a $U(1)_D$ gauge field $\zp_\mu$. The mass of the particle is provided by an abelian Brout-Englert-Higgs effect, triggered by an additional $U(1)_D$ scalar field $\varphi_D$. In total this allows for three new parameters of the theory. The dark charge $\ed$ and two parameters in the potential of the scalar field. However, the latter two can be varied independently to set the mass $\mzp$ of the dark photon and the mass of the scalar field. Thus, we take $\mzp$ as a free parameter of the theory. For the coupling to the SM we consider a kinetic mixing portal 
    \begin{equation}\label{eq:dark_kinetic_mixing}
        \mathcal{L}_\mathrm{mix} = \frac{\cmix}{\cos{\Omega_W}} Z^{\prime\,\mu\nu}B^\mathrm{SM}_{\mu\nu}
    \end{equation}
    with $Z^{\prime\,\mu\nu}$ and $B^\mathrm{SM}_{\mu\nu}$ the field strength tensors of the dark photon and SM Hypercharge. The parameter $\Omega_W$ denotes the Weinberg mixing angle and $\cmix$ is a real constant parametrizing the strength of the kinetic mixing. 

\subsubsection{Charge assignments} \label{sec:charge_assignments}
    The simplest way to couple the dark photon to the dark fermions is by gauging a suitable 1-parameter subgroup of the flavour symmetry $G_F$. This adds a coupling term between the dark photon and the dark electromagnetic current to the Lagrangian 
    \begin{equation}
        \mathcal{L}_{\Psi \zp} = -i \ed \Psi^\dagger \overline{\sigma}^\mu \Q \Psi \zp_\mu \,,
    \end{equation}
    which explicitly breaks the global $O(2\nf)$ symmetry. The charge assignment matrix $\Q$ is determined by the generator of the gauged 1-parameter subgroup of the flavour symmetry.     
    Since we consider vector-like dark quarks, one can only consider gauging part of the unbroken subgroup $H_F$. As a side-effect, we obtain that the $U(1)_D$ is consistent i.e. we do not have to worry about $[U(1)_D]^3$ triangle gauge anomalies as $H_F$  is anomaly free embedded in $G_F$. 
    One way to choose the charge assignments was presented in \cite{Hochberg:2015vrg}.  There the authors use the fact that $SO(2\nf) $ contains a  $U(\nf)\cong U(1)_B\times SU(\nf)_I$ subgroup. Gauging the $U(1)_B$ generator ensures that the pions still transform under a non-abelian $SU(\nf)_I$ symmetry. We choose this $U(1)_B$ generator to be the charge matrix which in the Nambu-Gorkov basis is given as 
    \begin{equation*}
        \Q = \frac{1}{\sqrt{4\nf}} \,\mathrm{diag}(\underbrace{1, \cdots}_{\nf},\underbrace{-1, \cdots}_{\nf}).
    \end{equation*}
    The reminiscent $SU(N_F)_I \subset U(\nf)$ global flavour symmetry prevents the dark pions from decaying into the SM. In the case of $N_F = 2$, the non-abelian symmetry $SU(\nf)_I$ acts on the Dirac quarks in the same way as Isospin in standard QCD. 
    This can best be seen by using $SO(4) \cong SU(2)_I\times SU(2)_B$, where the left symmetry acts on the flavour indices of the Dirac fermions $q^{(j)}$ as a left multiplication with an $SU(2)_I$ matrix in the fundamental representation. The 1-parameter subgroup generated by $\Q$ in $SU(2)_B$ acts analogous to the Baryon number symmetry in QCD, when translated back to the Dirac formulation. Thus, the above charge assignment corresponds to charging Baryon number symmetry, leaving Isospin unbroken. This interpretation was also adopted in \cite{DeGrand:2015lna}, providing more details on the action of $SU(2)_I$ on the two Dirac fermions. Finally let us comment on the uniqueness of this assignment. In order to guarantee the stability of the dark pions, one has to look for a charge assignment such that the pion currents are free of anomalies. This can be guaranteed demanding that the charge assignment satisfies 
    \begin{equation} \label{eq:anomaly_cancelation_condition}
        \Q^2 \propto \unity.
    \end{equation}
    Due to $\Q$ being traceless\footnote{The traceless property is also required to avoid gravitational anomalies.}, this condition strongly restricts the eigenvalues of the charge assignment and in fact renders above unique charge assignment up to a change of basis. 
    
    For general $\nf$ the pions split into a charged and a neutral multiplet under $SU(\nf)_I$, furnishing the symmetric and the adjoint representation of $SU(\nf)_I$ \cite{Hochberg:2015vrg}. A convenient choice of $SU(2\nf)$ generators, compatible with all these symmetry structures is presented in appendix \ref{sec:generators_of_su2nf}. Any kind of $U(1)_D$ charge assignment will always break the discrete $\mathbb{Z}_{2}$ symmetry explicitly.

\subsection{Light dark mesons states} \label{sec:dark_particles}
    For dark matter phenomenology we identify the pNGbs of the spontaneously broken (approximate) global chiral symmetry which are the lightest states in the physical spectrum that dominate the low energy behaviour of the theory as dark matter candidates. However, it has been shown that the interesting domain for dark matter phenomenology in parameters space prefers large number of colour degrees of freedom \cite{Hochberg:2014kqa} and typically lies close to region where other states e.g. vector mesons, become important for phenomenology. In the following we classify all these states with respect to parity and their flavour multiplet structure. The flavour symmetry for an isolated dark sector is given by $O(2\nf)$. After coupling to the dark photon the global symmetry is $SU(\nf)_I \times U(1)_B$. The representations of $U(1)_B$ may be used to classify the charge assignments under the $U(1)_D$ gauge symmetry. 

    \begin{figure}[]
        \centering
        \resizebox{0.7\textwidth}{!}{\tikzset{every picture/.style={line width=0.75pt}} 

\begin{tikzpicture}[x=0.75pt,y=0.75pt,yscale=-1,xscale=1]

\draw   (216,184) .. controls (216,181.79) and (217.79,180) .. (220,180) -- (292,180) .. controls (294.21,180) and (296,181.79) .. (296,184) -- (296,196) .. controls (296,198.21) and (294.21,200) .. (292,200) -- (220,200) .. controls (217.79,200) and (216,198.21) .. (216,196) -- cycle ;
\draw  [fill={rgb, 255:red, 247; green, 247; blue, 247 }  ,fill opacity=1 ] (216,214) .. controls (216,211.79) and (217.79,210) .. (220,210) -- (232,210) .. controls (234.21,210) and (236,211.79) .. (236,214) -- (236,226) .. controls (236,228.21) and (234.21,230) .. (232,230) -- (220,230) .. controls (217.79,230) and (216,228.21) .. (216,226) -- cycle ;
\draw  [fill={rgb, 255:red, 215; green, 215; blue, 215 }  ,fill opacity=1 ] (246,214) .. controls (246,211.79) and (247.79,210) .. (250,210) -- (262,210) .. controls (264.21,210) and (266,211.79) .. (266,214) -- (266,226) .. controls (266,228.21) and (264.21,230) .. (262,230) -- (250,230) .. controls (247.79,230) and (246,228.21) .. (246,226) -- cycle ;
\draw  [fill={rgb, 255:red, 167; green, 167; blue, 167 }  ,fill opacity=1 ] (276,214) .. controls (276,211.79) and (277.79,210) .. (280,210) -- (292,210) .. controls (294.21,210) and (296,211.79) .. (296,214) -- (296,226) .. controls (296,228.21) and (294.21,230) .. (292,230) -- (280,230) .. controls (277.79,230) and (276,228.21) .. (276,226) -- cycle ;
\draw   (26,66) .. controls (26,51.64) and (37.64,40) .. (52,40) -- (140,40) .. controls (154.36,40) and (166,51.64) .. (166,66) -- (166,144) .. controls (166,158.36) and (154.36,170) .. (140,170) -- (52,170) .. controls (37.64,170) and (26,158.36) .. (26,144) -- cycle ;
\draw   (76,188) .. controls (76,183.58) and (79.58,180) .. (84,180) -- (108,180) .. controls (112.42,180) and (116,183.58) .. (116,188) -- (116,212) .. controls (116,216.42) and (112.42,220) .. (108,220) -- (84,220) .. controls (79.58,220) and (76,216.42) .. (76,212) -- cycle ;
\draw  [fill={rgb, 255:red, 217; green, 217; blue, 217 }  ,fill opacity=1 ] (36,56) .. controls (36,52.69) and (38.69,50) .. (42,50) -- (150,50) .. controls (153.31,50) and (156,52.69) .. (156,56) -- (156,74) .. controls (156,77.31) and (153.31,80) .. (150,80) -- (42,80) .. controls (38.69,80) and (36,77.31) .. (36,74) -- cycle ;
\draw  [fill={rgb, 255:red, 160; green, 160; blue, 160 }  ,fill opacity=1 ] (36,96) .. controls (36,92.69) and (38.69,90) .. (42,90) -- (150,90) .. controls (153.31,90) and (156,92.69) .. (156,96) -- (156,114) .. controls (156,117.31) and (153.31,120) .. (150,120) -- (42,120) .. controls (38.69,120) and (36,117.31) .. (36,114) -- cycle ;
\draw  [fill={rgb, 255:red, 247; green, 247; blue, 247 }  ,fill opacity=1 ] (36,136) .. controls (36,132.69) and (38.69,130) .. (42,130) -- (150,130) .. controls (153.31,130) and (156,132.69) .. (156,136) -- (156,154) .. controls (156,157.31) and (153.31,160) .. (150,160) -- (42,160) .. controls (38.69,160) and (36,157.31) .. (36,154) -- cycle ;
\draw  [fill={rgb, 255:red, 217; green, 217; blue, 217 }  ,fill opacity=1 ] (81,191) .. controls (81,187.69) and (83.69,185) .. (87,185) -- (105,185) .. controls (108.31,185) and (111,187.69) .. (111,191) -- (111,209) .. controls (111,212.31) and (108.31,215) .. (105,215) -- (87,215) .. controls (83.69,215) and (81,212.31) .. (81,209) -- cycle ;
\draw   (206,66) .. controls (206,51.64) and (217.64,40) .. (232,40) -- (340,40) .. controls (354.36,40) and (366,51.64) .. (366,66) -- (366,144) .. controls (366,158.36) and (354.36,170) .. (340,170) -- (232,170) .. controls (217.64,170) and (206,158.36) .. (206,144) -- cycle ;
\draw  [fill={rgb, 255:red, 217; green, 217; blue, 217 }  ,fill opacity=1 ] (216,56) .. controls (216,52.69) and (218.69,50) .. (222,50) -- (350,50) .. controls (353.31,50) and (356,52.69) .. (356,56) -- (356,74) .. controls (356,77.31) and (353.31,80) .. (350,80) -- (222,80) .. controls (218.69,80) and (216,77.31) .. (216,74) -- cycle ;
\draw  [fill={rgb, 255:red, 160; green, 160; blue, 160 }  ,fill opacity=1 ] (216,136) .. controls (216,132.69) and (218.69,130) .. (222,130) -- (250,130) .. controls (253.31,130) and (256,132.69) .. (256,136) -- (256,154) .. controls (256,157.31) and (253.31,160) .. (250,160) -- (222,160) .. controls (218.69,160) and (216,157.31) .. (216,154) -- cycle ;
\draw  [fill={rgb, 255:red, 247; green, 247; blue, 247 }  ,fill opacity=1 ] (316,136) .. controls (316,132.69) and (318.69,130) .. (322,130) -- (350,130) .. controls (353.31,130) and (356,132.69) .. (356,136) -- (356,154) .. controls (356,157.31) and (353.31,160) .. (350,160) -- (322,160) .. controls (318.69,160) and (316,157.31) .. (316,154) -- cycle ;
\draw  [fill={rgb, 255:red, 217; green, 217; blue, 217 }  ,fill opacity=1 ] (266,96) .. controls (266,92.69) and (268.69,90) .. (272,90) -- (300,90) .. controls (303.31,90) and (306,92.69) .. (306,96) -- (306,114) .. controls (306,117.31) and (303.31,120) .. (300,120) -- (272,120) .. controls (268.69,120) and (266,117.31) .. (266,114) -- cycle ;

\draw (16,12) node [anchor=north west][inner sep=0.75pt]   [align=left] {\textbf{pseudo-scalar mesons}};
\draw (228,12) node [anchor=north west][inner sep=0.75pt]   [align=left] {\textbf{vector mesons}};
\draw (307,184.4) node [anchor=north west][inner sep=0.75pt]  [font=\footnotesize]  {$O( 4) \ \mathrm{Classification}$};
\draw (307,215.4) node [anchor=north west][inner sep=0.75pt]  [font=\footnotesize]  {$SU_{I}( 2) \times U_{B}( 1) \ \mathrm{Classification}$};
\draw (89,190.4) node [anchor=north west][inner sep=0.75pt]    {$\eta '$};
\draw (47,56.4) node [anchor=north west][inner sep=0.75pt]    {$\tilde{\pi }^{1}$};
\draw (86,56.4) node [anchor=north west][inner sep=0.75pt]    {$\tilde{\pi }^{2}$};
\draw (125,56.4) node [anchor=north west][inner sep=0.75pt]    {$\tilde{\pi }^{3}$};
\draw (47,97.4) node [anchor=north west][inner sep=0.75pt]    {$\tilde{\pi }^{4}$};
\draw (86,97.4) node [anchor=north west][inner sep=0.75pt]    {$\tilde{\pi }^{5}$};
\draw (125,97.4) node [anchor=north west][inner sep=0.75pt]    {$\tilde{\pi }^{6}$};
\draw (47,137.4) node [anchor=north west][inner sep=0.75pt]    {$\tilde{\pi }^{7}$};
\draw (86,137.4) node [anchor=north west][inner sep=0.75pt]    {$\tilde{\pi }^{8}$};
\draw (125,137.4) node [anchor=north west][inner sep=0.75pt]    {$\tilde{\pi }^{9}$};
\draw (226,57.4) node [anchor=north west][inner sep=0.75pt]    {$\tilde{\omega }^{10}$};
\draw (277,58.4) node [anchor=north west][inner sep=0.75pt]    {$\tilde{\omega }^{11}$};
\draw (323,58.4) node [anchor=north west][inner sep=0.75pt]    {$\tilde{\omega }^{12}$};
\draw (226,136.4) node [anchor=north west][inner sep=0.75pt]    {$\tilde{\rho }^{10}$};
\draw (323,137.4) node [anchor=north west][inner sep=0.75pt]    {$\tilde{\rho }^{12}$};
\draw (276,96.4) node [anchor=north west][inner sep=0.75pt]    {$\tilde{\rho }^{13}$};
\draw (219,213.4) node [anchor=north west][inner sep=0.75pt]    {$+$};
\draw (279,213.4) node [anchor=north west][inner sep=0.75pt]    {$-$};
\draw (250,214.4) node [anchor=north west][inner sep=0.75pt]    {$0$};

\end{tikzpicture}}
        \caption{Classification of all light states relevant for DM phenomenology with respect to parity and the global symmetries $O(4)$ and $SU(2)_I\times U(1)_B$. The gray scale indicates the charge of the particles under $U(1)_D$ within an isospin multiplet. The states are denoted in the eigenbasis of the charge operator $\Q$}
        \label{fig:classification_of_particles}
    \end{figure}
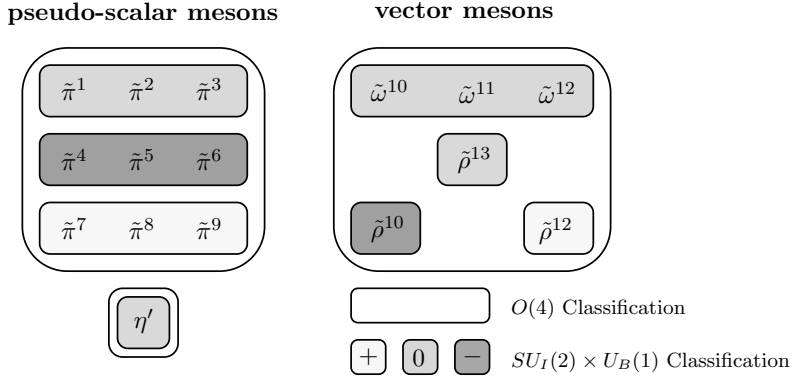

\subsubsection{Pseudo-scalar mesons}
    The vacuum expectation values of the commutator of the pseudo-scalar operators $\mathcal{O}_a^\mathrm{PS}$ in \eqref{eq:pseudo_scalar_bilinear} with the chiral charge operators associated to the broken symmetries turn out to be proportional to the chiral condensate $\qcon$. For $\nf = 2$ this indicates the presence of ten states in the Nambu-Goldstone phase of the theory \cite{Pich:2018ltt}, of which nine may be identified as the pNGb's of the symmetry breaking pattern $\liea{su}(4)\rightarrow \liea{so}(4)$. In analogy to QCD we denote these as dark pions $\pid_a$.  
    
    The tenth state, corresponding to $\mathcal{O}_0^\mathrm{PS}$, is related to the anomalous $U(1)_A$ and remains massive even in the chiral limit. This is analogous to QCD and can be seen from the PCBC relation \eqref{eq:pcbc_eta} in which the axial anomaly sources non-conservation of the associated current $j_0^\mu$. Hence, we expect this particle to be heavier than the dark pions in general. The precise mass, and mass splitting to the pions, needs to be calculated with the help of non-perturbative methods e.g. lattice field theory or functional methods.
    Nevertheless, in contrast to real world QCD, the relative mass splitting $\Delta m_\etap^2 / m_\pid^2$ between $\etap$ and $\pid$ might be small for mass-degenerate dark quarks\footnote{Even for mass non-degenerate dark quarks, these arguments should hold, since the mass-splitting should be small in order to make the dark pions sufficiently meta-stable.} and large $\nc$ arguments may apply, suppressing the gluonic contribution to the $\etap$ mass. The first means that a contribution from a heavy strange-quark-like state is absent, while the latter amounts to $\etap$ being an effective tenth pNGb state\footnote{The full $U(2\nf)$ can not be expected to be restored in the large $\nc$ limit because axial anomaly \eqref{eq:anomaly_functional_axial_anomaly} is not affected by the large $\nc$ limit. However, the topological charge density in the local current operator equation \eqref{eq: eta_pcac_in_large_nc} may be effectively vanishing in this limit. Since the operator identities for $\mathcal{O}_0^\mathrm{PS}$ with the currents and chiral condensate are identical in structure to the rest of the pNGb's, the only difference is the non-conservation of the current in the first place. If this contribution is suppressed in the large $\nc$ limit, a treatment as an effective pNGB appears valid.} in an appropriate large $\nc$ limit, explained further below. 
    
    Note that operators $\mathcal{O}_a^\mathrm{PS}$ are all hermitian and hence not all of them can have a defined charge under $U(1)_D$, since not all dark pions are neutral. For some calculations it is useful to adopt a basis $\Tilde{\pid}_a$ of dark pion states that are also eigenstates of the charge assignment operators $\Q$. In the basis chosen\footnote{See appendix \ref{sec:generators_of_su2nf} for more details.} this can be achieved by the following complex linear combination 
    \small
    \begin{align}
        \left(\begin{array}{c}
            \ket{\pic_1(p)} \\
            \ket{\pic_2(p)} \\
            \ket{\pic_3(p)} \\
            \ket{\pic_4(p)} \\
            \ket{\pic_5(p)} \\
            \ket{\pic_6(p)} \\
            \ket{\pic_7(p)} \\
            \ket{\pic_8(p)} \\
            \ket{\pic_9(p)} \\
        \end{array}\right)
        =    \frac{1}{\sqrt{2}}
        \left(
            \begin{array}{ccccccccc}
             \sqrt{2} & 0 & 0 & 0 & 0 & 0 & 0 & 0 & 0 \\
             0 & \sqrt{2} & 0 & 0 & 0 & 0 & 0 & 0 & 0 \\
             0 & 0 & \sqrt{2} & 0 & 0 & 0 & 0 & 0 & 0 \\
             0 & 0 & 0 & \,\,1\,\, & 0 & 0 & -i & 0 & 0 \\
             0 & 0 & 0 & 0 & \,\,1\,\, & 0 & 0 & -i & 0 \\
             0 & 0 & 0 & 0 & 0 & \,\,1\,\, & 0 & 0 & -i \\
             0 & 0 & 0 & \,\,1\,\, & 0 & 0 & i & 0 & 0 \\
             0 & 0 & 0 & 0 & \,\,1\,\, & 0 & 0 & i & 0 \\
             0 & 0 & 0 & 0 & 0 & \,\,1\,\, & 0 & 0 & i \\
            \end{array}
            \right)
        \left(\begin{array}{c}
            \ket{\pid_1(p)} \\
            \ket{\pid_2(p)} \\
            \ket{\pid_3(p)} \\
            \ket{\pid_4(p)} \\
            \ket{\pid_5(p)} \\
            \ket{\pid_6(p)} \\
            \ket{\pid_7(p)} \\
            \ket{\pid_8(p)} \\
            \ket{\pid_9(p)} \\
        \end{array}\right).
    \end{align}
    \normalsize
    The normalisation of the matrix is chosen such that the matrix preserves the normalisation of the pion states. The $\etap$ state is neutral and hence already a charge eigenstate.
    
\begin{figure}[]
        \centering
        \resizebox{0.65\textwidth}{!}{\tikzset{every picture/.style={line width=0.75pt}} 

\begin{tikzpicture}[x=0.75pt,y=0.75pt,yscale=-1,xscale=1]

\draw   (180,84) .. controls (180,81.79) and (181.79,80) .. (184,80) -- (216,80) .. controls (218.21,80) and (220,81.79) .. (220,84) -- (220,96) .. controls (220,98.21) and (218.21,100) .. (216,100) -- (184,100) .. controls (181.79,100) and (180,98.21) .. (180,96) -- cycle ;
\draw   (177,19) .. controls (177,14.58) and (180.58,11) .. (185,11) -- (209,11) .. controls (213.42,11) and (217,14.58) .. (217,19) -- (217,43) .. controls (217,47.42) and (213.42,51) .. (209,51) -- (185,51) .. controls (180.58,51) and (177,47.42) .. (177,43) -- cycle ;
\draw  [draw opacity=0][fill={rgb, 255:red, 217; green, 217; blue, 217 }  ,fill opacity=1 ] (182,22) .. controls (182,18.69) and (184.69,16) .. (188,16) -- (206,16) .. controls (209.31,16) and (212,18.69) .. (212,22) -- (212,40) .. controls (212,43.31) and (209.31,46) .. (206,46) -- (188,46) .. controls (184.69,46) and (182,43.31) .. (182,40) -- cycle ;
\draw   (10,38) .. controls (10,22.54) and (22.54,10) .. (38,10) -- (132,10) .. controls (147.46,10) and (160,22.54) .. (160,38) -- (160,122) .. controls (160,137.46) and (147.46,150) .. (132,150) -- (38,150) .. controls (22.54,150) and (10,137.46) .. (10,122) -- cycle ;
\draw  [draw opacity=0][fill={rgb, 255:red, 217; green, 217; blue, 217 }  ,fill opacity=1 ] (26,26.8) .. controls (26,23.04) and (29.04,20) .. (32.8,20) -- (53.2,20) .. controls (56.96,20) and (60,23.04) .. (60,26.8) -- (60,47.39) .. controls (60,51.14) and (56.96,54.19) .. (53.2,54.19) -- (32.8,54.19) .. controls (29.04,54.19) and (26,51.14) .. (26,47.39) -- cycle ;
\draw  [draw opacity=0][fill={rgb, 255:red, 217; green, 217; blue, 217 }  ,fill opacity=1 ] (26,66.8) .. controls (26,63.04) and (29.04,60) .. (32.8,60) -- (53.2,60) .. controls (56.96,60) and (60,63.04) .. (60,66.8) -- (60,133.2) .. controls (60,136.96) and (56.96,140) .. (53.2,140) -- (32.8,140) .. controls (29.04,140) and (26,136.96) .. (26,133.2) -- cycle ;
\draw  [draw opacity=0][fill={rgb, 255:red, 217; green, 217; blue, 217 }  ,fill opacity=1 ] (70,66.8) .. controls (70,63.04) and (73.04,60) .. (76.8,60) -- (97.2,60) .. controls (100.96,60) and (104,63.04) .. (104,66.8) -- (104,133.2) .. controls (104,136.96) and (100.96,140) .. (97.2,140) -- (76.8,140) .. controls (73.04,140) and (70,136.96) .. (70,133.2) -- cycle ;
\draw  [draw opacity=0][fill={rgb, 255:red, 217; green, 217; blue, 217 }  ,fill opacity=1 ] (110,26.8) .. controls (110,23.04) and (113.04,20) .. (116.8,20) -- (137.2,20) .. controls (140.96,20) and (144,23.04) .. (144,26.8) -- (144,133.2) .. controls (144,136.96) and (140.96,140) .. (137.2,140) -- (116.8,140) .. controls (113.04,140) and (110,136.96) .. (110,133.2) -- cycle ;
\draw  [draw opacity=0][fill={rgb, 255:red, 217; green, 217; blue, 217 }  ,fill opacity=1 ] (70,26.84) .. controls (70,23.06) and (73.06,20) .. (76.84,20) -- (133.16,20) .. controls (136.94,20) and (140,23.06) .. (140,26.84) -- (140,47.35) .. controls (140,51.13) and (136.94,54.19) .. (133.16,54.19) -- (76.84,54.19) .. controls (73.06,54.19) and (70,51.13) .. (70,47.35) -- cycle ;
\draw  [draw opacity=0][fill={rgb, 255:red, 217; green, 217; blue, 217 }  ,fill opacity=1 ] (180,114) .. controls (180,111.79) and (181.79,110) .. (184,110) -- (216,110) .. controls (218.21,110) and (220,111.79) .. (220,114) -- (220,126) .. controls (220,128.21) and (218.21,130) .. (216,130) -- (184,130) .. controls (181.79,130) and (180,128.21) .. (180,126) -- cycle ;

\draw (231,86.4) node [anchor=north west][inner sep=0.75pt]  [font=\footnotesize]  {$O( 4) \ \mathrm{Classification}$};
\draw (231,116.4) node [anchor=north west][inner sep=0.75pt]  [font=\footnotesize]  {$O( 2) \times O( 2) \ \mathrm{Classification}$};
\draw (190,22.4) node [anchor=north west][inner sep=0.75pt]    {$\eta '$};
\draw (35,28.4) node [anchor=north west][inner sep=0.75pt]    {$\tilde{\pi }^{1}$};
\draw (76,28.59) node [anchor=north west][inner sep=0.75pt]    {$\tilde{\pi }^{2}$};
\draw (115,28.59) node [anchor=north west][inner sep=0.75pt]    {$\tilde{\pi }^{3}$};
\draw (35,71.4) node [anchor=north west][inner sep=0.75pt]    {$\tilde{\pi }^{4}$};
\draw (77,71.59) node [anchor=north west][inner sep=0.75pt]    {$\tilde{\pi }^{5}$};
\draw (118,69.59) node [anchor=north west][inner sep=0.75pt]    {$\tilde{\pi }^{6}$};
\draw (34,111.59) node [anchor=north west][inner sep=0.75pt]    {$\tilde{\pi }^{7}$};
\draw (77,111.59) node [anchor=north west][inner sep=0.75pt]    {$\tilde{\pi }^{8}$};
\draw (118,111.59) node [anchor=north west][inner sep=0.75pt]    {$\tilde{\pi }^{9}$};

\end{tikzpicture}}
        \caption{Classification of the pseudo scalar mesons in presence of an explicit mass-split $m_2-m_1 = \Delta m$ of the quark current masses.}
        \label{fig:classification_of_mass_split_pions}
    \end{figure}
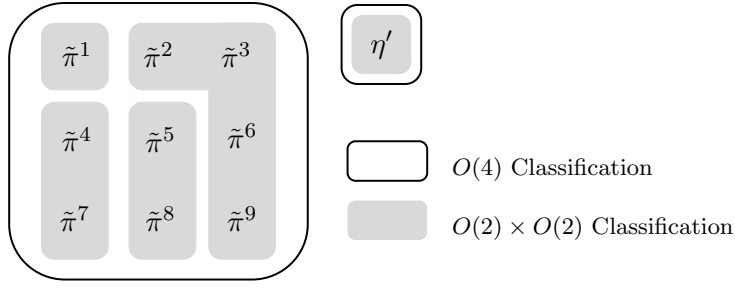

    In the case of an explicit mass splitting $m_1-m_2 = \Delta m$ the dark pions in isolation arrange in multiplets under $O(2)\times O(2)$, as summarised in figure \ref{fig:classification_of_mass_split_pions}. The singlet dark pion is not protected by any flavour symmetry and hence may decay in presence of mediator. In order to avoid problems with dark matter stability, we focus on the mass degenerate case.  
    
\subsubsection{Large $\nc$ considerations for $\etap$}
    While the existence of $\etap$ in our setup has previously been established in section \ref{sec:dark_particles}, whether it will ever become light enough to matter for phenomenological purposes is unclear. Such investigations can be performed on lattice, however it is out of scope for our current work. We would instead like to develop an expectation about whether $\etap$ can become light using perturbative arguments. 

    Such approaches have been used in analysing real world QCD theories. In that case, in the 't Hooft large $\nc$ limit \cite{tHooft:1973alw}, the contributions of the axial anomaly \eqref{eq:pcbc_eta} are suppressed by a factor $1/\nc$ \cite{Witten:1979vv} and hence the $\eta'$ state in QCD becomes massless in the chiral limit for $\nc\rightarrow \infty$. Large $\nc$ considerations have been useful to investigate potentially non-perturbative features of QCD, which are not accessible in a small-coupling perturbative approach \cite{Lucini:2012gg}. However, results like quark loop suppression leading to a geometric classification of classes of diagrams 
    heavily depend on the fact that the quarks transform in the fundamental representation of $SU(\nc)$. 
    
    The main argument towards this is to understand whether the second term representing gluodynamic contribution in \eqref{eq:pcbc_eta} can become arbitrarily small in large $\nc$ limit. This is however a non-trivial question given that the running of $\gd$ depends on $\nc$. Similar to the original discussion by 't Hooft, we resort to writing $\gd$ in terms of $\lambda = \beta_0 \gd^2$ and subsequently \eqref{eq:pcbc_eta} becomes 
    \begin{equation} \label{eq: eta_pcac_in_large_nc}
        \partial_\mu j^\mu_0 =  - i\,m\, \mathcal{O}^{\mathrm{PS}}_0 - \frac{T_\R}{\beta_0}\;\lambda\,\frac{\epsilon^{\mu\nu\rho\sigma}\delta_{\alpha\beta}}{16\sqrt{2}\pi^2} \Fc_{\mu\nu}^\alpha \Fc_{\rho\sigma}^\beta .
    \end{equation}
    Here $\beta_0$ (and $\beta_1$ below) denote the renormalisation scheme independent one- (and two-) loop coefficients \cite{Sannino:2009aw} of the $\beta$-function for the strong dark coupling $\gd$. Eqn. \eqref{eq: eta_pcac_in_large_nc} allows to analyse the large $\nc$ behaviour in terms of $T_\R/\beta_0\;\lambda$. The value of $\lambda$ is determined  by the renormalisation group equation from an initial value $\lambda_0$ at a UV cutoff 
    \begin{equation}\label{eq:beta_function_lambda}
        \beta(\lambda) = -\frac{2}{(4\pi)^2} \lambda^2 - \frac{2}{(4\pi)^5} \frac{\beta_1}{\beta_0^2}\lambda^4 + \dots \, ,
    \end{equation}
    where dots denote higher-loop contributions.
    For an asymptotically free threoy in absence of Banks-Zaks fixed point, the coefficient $\beta_1/\beta^2_0$ in \eqref{eq:beta_function_lambda} becomes a constant in the large $\nc$ limit and thus the running of $\lambda$ does not have any additional $\nc$ dependence up to two loops. Its value can thus be considered to be almost $\nc$ independent for sufficiently large number of colours. Given the explicit expression of $\beta_0$, for the second term in \eqref{eq: eta_pcac_in_large_nc} to vanish in large $\nc$ limit, 
    \begin{equation} \label{eq:large_nc_criterion}
        \frac{T_\R}{c_{\mathrm{adj}}} \xrightarrow[\nc\rightarrow\infty]{} 0
    \end{equation}
    is necessary. This criterion can be checked on purely representation theoretical grounds. Table \ref{tab:properties of representation} in the appendix shows that only the fundamental representations of the classical groups feature a light dark $\etap$ state in the large $\nc$ limit. Luckily, a lot of the standard treatments from large $\nc$ real world QCD remain valid for these theories. Essentially, all techniques and results for the lowest order expansion in terms of $1/\nc$ can be assumed to remain valid also for (pseudo-)real theories. This can be understood by the following argument. The fact that the fermions transform in the fundamental (or vector) representation of the gauge group, allows a geometric classification of Feynman diagrams. The difference between the complex and the real case occurs due to the additional reality condition imposed on the colour matrices. For the complex case only oriented geometries are allowed, while for the real case there may also be non-orientable geometric structures. However, such are typical higher genus surfaces and thus contribute only to higher order in $1/\nc$ \cite{tHooft:2002ufq}.

\subsubsection{Vector mesons}
    For the region $\mpi / \fpi > 4$, which is required for these models to successfully address the dark matter problem \cite{Hochberg:2014kqa}, the vector mesons are expected to be close to the two pion threshold $\mv \approx 2\mpi$. This is important since for $\mv < 2\mpi$, the vector mesons are stable in the isolated theory. When coupled to the SM, they can decay via the dark portal and hence take part in the cosmic depletion process \cite{Bernreuther:2023kcg, Bernreuther:2019pfb, Berlin:2018tvf, Choi:2018iit}. Adding vector mesons may also help improve predictability of the low energy effective theory for $\mpi / \fpi \approx 4\pi$ \cite{Harada:2003jx, Choi:2018iit}.
    
    A full classification of these states in the case of $\nf = 2$ can be found in figure \ref{fig:classification_of_particles}. We note that the parameters $\mv$, $\mpi$ and $\fpi$ are not independent, but are related by the underlying UV theory. Thus, $\mv$ should be determined for example as a function of $\mpi$ and $\fpi$ by the use of lattice studies. Bilinear interpolating operators, with a significant overlap with the vector meson states are provided in appendix \ref{sec:interpolating_linear_operators}. These are useful for investigations using non-perturbative techniques.

\section{Long range description}
    \label{sec:long_range_description}
        We turn towards the low energy effective description of the relevant degrees of freedom discussed in the previous section. The Lagrangian of vector mesons and pions is constructed via the hidden local symmetry (HLS) \cite{Bando:1987br, Harada:2003jx, Georgi:1989xy, Casalbuoni:1986vq, Casalbuoni:1985kq, Casalbuoni:1988xm, Appelquist:1999dq}. This approach was shown to be equivalent to many other approaches at the level of on-shell tree-level amplitudes, but has the advantageous feature of allowing a well-defined derivative expansion of the effective Lagrangian \cite{Harada:2003jx}. This allows a consistent truncation of the low energy theory. Especially, when fixing the HLS gauge, the model is equivalent to the non-linear $\Sigma$-model, which we will refer to as Callan-Coleman-Wess-Zumino (CCWZ) model \cite{Callan:1969sn,Coleman:1969sm}. We note here that it is also possible to include axial-vectors within the  generalised HLS formalism \cite{Bando:1987br}. However, we do not focus on them here as they are heavier than vector mesons by about a factor of $\sqrt{2}$ \cite{Hietanen:2012sz}.  Taking also into account large $\nc$ arguments we will consistently include the $\etap$ meson into the effective theory. The dark photon is introduced by gauging part of the unbroken flavour symmetry, exactly in the same way as it was done in the UV. Furthermore, the general language adopted by \cite{Bando:1987br} turns out to be well suited for the description of the anomalous part of the action i.e. the Wess-Zumino-Witten term.
    
    For the following it will be convenient to add scalar and vector source terms to the UV Lagrangian \eqref{eq:dark_strong_quark_lagrangian_nambu_gorkov}, which transform such that the UV Lagrangian is invariant under local $SU(2\nf)$ transformation. The UV Lagrangian \eqref{eq:dark_strong_quark_lagrangian} is modified to
    \begin{equation} \label{eq:uv_source_lagrangian}
        \mathcal{L}^\mathrm{UV}_\mathrm{q;\,Ext} =  i\Psi^\dagger \sigmab^\mu\D{R}{\Ac}\Psi + i\Psi^\dagger \overline{\sigma}^\mu \Af_\mu\Psi - \frac{1}{2} \left(\Psi^\dagger \ESmat \ECmat  \fs^* \Psi^* - \Psi^\top \ESmat^* \ECmat^*  \fs \Psi  \right).
    \end{equation}
    With the help of the ``spurion-fields" $\Af_\mu$ and $\fs$, it will be possible to easily include effects of the mass term and the dark photon via setting $\fs = M$ and $\Af_\mu = -i \ed\zp_\mu \mathcal{Q}$. Their transformation behaviour under the local symmetry is summarised in table \ref{tab:transformation_properties_of_HLS_fields}. With all these ingredients we may formulate a low energy effective description of all the relevant states involved in the phenomenologically interesting processes of these dark matter models discussed in section \ref{sec:dark_particles}. This procedure is well known and was studied in depth for $SU(N)$ theories \cite{Meissner:1987ge, Scherer:2012xha}. The HLS approach was originally formulated for general coset spaces as well \cite{Bando:1987br} and several useful results for chiral perturbation theory of general coset spaces exist \cite{Bijnens:2011fm}. The purpose of the following is not to reinvent these results, but to bring them together in the context of strongly interacting dark matter to provide a solidly worked out framework, ready to be used by phenomenologists.

\subsection{Hidden local symmetry Lagrangian}
\label{sec:HLS_lagrangian}
    \begin{table}
        \centering
        \renewcommand{\arraystretch}{1.6}
        \begin{tabular}{c|c|c}
             Field    &$G_{F, \,\mathrm{local}}^\mathrm{HLS}\times H_{F, \,\mathrm{local}}^\mathrm{HLS}$ & $\Par$  \\ \hline
             $\gamma$ & $ U_g \gamma U_h^\dagger $ & $ \gamma^\dagger(t, \Vec{-x})$ \\
             $\Av_\mu$& $ U_h \Av_\mu U_h^\dagger + U_h\partial_\mu U_h^\dagger$ & $ g_{\mu\mu}\,\Nar\left(\Av_\mu(t, \Vec{-x})\right)$ \\
             $\Af_\mu$& $ U_g \Af_\mu U_g^\dagger + U_g\partial_\mu U_g^\dagger$ & $ g_{\mu\mu}\,\Nar\left(\Af_\mu(t, \Vec{-x}\right)$ \\
             $\fs    $& $ U_g \fs U_g^\top$        & $ \fs $ 
        \end{tabular}
        \caption{Summary of the transformation behaviour of the building blocks for the HLS approach. Here $U_g(x)\in G_{F, \,\mathrm{local}}^\mathrm{HLS}$, $U_h(x) \in H_{F, \,\mathrm{local}}^\mathrm{HLS}$ and $\Nar$ is the naive parity operation defined in \eqref{eq:lie_algebra_involutive_automorphism}. The action of charge conjugation $\CC$ is already included in the flavour symmetry. }
        \label{tab:transformation_properties_of_HLS_fields}
    \end{table}
    We start by discussing a Lagrangian, describing the interaction between the dark pions $\pi$, dark mesons $\rho$, $\omega$ and the dark photon $\zp$. For this we use the framework of HLS \cite{Bando:1987br}, very  successfully applied to real world QCD. The building blocks of the HLS approach are matrix-valued fields, transforming in a linear representation of the group $G_{F, \,\mathrm{local}}^\mathrm{HLS} \times H_{F, \,\mathrm{local}}^\mathrm{HLS}$. In the HLS approach $H_F = SO(2\nf)$ is always considered as local. Since we want to make contact to the external sources via the spurion field $\Af_\mu$, we also consider $G_F=SU(2\nf)$ as a local symmetry. If we do not care about the gauging of the chiral symmetry in the UV, then we can take $G_F$ as global symmetry. The vector particles $\rho$, $\omega$ and $\zp$ are modeled with the help of matrix-valued vector fields. For the dark vector mesons we use the gauge field $\Av_\mu = -i \gv \Av_\mu^A T_A^\F$ related to the local group $H_{F, \,\mathrm{local}}^\mathrm{HLS}$. The constant $\gv$ is a yet unspecified parameter of the theory, related to the interaction strength of the vector mesons and thus to the underlying strong interaction of the dark sector.  The external sources $\Af_\mu$ are implemented as the gauge fields of $ G_{F, \,\mathrm{local}}^\mathrm{HLS}$, and may be used to include the dark photon by setting $\Af_\mu = -i \ed\zp_\mu \mathcal{Q}$. 
    
    In order to introduce the pions we introduce a $G_F$-valued scalar field $\gamma$, transforming in a bi-fundamental representation of the HLS group. The transformation behaviour of all the fields are summarised in table~\ref{tab:transformation_properties_of_HLS_fields}. Taking into account the splitting $\liea{g}_F = \liea{h}_F \oplus \ks$, we may always decompose \cite{Callan:1969sn, Coleman:1969sm}
    \begin{equation} \label{eq:pion_compensator_factorization}
        \gamma = e^{-\ngb} e^{-\sigma}
    \end{equation} 
    such that $\ngb \in \ks$ and $\sigma \in \liea{h}_F$. Due to their transformation behaviour, the fields $\ngb$ may now be interpreted as the Nambu-Goldstone bosons of the spontaneously broken global symmetry. Thus, when re-scaling the components of $\ngb$ by an appropriated dimensional constant $\fpi$, one may interpret them as dark pions according to 
    \begin{equation} \label{eq:interpret_ngb}
        \ngb = -i \frac{\pid^a}{\fpi} T_a^\F.
    \end{equation}
    The compensator fields $\sigma$ do not have a direct interpretation as scalar fields on their own and are best removed by fixing a unitary gauge for the fields $\Av_\mu$ via the HLS gauge-fixing condition 
    \begin{equation}\label{eq:HLS_gauge_fixing_condition}
        e^{-\sigma} = \unity.
    \end{equation}
    In order to preserve this condition under an arbitrary $G_{F, \,\mathrm{local}}^\mathrm{HLS}$ transformation $U_g(x)$, one must also add a compensating $H_{F, \,\mathrm{local}}^\mathrm{HLS}$ transformation $U_h(x) = U_h[U_g(x),\pi(x)]$, which depends on $U_g(x)$ and $\pi(x)$. Hence this breaks the HLS group $G_{F, \,\mathrm{local}}^\mathrm{HLS} \times H_{F, \,\mathrm{local}}^\mathrm{HLS}$ down to a non-linear realised subgroup $G_{F,\;\mathrm{local}}^\mathrm{CCWZ}$. This non-linear representation is exactly the transformation in the CCWZ construction \cite{Callan:1969sn, Coleman:1969sm} i.e. the non-linear $\Sigma$-model, fortifying the interpretation of $\ngb$ as the Nambu-Goldstone bosons. In fact it was demonstrated that integrating out the vector meson fields $\Av_\mu$ with their equation of motion, after the HLS gauge-fixing, renders the HLS equivalent to the non-linear $\Sigma$-model \cite{Bando:1987br}. The HLS symmetry is used mainly as an organizational tool, allowing for consistent truncation \cite{Georgi:1989xy} in terms of a derivative expansion. 
    
    In order to construct the Lagrangian, it is useful to combine the fields $\gamma$, $\Av$ and $\Af$, as well as potential derivatives thereof, into terms with simple transformation behaviour.  From the quantity $\gamma$ we can construct the Maurer-Cartan form $\mcf_\mu$ and the gauge-field $\Afh_\mu$ 
    \begin{align}
        \mcf_\mu &= \gamma^\dagger \partial_\mu \gamma \\
        \Afh_\mu &= \gamma^\dagger \Af_\mu \gamma.
    \end{align}
   Further we define the combined quantity
    \begin{equation}\label{eq:shifted_maurer_cartan_form}
        \mcfh_\mu = \mcf_\mu + \Afh_\mu 
    \end{equation}
    which transforms as $\mcfh_\mu \mapsto U_h\mcfh_\mu U_h^\dagger + U_h \partial_\mu U_h^\dagger$ and is thus invariant under $G_{F, \,\mathrm{local}}^\mathrm{HLS}$. 
    The quantity $\mcfh_\mu$ is $\liea{g}_F$-valued. The coset space $G_F/H_F$ may be split into 
    \begin{equation}
        \mcfh_\mu = \mcfh_{h;\mu} +\mcfh_{k;\mu}
    \end{equation}
    by using the parity operator $\Nar$ to project out its component on $\liea{h}_F$ and $\ks$. While $\mcfh_{h;\mu}$ transforms the same as $\mcfh$, we have that $\mcfh_{k;\mu}$ transforms in the adjoint of $H_{F, \,\mathrm{local}}^\mathrm{HLS}$. 
    If we further subtract the field $\Av_\mu$, the quantity $\mcfh_{h;\mu} - \Av_\mu$ also transforms in the adjoint of $H_{F, \,\mathrm{local}}^\mathrm{HLS}$. From these quantities we can now build all local terms that are invariant under the HLS, parity and charge counjugation. Hence, we use them to build the low energy effective Lagrangian. By using the derivative expansion of the HLS approach \cite{Harada:2003jx} we can classify sub-leading contributions in the Lagrangian. In this counting scheme a derivative i.e. external momentum $p$ is considered to be a small quantity $\delta \sim p$. The couplings of the HLS gauge fields are assumed to be smaller or at most of the same order, e.g. $\delta  \sim \gv \sim \ed$. In fact, in order to include the dark photon by gauging the flavour symmetry in the effective Lagrangian of the strong dark interactions, one implicitly assumes that one may treat $\zp$ as a small perturbation in the IR. Hence $\ed \ll \gv \sim \delta$. The success of the counting scheme is rooted in the gauge structure of the HLS approach \cite{Georgi:1989xy}. The lowest order ($\mathcal{O}(\delta^2)$) Lagrangian in the chiral limit, i.e. $\fs = 0$, is given by \cite{Bando:1987br} 
    \begin{equation} \label{eq:HLS_lagrangian}
        \mathcal{L}^\mathrm{IR;(2)}_\mathrm{HLS} = -\fpi^2 \Tr{\mcfh_{k;\,\mu}\, \mcfh_{k}^\mu} -\clhs \fpi^2 \Tr{\left(\mcfh_{h;\,\mu} -V_\mu\right)\left(\mcfh_{h}^\mu -V^\mu\right)}.
    \end{equation}
    The prefactor $-\fpi^2$ of the first term ensures canonical normalisation of the pion fields. The parameter $\clhs$ is a dimensionless, undetermined parameter of the theory. The Lagrangian obtained is the most general HLS result. It may be convenient to rewrite the Lagrangian as\footnote{This result is independent of the condition \eqref{eq:HLS_gauge_fixing_condition}. The HLS gauge is only fixed for the expansion in terms of the pions. }
    \begin{align}
        \label{eq:HLS_lagrangian_1}
        \mathcal{L}^\mathrm{IR;(2)}_\mathrm{HLS} 
        =& -\fpi^2 \Tr{\mcf_{k;\,\mu}\mcf_{k}^{\mu}}\\
        \label{eq:HLS_lagrangian_2}
        & -\fpi^2 \Tr{\Afh_{k;\,\mu} \Afh^\mu_k + 2\mcf_{k;\mu}\Afh^\mu_k}\\
        \label{eq:HLS_lagrangian_3}
        & -\clhs\fpi^2 \Tr{\mcf_{h;\,\mu}\mcf_{h}^{\mu} + \Av_\mu\Av^\mu - 2 \Av_{\mu}\mcf_{h}^{\mu}}\\
        \label{eq:HLS_lagrangian_4}
        & -\clhs\fpi^2 \Tr{\Afh_{h;\,\mu} \Afh^\mu_h + 2 \,\mcf_{h;\mu}\Afh^\mu_h  - 2V_\mu\Afh^\mu_h}.
    \end{align}
    In this form one can read off the Lagrangian for several special cases. 
    If we would like to look at the dark sector in isolation i.e. if we have no dark-photon field $\zp$, we simply neglect the terms \eqref{eq:HLS_lagrangian_2} and \eqref{eq:HLS_lagrangian_4}, since they vanish in the decoupling limit $\Af_\mu \rightarrow 0$.
    In the case one wants to do dark matter phenomenology without the vector mesons one simply needs to integrate them out by using their equation of motion $V = \mcfh_{h;\,\mu}$. This leads to vanishing of the terms \eqref{eq:HLS_lagrangian_3} and \eqref{eq:HLS_lagrangian_4}. Thus, in the following, this can always be accounted for by setting $\clhs = 0$ in the results that follow. The results then, after enforcing the condition \eqref{eq:HLS_gauge_fixing_condition}, coincide with the CCWZ construction \cite{Bando:1987br}.
    Of course, if one wants to treat the vector mesons $\Av_\mu$ or the vector sources $\Af_\mu$ as dynamical fields, one should also include their kinetic terms in the Lagrangian
    \begin{equation}
        \mathcal{L}^\mathrm{IR;(2)}_\mathrm{HLS,YM} = -\frac{1}{4} \Fv_{\mu\nu}^A\Fv^{\mu\nu}_A -\frac{1}{4}\Ff_{\mu\nu}^\alpha\Ff^{\mu\nu}_\alpha.
    \end{equation}
    Indeed, these appear at order $\mathcal{O}(\delta^2)$ in the HLS counting scheme. It is a central assumption of HLS that the kinetic term for the vector mesons is dynamically generated by quantum effects of the underlying strong dynamics \cite{Bando:1987br}. 
    Integrating the vector mesons out with the equation of motion and keeping terms up to a consistent order in the HLS counting scheme, reproduces higher order terms\footnote{At least that is the case in the case of $SU(N)$ theories with  $\nf$ fundamental fermions as considered in \cite{Harada:2003jx}. Based on the symmetry structure we would expect exactly the same result in the (pseudo-)real case. In the complex case, the obtained LECs from integrating out the vector mesons to a large extent saturate the experimental values. Such a statement can of course not be made in the present case, however it supports the use of HLS as an appropriate low energy effective description.} of the CCWZ construction~\cite{Harada:2003jx}. 
    
    The HLS construction prevents us from introducing an explicit mass term for the vector mesons. However, such a term appears dynamically, and is related to two of the free parameters of the theory, the constant $\clhs$ and the coupling $\gv$. This can be seen by performing a chiral expansion of the Lagrangian \eqref{eq:HLS_lagrangian}.
    For technical details on the expansion see appendix \ref{sec:chiral_expansion_appendix}, which provides a convenient framework for performing the chiral expansion, using the properties of the symmetric splitting $\liea{g}_F = \liea{h}_F + \ks$.
    The lowest order expansion and truncation, describing all tree-level processes involving at most four dark pions is given by 
    \begin{align}
        \label{eq:HLS_lagrangian_1_expanded}
        \mathcal{L}^\mathrm{IR;(2)}_\mathrm{HLS} 
        &=
        \frac{1}{2} \delta_{ab} \; \partial_\mu \pi^a \partial_\mu \pi^b 
        + g_{4\pid}\,C_{abcd}\, \pid^a\partial_\mu \pid^b\, \pid^c \partial^\mu \pid^d \\
        \label{eq:HLS_lagrangian_2_expanded}
        &
        + g_{\zp\pid\pid}\,C_{qab}\,\zp_\mu\, \pid^a\partial_\mu\pid^b
        + g_{\Av\pid\pid}\,C_{Aab}\,\Av_\mu^A\, \pid^a\partial_\mu\pid^b \\
        \label{eq:HLS_lagrangian_3_expanded}
        &
        +\frac{\mv^2}{2} \delta_{AB}\Av^A_\mu\Av^{B\mu} 
        +\frac{\mv^2 r^2}{2} \zp_\mu{\zp}^\mu -\mv^2 r \Av^A_\mu{\zp}^\mu \Q_{qA} \\
        \label{eq:HLS_lagrangian_4_expanded}
        & 
        + g_{\zp4\pid}\, C_{aqbcd}\, \pid^a\zp_\mu\pid^b\pid^c\,\partial^\mu\pid^d 
        +g_{\Av4\pid}\, C_{aAbcd}\, \pid^a\Av^A_\mu\pid^b\pid^c\,\partial^\mu\pid^d\\
        \label{eq:HLS_lagrangian_5_expanded}
        &
        + g_{\zp\zp\pid\pid}\, C_{aqbq}\, \pid^a\zp_\mu\pid^b{\zp}^\mu
        - g_{\Av\zp\pid\pid}\, C_{ABab}\, \pid^a\Av_\mu^A \pid^b{\zp}^\mu \\
        \label{eq:HLS_lagrangian_6_expanded}
        &
        + g_{\zp\zp4\pid}\, C_{abqcdq}\,\pid^a\pid^b\zp_\mu\pid^c\pid^d{\zp}^\mu \,
        - g_{\zp\Av4\pid}\, C_{abAcdq}\,\pid^a\pid^b\Av^A_\mu\pid^c\pid^d{\zp}^\mu\\
        \nonumber&+\mathcal{O}(\pid^6;\delta^2).
    \end{align}
    Here the indices $a,b,c, \dots$ sum over the broken generators of $\liea{g}_F$ e.g. $a = 1,2,\dots,9$. The indices $A,B,\dots$ sum over the unbroken generators, e.g. $A = 10,11,\dots,15$.
    The index $q$ denotes the index of the generator that is proportional to the charge assignment matrix ~$\Q$. In the case discussed $q = 13$. 
    The coefficient matrices can be expressed as traces over generators
    \begin{align}
        \Q_{qN} &= 2\Tr{\Q T^\F_N} \\
        C_{NMK}    &= -2i\Tr{T_N^\F\combr{T_M^\F,T_K^\F}} \\
        C_{NMKL}   &= - 2\Tr{\combr{T_N^\F,T_M^\F}\combr{T_K^\F,T_L^\F}}\\
        C_{NMKLH}  &=  2i\Tr{\combr{T_N^\F,T_M^\F}\combr{T_k^F,\combr{T_L^\F,T_H^\F}}}\\
        C_{NMKLHI} &= 2\Tr{\combr{T_N^\F,\combr{T_M^F,T_K^\F}}\combr{T_L^F,\combr{T_H^\F,T_I^\F}}}
    \end{align}
    and can all be reduced to contractions of the structure constants $C_{NMK}$ of $\liea{g}_F$. Here the indices $N,M,K,\dots$ run over all generators e.g. $N = 1,2,\dots, 15$. In general, a quantity with an index $q$ can be expressed by contracting it with $\Q_{qN}$ e.g. $C_{qMK}  = \sum_N \Q_{qN} C_{NMK} $.
    The coupling constants are given in terms of the four parameters $\clhs,\gv,\fpi,\ed$.
    {\renewcommand{\arraystretch}{3}
    \begin{alignat}{2}
        g_{4\pid} &= \frac{4-3\clhs}{24\fpi^2}  \\
        r &= \frac{\ed}{\gv}   \quad \quad \quad \quad \quad \quad\quad\quad \quad\quad\;
        \quad\quad \mv^2 &&= \clhs\fpi^2\gv^2 \label{eq: KSFR_SO}\\
        g_{\zp\pid\pid} &= \ed\frac{2-\clhs}{2}  \quad \quad \quad \quad \quad\quad\quad\quad
        g_{\Av\pid\pid} &&= \clhs\frac{\gv}{2}\quad\quad \\
        g_{\zp4\pid} &= \frac{\ed}{24\fpi^2}\left(7 \clhs-12\right)\quad \quad\quad\quad\quad  
        g_{\Av4\pid} &&= \clhs\frac{\gv}{24\fpi^2}\quad\quad  \\
        g_{\zp\zp\pid\pid} &=  \ed^2\frac{(\clhs-1)}{2} \quad\quad\;\,\quad\quad\quad\quad
        g_{\zp\Av\pid\pid} &&= \clhs\frac{\gv\ed}{2} \quad\quad \\
        g_{\zp\zp4\pid} &= \ed^2\frac{(\clhs-1)}{6\fpi^2} \quad\quad\quad\quad\quad\quad\;\,
        g_{\Av\zp4\pid} && = \clhs\frac{\gv\ed}{24\fpi^2} 
        \label{eq: clhs_fixing}
    \end{alignat}}

    which are very similar to that of the QCD, e.g. \eqref{eq: KSFR_SO} is analogue of the QCD KSRF relation. To our knowledge, this relation has not yet been tested on lattice for $SO(\nc)$ gauge group.
    Of the quantities $\clhs,\gv,\fpi$, only one is a free parameter of the theory, being related to the others via non-trivial relations, determined by the UV theory.  Relations among these might be studied for the dark sector in isolation on the lattice e.g. by studying KSRF relation~\cite{Kawarabayashi:1966kd, Riazuddin:1966sw} (see e.g. \cite{DeGrand:2019vbx} for a discussion in context of $SU(\nc)$ theories). Interestingly, from the knowledge of $\clhs$ in isolation, it seems one can also infer information about the interaction of the dark hadronic sector with dark electromagnetism. As a phenomenological guideline on what value we can expect for the dimensionless quantity $\clhs$, note that $g_{\zp \pid\pid} = 0$ for $\clhs = 2$. Hence, the dark pion form factor is dominated by the contributions of a neutral vector meson, interacting with the pions and subsequently oscillating into a dark photon \cite{Bando:1987br}. This can be seen as a realization of vector meson dominance (VMD)  and indeed for this parameter the Lagrangian \eqref{eq:HLS_lagrangian_1_expanded}-\eqref{eq:HLS_lagrangian_3_expanded} reproduces the phenomenological Lagrangian of VMD \cite{OConnell:1995nse}. $\clhs = 2$ also corresponds to the choice in QCD and ensures coupling universality $\gv = g_{\Av\pid\pid}$.
    
    All phenomena related to $\pid\pid \rightarrow \pid\pid$ at tree-level can successfully be treated with the first three lines \eqref{eq:HLS_lagrangian_1_expanded}-\eqref{eq:HLS_lagrangian_3_expanded} of the Lagrangian. Eq. \eqref{eq:HLS_lagrangian_4_expanded}-\eqref{eq:HLS_lagrangian_6_expanded} can only contribute via loops and thus are suppressed by an additional factor $p^2 = \delta^2$. Hence, when only considering $\pid\pid \rightarrow \pid\pid$ processes, we could actually neglect these from the Lagrangian.
    However, it should be noted that semi-annihilation processes, like $\pid\pid\pid \rightarrow \Av\pid$ or $\pid\pid\pid \rightarrow \zp\pid$ described by \eqref{eq:HLS_lagrangian_4_expanded}, enter at the same order $\mathcal{O}(\delta^2)$. Such processes may affect the cosmological depletion of dark matter if $\mv<2\mpi$ \cite{Berlin:2018tvf} or $\mzp\sim2\mpi$, since these terms can also provide number changing processes in the dark sector, which might contribute to the freeze out of the dark sector species \cite{Bernreuther:2023kcg}.
    
\subsubsection{Contributions of explicit symmetry breaking}  
        So far we have considered the chiral limit. In order to take into account the effects of the explicit symmetry breaking by a mass term, e.g. $\fs \neq 0$, it is useful to work with a field variable $\Sigma$, that transforms linearly under the group $G_{F,\,\mathrm{global}}^\mathrm{CCWZ}$. Such a field can be build from $\gamma$, because the coset space is symmetric \cite{Callan:1969sn, Coleman:1969sm}
        \begin{equation}\label{eq:linear_coset_representative}
            \Sigma := \gamma \EFmat \gamma^\top.
        \end{equation}  
        This quantity transforms as $\Sigma \rightarrow U_g\Sigma U_g^\top$ under arbitrary HLS transformations and does not see anything of $H_{F, \,\mathrm{local}}^\mathrm{HLS}$. Hence, it transforms linearly under $G_{F,\,\mathrm{global}}^\mathrm{CCWZ}$ after HLS gauge fixing. 
        Now, if we introduce the condensate matrix
        \begin{equation}
            \{\qconm\}^k_{l} = \Exv{\Psi^{(k)\dagger} \ESmat\ECmat\EFmat_{lm}\Psi^{(m)*}} -  \Exv{\Psi^{(k)\top}\ECmat\ESmat\EFmat_{lm}\Psi^{(m)}},
        \end{equation}
        and if $\Tr{\qconm} = \qcon \neq 0$, the remaining unbroken symmetry dictates $\qconm \propto \EFmat$. Hence, up to a normalisation, we may interpret $\Sigma$ as fluctuation around the chiral condensate \cite{Kogut:2000ek}, parameterized via the dark pion fields. In the ground state it should hold
        \begin{equation}
            \Exv{\Sigma} = \Sigma[\pid = 0] = \EFmat.
        \end{equation}
        Now again, we consider all local terms, compatible with Lorentz-symmetry, HLS and parity.
        Taking also into account the spurion field $\fs$, and following the previous discussion, we obtain only one new term at lowest order
        \begin{equation}\label{eq:explicit_mass_breaking_contribution}
            \mathcal{L}_M^{IR;(2)} = C_\fs \left(\Tr{X^\dagger\Sigma} + \Tr{X\Sigma^\dagger}\right).
        \end{equation}
        Setting $\fs = m\EFmat$, and demanding $\frac{\delta Z^\mathrm{UV}}{\delta m} = \frac{\delta Z^\mathrm{IR}}{\delta m}$, with $Z^{\mathrm{UV}/\mathrm{IR}}$ the partition function in the UV and IR, one obtains  $16 C_\fs = \qcon$ \cite{Kogut:2000ek, Scherer:2012xha}. 
        Expanding the HLS-gauge-fixed action to lowest order yields a mass term for the dark pions and four pion contact interactions. The pion mass is given by  
        \begin{equation}\label{eq:gmor_pions}
            \mpi^2 = \frac{m\qcon}{4\fpi^2}.
        \end{equation}
        This is the analog of the Gell-Mann-Oakes-Renner (GMOR) relation from QCD. Within the HLS formalism additional mass terms corresponding to explicit breaking of HLS symmetry breaking can appear (see e.g. \cite{Bennett:2017kga}). However they occur at higher order and we do not include them here.
        
        The overall four-point interactions among the dark pions become modified by a contribution involving the totally symmetric\footnote{The round brackets denote total symmetrization i.e. $C_{(i_1,\dots,i_n)} = \frac{1}{n!} \sum_{\sigma \in S^n} \,C_{i_{\sigma(1)},\dots,i_{\sigma(n)}}$. $S^n$ denotes the group of all permutations of $n$ objects.} coefficients $S_{abcd} = \Tr{T_{(a}^\F T_b^\F T_c^\F T_{d)}^\F}$. The explicit form of the resulting LO Lagrangian is given by
        \begin{equation}
            \label{eq:interactions_pions}
           \mathcal{L}_{4\pid}^{IR;(2)} = g_{4\pid}\,C_{abcd}\, \pid^a  \pid^c \,\partial_\mu \pid^b\,\partial^\mu \pid^d + \frac{\mpi^2}{3\fpi^2}S_{abcd}\,\pi^a\pi^b\pi^c\pid^d. 
        \end{equation}
        It is important here to note that the four pion vertex is dependent on the HLS free parameter $\clhs$ through the coefficient $g_{4\pi}$. In order to compute physical $4\pi$ scattering amplitudes with $\clhs \neq 0$ one must also take into account the associated vector meson terms to obtain consistent results.
        If one intends to make contact with the formalism in \cite{Hochberg:2014kqa, Kogut:2000ek}, one may express \eqref{eq:HLS_lagrangian_1} and \eqref{eq:HLS_lagrangian_2} in terms of a covariant derivative of $\Sigma$ by using \begin{equation}
            \Tr{\mcfh_{k;\,\mu}\mcfh_{k}^\mu}= -\frac{1}{4} \Tr{(\partial_\mu\Sigma + \Af_\mu\Sigma + \Sigma\Af_\mu^\top)(\partial_\mu\Sigma + \Af^\mu\Sigma + \Sigma\Af^{\mu\,\top})^\dagger}.
        \end{equation}
        The Lagrangian derived so far exhibits an additional, non-physical symmetry, given by the naive parity transformation, described by acting with $\Nar$ from \eqref{eq:lie_algebra_involutive_automorphism} on the $\pid$-fields, without changing the spacetime argument. This prevents the occurrence of processes with an odd number of dark pions. Hence, the Lagrangian so far will not feature a five pion vertex.

\subsection{Wess-Zumino-Witten action} \label{sec:wzw_action}
    In the following, we turn our attention to terms in the low-energy effective action, which violate naive-parity, while respecting all other physical symmetries. These allow for processes involving an odd-number of dark pions. They slipped our attention so far, because they are of higher order. They are however required for the low energy theory to match the 't Hooft anomaly structure of the underlying UV theory \cite{tHooft:1980xss}. 
    
    All terms constructed so far are ``non-anomalous'' i.e. they are invariant under\footnote{The global symmetry in the 't Hooft argument is not the hidden local symmetry group $G_{F, \,\mathrm{local}}^\mathrm{HLS}$, but the diagonal subgroup $G_{F,\,\mathrm{local}}^\mathrm{CCWZ}$, which remains after gauge-fixing $H_F^\mathrm{HLS}$. The dark pions transform in a non-linear realization of this group, which is essential for the anomaly matching argument.} the fully gauged $G_{F,\,\mathrm{local}}^\mathrm{CCWZ}$ symmetry. Hence, they cannot satisfy the anomaly equation discussed below. Thus, we select this subset of higher order terms to be part of the low energy effective theory. These terms go under the name Wess-Zumino-Witten (WZW) terms~\cite{Wess:1971yu, Witten:1983tw}. A standard construction and classification of such terms \cite{DHoker:1994rdl, DHoker:1995mfi, Brauner:2018zwr}, originating from an elegant geometric interpretation \cite{Witten:1983tw} exists. 
    This classification applies for a large class of theories of fields $\gamma: S^4 \rightarrow G_F/H_F$, where $G_F$ is compact, $H_F$ a Lie-subgroup and $G_F/H_F$ a connected, homogeneous space satisfying the topological condition $\hg_4(G_F/H_F) = 0$. Here $\hg_4(X)$ denotes the so-called ``fourth homotopy group'' of a topological space $X$. However, if this condition is not satisfied, the geometric classification is inconclusive, as is pointed out\footnote{Contrary to the claim in \cite{DHoker:1994rdl}.} in \cite{Davighi:2018inx}. In the case of interest $\hg_4(SU(4)/SO(4)) \neq 0$, hence the condition is not satisfied\footnote{That this might be a problem was also already remarked in \cite{Brauner:2018zwr}.}. More details on this can be found in Appendix \ref{sec:homotpy_groups_of_coset_space}. 
    
    In order to proceed, we retreat to the original argument given by Wess and Zumino  \cite{Wess:1971yu}, which was later generalised \cite{Chu:1996fr} and especially works for arbitrary compact $G_F$, broken to an anomaly free subgroup $H_F$ such that $G_F/H_F$ is connected\footnote{For a more modern WZW construction see also \cite{Freed:2006mx, Lee:2020ojw}.}. If $\hg_4(G_F/H_F) = 0$ is satisfied, the construction can be shown to be consistent with the geometric one \cite{Chu:1996fr}. 
    As a side effect of this construction the coefficient of the WZW term in the low energy effective action is simultaneously determined from the anomaly matching argument. We find that for the real representation SIMP models discussed in \cite{Hochberg:2014kqa}, the coefficient of the $3\to 2$ pion scattering vertex is overestimated by a factor two. This was realised also with a geometrical argument in \cite{Kamada:2017tsq}. 
    
    In the following it will be useful to stick to the language of Lie-algebra valued differential forms. Especially, we use the gauge connection 1-forms $\Af = \Af_\mu \d x^\mu$ and $\Ac = \Ac_\mu \d x^\mu$, which are matrix-valued differential forms. The method used makes explicit use of the non-linear transformation behaviour of the pions and requires to gauge-fix\footnote{If  $\hg_4(G_F/H_F) = 0$, a solution to the anomaly equation may be constructed via the methods presented in \cite{Brauner:2018zwr}, without enforcing the HLS-gauge fixing. Hence, this seems to be a technical issue.} the HLS according to \eqref{eq:HLS_gauge_fixing_condition}.   
    
    \subsubsection{A solution of the 't Hooft anomaly equation without dark vector mesons}
    Starting point is the gauging of the flavour symmetry $G_F = SU(2\nf)$ of the Lagrangian \eqref{eq:dark_strong_quark_lagrangian} in the chiral limit, to obtain an action $S_\mathrm{q, cov.}^{UV}[q, \Ac+\Af]$. Here $\Ac+\Af$ denotes the gauge connection of $G_C\times G_F^\mathrm{CCWZ}$. The obtained action coincides with the one if we would have used \eqref{eq:uv_source_lagrangian} with $\fs=0$. A general gauge transformation is parameterized by a gauge transition function $\pepsilon: \mathcal{M}\rightarrow \liea{g}_C \oplus \liea{g}_F$, which might be split according to $\pepsilon = \pepsilon^C + \pepsilon^\F$. Gauge transformations $U^{C/\F} := e^{-\pepsilon^{C/\F}}$, belonging to either $G_F$ or $G_C$, commute with all transformations of the other type. A general gauge transformation is given by $\Ac+\Af \rightarrow (\Ac+\Af)' = \Ac'+\Af'$, where 
    \begin{align}
        \Ac' &= U^C \Ac U^{C\dagger} + U^C\mathrm{d}U^{C\dagger} = \Ac + \delta_{\pepsilon}^C\Ac+ \dots \\
        \Af' &= U^F \Af U^{F\dagger} + U^F\mathrm{d}U^{F\dagger} = \Af + \delta_{\pepsilon}^F\Af + \dots\, .
    \end{align}
    Next we introduce  a functional $\widetilde{W}[\Ac+ \Af]$ via the partition function 
    \begin{equation}
        \widetilde{Z}[\Ac+\Af] = e^{\i \widetilde{W}[\Ac+\Af]} = \int \mathcal{D}q\mathcal{D}\overline{q}~e^{\i S_\mathrm{s,cov.}^{UV}[q,\Ac+\Af]}.
    \end{equation}
    If the theory has an anomaly, the functional $\widetilde{W}$ is not invariant under gauge-transformations and the anomaly functional is exactly given by the gauge variation of $\widetilde{W}$ i.e. 
    \begin{equation}\label{eq:definition_anomaly_functional}
        \widetilde{W}[\Ac'+\Af'] - \widetilde{W}[\Ac+\Af] = \mathcal{A}[\pepsilon, \Ac+\Af].    
    \end{equation}
    At this stage the fields $\Ac$ and $\Af$ are classical background fields without any dynamics. In order to interpret $\Ac_\mu^\alpha$ as the dark gluon fields we need to add a Yang-Mills term to the action $S_\mathrm{q, cov.}^{UV}$ and path-integrate over the $\Ac$-fields. The path-integral 
    \begin{equation}
    \label{eq:partition_function_uv}
        Z_{UV}[\Af] = e^{\i W_{UV}[\Af]} 
        = \int \mathcal{D}A~e^{i\widetilde{W}[A+B]+ \i S_{YM}[A]}
    \end{equation}
    is only well defined if $\widetilde{W}[\Ac+\Af]$ is invariant under $G_C$ gauge transformations. Since the representation $\R$ of $G_C$ is real and all the generators of $G_F$ and $G_C$ are traceless, the associated anomaly vanishes. Hence, gauge invariance under $G_C$ is guaranteed. Gauge variations of $W_{UV}[\Af]$ associated with $G_F$ on the other hand may produce an anomaly, which is proportional to $\tr{T^F_N \acombr{T^\F_K,T^\F_L}}$. There is no reason why this anomaly should be absent and as it turns out for $G_F = SU(2\nf)$ it is not. We may now proceed in the same fashion in the low energy regime. For this we neglect for now the vector mesons $\Av$ and start only with the flavour gauged CCWZ Lagrangian i.e. $S^{IR}_\mathrm{cov.}[\xi,\Af] = S^{IR; (2)}_\mathrm{HLS}[\xi,\Af; \clhs=0]$. We consider the HLS to be gauge fixed and $\gamma$ to transform non-linear under the flavour symmetry $G_{F,\,\mathrm{local}}^\mathrm{CCWZ}$. We hence always take $\gamma = \exp{-\xi}$. The partition function gives us
    \begin{equation} \label{eq:partition_function_ir}
        Z_{IR}[\Af]  = e^{\i W_{IR}[\Af]} 
        = \int \mathcal{D}\xi~e^{\i S^ {IR}_\mathrm{cov.}[\xi,\Af]}.
    \end{equation}
    Note that, $\Af$ is a generic $\liea{g}_F$-valued non-abelian gauge-connection, not only the dark photon.
    According to the anomaly matching argument \cite{tHooft:1980xss}, the IR theory must reproduce the same anomaly under a $G_F$ gauge variation as the theory in the UV i.e. $\delta_{\pepsilon}^\F W_{IR}[\Af] \overset{!}{=} \mathcal{A}[\pepsilon^\F, \Af]$. 
    However, the action $S^{IR; (2)}_\mathrm{HLS}[\xi,\Af; \clhs=0]$ is gauge invariant and thus gives $\delta_{\pepsilon}^F W_{IR}[\Af] = 0$. From this we can conclude that so far we miss a part in the low-energy effective description, the so-called Wess-Zumino-Witten term. 
    In order to satisfy the anomaly matching condition we add to $ S^{IR}_\mathrm{cov.}[\xi,\Af]$ an action $S_{WZW}[\xi,\Af]$, which satisfies the following anomaly equation 
    \begin{equation} \label{eq:thooft_anomaly_equation}
        \delta_{\pepsilon}^\F S_{WZW}[\xi,\Af] = \mathcal{A}[\pepsilon^\F, \Af].
    \end{equation}
    Algebraically, the gauge variation operator $\delta_{\pepsilon}^\F$ acts  as a derivative and its action on the Nambu-Goldstone bosons is defined via
    \begin{equation}
        e^{\delta_{\pepsilon}^F}e^{-\xi} = e^{-\pepsilon^\F}e^{-\xi}e^{\lambda}
    \end{equation}
    where $\lambda = \lambda[\xi,\pepsilon] \in \mathfrak{h}_F$. In \cite{Chu:1996fr} it was proven that such an action exists if none of the unbroken currents, associated with symmetry transformations of $H_F$, are anomalous in presence of arbitrary background gauge fields $\Af$. This condition may be expressed as
    \begin{equation} \label{eq: bardeen_condition}
        \forall \pepsilon \in \mathfrak{h}_F: \mathcal{A}[\pepsilon, \Af] = 0.
    \end{equation}
    Can and should we, in our theory, impose this additional constraint on the anomaly? It is well known that when calculating the anomaly from triangle diagrams, the freedom to choose a regularisation scheme affects the anomaly. This choice can be used to put the anomaly in certain currents, for example the broken currents, such that the unbroken ones are free of anomalies \cite[Chpt. 22]{Weinberg:1996kr}. This freedom can be used to enforce this additional condition. The question if we should impose the condition depends on the physical interpretation of the fields $\Af$. If we want to interpret\footnote{Or at least a subset of $\Af$, corresponding to the correct one-parameter subgroup} them as the dark photon fields $Z'$, this condition is actually required by physics in order to obtain a well defined gauge theory for $Z'$. 
    If $\Af$ has no physical interpretation and is simply a background field that gets switched off later on in the calculation, we are free to choose any regularisation, so we can impose this condition freely, as long as we do it consistently. 
    Next we introduce a parameterized version of the shifted Maurer-Cartan form \eqref{eq:shifted_maurer_cartan_form}
    \begin{equation} \label{eq:schifted_parameterized_maurer_cartan_form}
        \mcfh_\tau := e^{-\tau \delta^\F_\xi}\Af = e^{-\tau \xi}\Af e^{\tau \xi} + e^{-\tau \xi}\mathrm{d}e^{\tau \xi} =: \Afh_\tau + \mcf_\tau
    \end{equation}
    with $\tau$ rescaling the pNGB fields.
    If the condition \eqref{eq: bardeen_condition} is satisfied, the Wess-Zumino-Witten action is given by \cite{Chu:1996fr}
    \begin{equation} \label{eq:wzw_action}
        S_{WZW}[\xi,B] = \int_0^1 \mathrm{d}\tau~ \mathcal{A}[\xi, \mcfh_\tau].
    \end{equation}
    Note that the Nambu-Goldstone fields $\xi$ do not only enter the first argument of the anomaly, but also via $\mcfh_\tau$. 
    As a last step we only need to determine the form of the anomaly in the UV. For this we make use of the Wess-Zumino consistency condition \cite{Wess:1971yu} and the Stora-Zumino descent equations \cite{Zumino:1983rz}. 
    The latter fix the anomaly up to the gauge variation of a local functional. Thus we have the following ansatz for the anomaly
    \begin{equation}
        \mathcal{A}[\epsilon, B] = \mathcal{N}\left(\mathcal{A}_0[\epsilon, B] + \delta_{\pepsilon}^F \mathcal{F}_{BC}[B]\right)
    \end{equation}
    where $\mathcal{N}$ is a normalisation and $\mathcal{A}_0[\pepsilon, B]$ is the canonical, consistent anomaly with imposed Bose symmetry \cite{Bilal:2008qx} given by
    \begin{equation} \label{eq:anomaly_functional_ansatz}
        \mathcal{A}_0[\pepsilon, B] = \int_\mathcal{M} \tr{\pepsilon~\mathrm{d}\left(B\mathrm{d}B+\frac{1}{2}B^3\right)}.
    \end{equation}
    Here the product of the differential forms is the exterior product, not to be confused with an ordinary matrix product. 
    The local functional $\mathcal{F}_{BC}$ acts as the analog of the Bardeen counter term \cite{Adler:1969er} and was determined in \cite{Chu:1996fr}. We will not state the expression for $\mathcal{F}_{BC}$, since further simplifications in the explicit expression arise due to parity. 
 
    As demonstrated in \cite{Chu:1996fr}, the ansatz for the anomaly \eqref{eq:anomaly_functional_ansatz} may be split into three parts 
    \begin{equation} \label{eq:anomaly_decomposition}
        \mathcal{A}[\pepsilon, B] = \mathcal{A}_-[\pepsilon_k, B] + \mathcal{A}_+[\pepsilon_k, B] + \mathcal{F}_R[\pepsilon_k, B] 
    \end{equation}
    where $\mathcal{A}_\pm[\pepsilon_k, B]$ signify  $\pm$ parity projections and $\mathcal{F}_R[\pepsilon_k, B] = 0$ for $SU(2\nf)/SO(2\nf)$.
    The functional $\mathcal{F}_R$ vanishes if the space $G_F/H_F$ is symmetric. For the other two parts\footnote{Note that our definition of parity is little bit different from the definition in \cite{Chu:1996fr}. Their definition of parity commutes with gauge-transformations. This is because they define it on the split components each. For us it holds $\Par\delta_\pepsilon^F = \delta_{\Par\pepsilon}^F \Par$. The validity of their arguments remain.} $\mathcal{A}_\pm[\pepsilon_k, \Par B] = \mp \mathcal{A}_\pm[\pepsilon_k, B]$ holds. Since spatial parity is a good symmetry of the quantum theory, one can determine from \eqref{eq:definition_anomaly_functional}-\eqref{eq:partition_function_uv} that $\mathcal{A}[\pepsilon, \Par B] =\mathcal{A}[\pepsilon, B]$, and hence $\mathcal{A}_+[\pepsilon_k, B] = 0$. The remaining term is given explicitly via 
    \begin{equation} \label{eq:anomaly_expression}
        \mathcal{A}_-[\pepsilon_k, B] = \mathcal{N} \int_\mathcal{M} 
        \Tr{\pepsilon_k\left(3 F_{B;h}^2 + F_{B;k}^2 - 4\left(B_k^2F_{B;h} + B_k F_{B;h} B_k + F_{B;h} B_k^2\right) + 8 B_k^4\right)}\\
    \end{equation}
    where $F_B = \mathrm{d}B + B^2 = F_{B;h} + F_{B;k}$ and $B=B_h+B_k$. Note that only $\pepsilon_k$ appears in \eqref{eq:anomaly_decomposition} and \eqref{eq:anomaly_expression}, since by construction $ \mathcal{A}[\pepsilon_h, B] = 0$ must hold, enforced by the counter term. 
    It is well known that the Wess-Zumino consistency condition is so restricting that it fixes the anomaly if only the quadratic coefficient of the anomaly is known. The ansatz \eqref{eq:anomaly_functional_ansatz} satisfies this condition and the only open parameter $\mathcal{N}$ determines the quadratic coefficient. We can calculate the coefficient of the contribution of $3\mathcal{N}\tr{\pepsilon_k F_{B;h}^2}$ to the anomaly from a perturbative, one-loop triangle diagram calculation, involving a broken and two unbroken currents. 
    However, for this calculation it is important to choose the regularisation consistently, since we imposed the condition \eqref{eq: bardeen_condition} \cite[Chpt. 22]{Weinberg:1996kr}.
    From this we obtain
    \begin{equation}
        \mathcal{N} = \frac{i\, d_\R}{24 \pi^2}.
        \label{eq:WZW_coeff}
    \end{equation}
    The dimensionality $d_\R$ of the gauge group representation enters because every gauge degree of freedom produces a copy of the flavour anomaly. 
    This result is consistent with the normalisation in \cite{Brauner:2018zwr}.
    When working in the vector representation of $SO(\nc)$, we obtain $d_\R = \nc$. The physical Wess-Zumino-Witten action may now be obtained by setting the fictitious gauge fields to a physical value. The ungauged Wess-Zumino-Witten action may be obtained in the decoupling limit\footnote{We assume this limit to be well defined and that the theory converges again to the one without background gauge fields.} $B\rightarrow 0$. In this limit 
    \begin{equation}
        \lim_{B\rightarrow 0} \mcfh_\tau(x) = e^{-\tau \xi(x)}\mathrm{d}e^{\tau \xi(x)} =: \Omega_\tau(x).
    \end{equation}
    Thus $F_\tau := F_B[\mcfh_\tau] = \mathrm{d}\mcfh_\tau + \mcfh_\tau^2 = 0$ and hence the only term in $\mathcal{A}[\ngb, \mcfh_\tau]$ surviving is involving $(\mcfh_\tau)_k^4 = (\mcf_\tau)_k^4$. The Wess-Zumino-Witten action is given by 
    \begin{equation} \label{eq:ungauged_wzw_action}
        S_{WZW}[\xi] = \lim_{B\rightarrow 0}S_{WZW}[\xi, B] = \frac{\i d_\R}{3\pi^2} \int_\mathcal{M} \int_0^1 \mathrm{d}\tau~\tr{\xi (\mcf_\tau)_k^4}.
    \end{equation}
    Expanding $\mcf_\tau$ to first order, it is possible to integrate out $\tau$ explicitly and one obtains a five point vertex involving the Nambu-Goldstone fields
    \begin{equation}
        S_{WZW}[\xi] \approx \frac{\i d_\R}{15\pi^2} \int_\mathcal{M} \tr{\xi \mathrm{d}\xi \mathrm{d}\xi \mathrm{d}\xi \mathrm{d}\xi}
    \end{equation}
    with $\xi = -i\pi/F_\pi$. This corresponds to half the vertex stated in \cite{Hochberg:2014kqa}.         

\subsubsection{General solution to the t'Hooft anomaly equation}
    The solution derived so far did not involve the vector mesons $\Av$. Since the anomaly equation \eqref{eq:thooft_anomaly_equation} is linear, adding the solution $S_\mathrm{WZW}$ to the HLS gauge-fixed action $S^{IR; (2)}_\mathrm{HLS}[\xi,\Af,\Av]$ now also reproduces the anomaly structure correctly, one might consider the issue resolved. 
    However, due to the linearity of \eqref{eq:thooft_anomaly_equation}, the solution is not unique and actually four more undetermined parameters appear, when including the vector mesons. 
    This is because we may construct four linearly independent operators\footnote{There two more terms one can construct, which are $\Tr{F_{\Afh,k}(\mcfh_h-V)^2}$ and $\Tr{F_{\Afh,k}\mcfh_k^2}$. Those can be shown to vanish using that $\Nar(AB) = (-)^{pq+1}\Nar(B)\Nar(A)$ and $Tr{AB} = \Tr{\Nar(A)\Nar(B)} = -\Tr{\Nar(AB)}$. Here $A$ is a $p$-form and $B$ a $q$-form. With the same relations the second equality sign in \eqref{eq:non_anomalous_wzw_3} and \eqref{eq:non_anomalous_wzw_4} can be proven.} at the same order as the WZW term, which are invariant under the HLS and spatial parity, but which explicitly break naive parity. They are
    \begin{align}
        \label{eq:non_anomalous_wzw_1}
        \mathcal{L}^\mathrm{IR, anom.}_{1} &= \Tr{\left(\mcfh_{h}-V\right)\mcfh_{k}^3}\\
        \label{eq:non_anomalous_wzw_2}
        \mathcal{L}^\mathrm{IR, anom.}_{2} &= \Tr{\left(\mcfh_{h}-V\right)^3\mcfh_{k}} \\
        \label{eq:non_anomalous_wzw_3}
        \mathcal{L}^\mathrm{IR, anom.}_{3} &= \Tr{F_\Av(\mcfh_h-V)\mcfh_k} = - \Tr{F_\Av\mcfh_k(\mcfh_h-V)}\\
        \label{eq:non_anomalous_wzw_4}
        \mathcal{L}^\mathrm{IR, anom.}_{4} &= \Tr{F_{\Afh;\,h}(\mcfh_h-V) \mcfh_k} = -\Tr{F_{\Afh;\,h}\mcfh_k (\mcfh_h-V)}.
    \end{align}
    We again used the language of $\liea{g}_F$-valued differential forms i.e. $\mcfh_{h} = \mcfh_{h;\,\mu}\d x^\mu$ and $\mcfh_{k} = \mcfh_{k;\,\mu}\d x^\mu$.
    Further $F_\Av = \frac{\Av_{\mu\nu}}{2}\d x^\mu \wedge\d x^\nu$, $F_{\Afh} = \gamma^\dagger F_\Af \gamma$ and $ F_\Af = \d \Af + \Af^2 = \frac{\Ff_{\mu\nu}}{2}\d x^\mu \wedge\d x^\nu $.
    Since, they can be added to the action without altering the anomaly structure and bare the same features as the WZW term otherwise, we should also add them to the action. The full generalised\footnote{One should remark that this solution is derived for HLS vector mesons in unitary gauge. It is no problem to generalize this result to arbitrary HLS gauges by using the five dimensional description, available if $\pi_4(G_F/H_F) = 0$. Thus, the requirement of a specific HLS gauge in the case of $\pi_4(G_F/H_F) \neq 0$ seems to be only a technical issue.} WZW action, as general $\mathcal{O}(\delta^4)$ solution to the anomaly equation \eqref{eq:thooft_anomaly_equation}, is hence given by\footnote{We note here that the gauged WZW for $Sp(4)$ theory obtained in \cite{Kulkarni:2022bvh} was derived under so called massive Yang-Mills approach and not under HLS. The two approaches are related to each other via specific choice of gauges~\cite{Bando:1987br}.}
    \begin{equation} \label{eq:generalized_wzw_action}
        S_{WZW} = \int_0^1 \mathrm{d}\tau~ \mathcal{A}[\xi, \mcfh_\tau] +  \jpc{ \frac{i\, d_\R}{8 \pi^2}} \sum_{i=1}^{4} C^\mathrm{anom.}_{i} 
        \int_\mathcal{M} \mathcal{L}^\mathrm{IR, anom.}_{i}.
    \end{equation}
    Phenomenological guidance for the values of the constants $C^\mathrm{anom.}_{i}$ can be obtained by VMD considerations. This is further discussed in section \ref{sec:dark_photon_in_ir}. 
    The general solution is a generalization of the result obtained in real world QCD \cite{Harada:2003jx}. We note that \cite{Fujiwara:1984mp} has same result as \cite{Harada:2003jx} with two superfluous terms. In our construction, the issues that lead to the superfluous terms are avoided automatically. Since the structure of this solution overall is exactly the same as in the one obtained in real world QCD, we expect all the statements in \cite{Fujiwara:1984mp} to remain true. For details see appendix \ref{sec:connection_to_qcd}.

\subsection{Taking into account $\eta'$}\label{sec:include_etap}

    In the cases where the $\etap$ particle is expected to be close in mass to the other dark pions, we use a combined approach of chiral perturbation theory and large $\nc$ arguments, in order to include it into the low energy effective description consistently. For this we followed the method of \cite{Herrera-Siklody:1996tqr}, coupling an external pseudo scalar spurion source $\theta$ to the topological charge $\Qtopo$ and adding it to the Lagrangian \eqref{eq:uv_source_lagrangian} in the UV, such that it counters the effects of the axial anomaly. This allows the usage of an effectively enlarged hidden local symmetry $\Tilde{G}_{F, \,\mathrm{local}}^\mathrm{HLS} \times \Tilde{H}_{F, \,\mathrm{local}}^\mathrm{HLS} = U(2\nf) \times O(2\nf)$ to model the IR Lagrangian. Fixing the spurion source to a vanishing value $\theta = 0$ then allows to take into account the effects of the axial anomaly systematically. 
    For this we first classify all terms of lowest order in $\mathcal{O}(\delta^2)$ and afterwards drop terms which are suppressed in the large $\nc$ limit. In $SU(\nc)$ gauge theories, the order $\mathcal{O}(1/\nc)$ in large $\nc$ suppression of each term in the effective action can be inferred by the counting rules developed in \cite{Kaiser:2000gs}. However, due to the geometric argument given in \cite{tHooft:2002ufq}, as already discussed in section \ref{sec:dark_particles}, we expect that the counting rules for the $SO(\nc)$ case work the same, as long as we do not go beyond leading order. 
    Below, we only discuss the derived Lagrangian, skipping a detailed derivation, since the treatment follows closely the one in \cite{Herrera-Siklody:1996tqr}, where the explicit application of the large $\nc$ counting rules, to drop suppressed contributions, was exemplified at the end of the article.

\subsubsection{Non-anomalous action}
    In order to include the $\etap$ as an effective pNGb in the large $\nc$ limit, a treatment which was well motivated in section \ref{sec:dark_particles}, one extends the field $\Tilde{\gamma}\in \Tilde{G}_F$ to be valued in the enlarged chiral group. Then
    \begin{equation}
        \Tilde{\Sigma} = \Tilde{\gamma} \EFmat \Tilde{\gamma}^\top
    \end{equation}
    transforms in linear under $\Tilde{G}_{F, \,\mathrm{local}}^\mathrm{HLS}$ and is ignorant to $\Tilde{H}_{F, \,\mathrm{local}}^\mathrm{HLS}$. Hence, $\Tilde{\Sigma}$ transforms also linear under $\Tilde{G}_{F, \,\mathrm{local}}^\mathrm{CCWZ} $. The other quantities, like $\Tilde{\mcf}$, are defined analogously as in section \ref{sec:HLS_lagrangian}. 
    All the information of $\etap$ is captured in the phase of $\Tilde{\gamma}$ or better
    \begin{equation}\label{eq:eta_definition}
        \frac{\etap}{\fetap} = \frac{-i}{\sqrt{\nf}} \ln{\det{\Tilde{\gamma}}}\,,
    \end{equation}
    where $\fetap$ is a constant of unit energy dimension in order to interpret $\etap$ as a proper scalar field, related to the effective pNGb. The transformation behaviour of $\etap$ may also be inferred from \eqref{eq:eta_definition}, transforming as a singlet under $O(2\nf)$ and being shifted by a phase under anomalous transformations in $U(2\nf)$. After HLS gauge-fixing, this allows to interpret 
    \renewcommand{\arraystretch}{1.25}
    \begin{equation} \label{eq:reinterpret_ngb}
        \ngb^a = \left\{\begin{array}{ll}
             \frac{\etap}{\fetap} & \text{if } a=0  \\
             \frac{\pid^a}{\fpi}  & a\neq 0
        \end{array}\right.
    \end{equation}
    \renewcommand{\arraystretch}{1}
    where in general $\fetap \neq \fpi$ are different decay constants. In fact, there is no symmetry structure for finite $\nc$ that might relate these two constants. In the EFT this manifests by the presence of an additional kinetic term\footnote{The traces would have vanished in the case without the $\etap$.} for the $\etap$
    \begin{equation} \label{eq:eta_kinetic_term}
        \Tilde{\mathcal{L}}^\mathrm{IR;(2)}_\mathrm{HLS} \supset \frac{\fetap^2 - \fpi^2}{2\fetap^2} \partial_\mu\etap\partial^\mu\etap =  \frac{\fpi^2- \fetap^2}{2\nf}\Tr{\Tilde{\mcf}_{k;\,\mu}}\Tr{\Tilde{\mcf}_{k}^\mu}
    \end{equation}
    independent of the $\pid$ kinetic term.
    We have used here that    \begin{equation}\label{eq:mcf_tilde}
        \Tilde{\mcf}_\mu = i\frac{\partial_\mu \etap}{\fetap} T_0^F + \mcf_\mu
    \end{equation}
    and $\Tr{\mcf_\mu} = 0$. 
    The LEC of the term \eqref{eq:eta_kinetic_term} is fixed by the canonical normalisation condition. Thus, if we would have set $\fetap = \fpi$ in \eqref{eq:reinterpret_ngb}, we would have ended up with an additional LEC that allowed us to again vary them independently. However, from large $\nc$ counting rules, we learn that flavour traces in the EFT are related to quark loops and contributions from such terms are suppressed stronger the more flavour traces appear \cite{tHooft:1980xss}. Hence, by comparison with the $\pid$ kinetic term, we estimate
    \begin{equation}
        \frac{\fetap^2 - \fpi^2}{2\nf \fpi^2} \sim \mathcal{O}(1/\nc) \xrightarrow[\nc\rightarrow0]{}0.
    \end{equation}
    At lowest order $\mathcal{O}(\delta^2)$, we obtain only one additional term dominant in $1/\nc$ suppression, 
    given by
    \begin{equation}
        \Tilde{\mathcal{L}}^\mathrm{IR;(2)}_\mathrm{HLS} \supset -\Delta \meta^2 \frac{\etap^2}{\fetap^2} = \frac{\Delta \meta^2}{\fetap^2 \nf}\ln{2\, \det{\Tilde{\gamma}}}.
    \end{equation}
    The LEC $\Delta \meta^2 $ parametrizes the mass of the $\etap$ in the chiral limit. This log-det formula for the mass term is also seen in QCD for the contributions of the axial anomaly \cite{Witten:1979vv}. From the large $\nc$ counting rules one can infer that 
    \begin{equation}
        \frac{\Delta \meta^2}{\fetap^2} \sim \mathcal{O}(1/\nc) \xrightarrow[\nc\rightarrow0]{}0,
    \end{equation}
    consistent with expectations from the Witten-Veneziano formula \cite{Witten:1979vv, Veneziano:1979ec}. Note that we did not put these terms by hand, but they are present at lowest order within a consistent low energy effective construction based on a combined derivative and large $\nc$ expansion \cite{Herrera-Siklody:1996tqr}.  
    Finally, we take into account what happens if we move away from the chiral limit by introducing quark masses. At lowest order in $1/\nc$, this results in the analog of \eqref{eq:explicit_mass_breaking_contribution} and is explicitly written as
    \begin{equation} \label{eq:explicit_mass_breaking_contribution_eta}
        \tilde{\mathcal{L}}_M^{IR;(2)} = C_\fs \left(\Tr{X^\dagger\tilde{\Sigma}} + \Tr{X\tilde{\Sigma}^\dagger}\right).        
    \end{equation}
    The interpretation of the associated low energy effective constant $C_\fs = \qcon / 16$ remains unchanged. It defines the pion mass $\mpi$ via \eqref{eq:gmor_pions}. From a lowest order expansion we obtain the following GMOR-like relation for the $\etap$ mass
    \begin{equation}
        \meta^2 = \mpi^2 \frac{\fpi^2}{\fetap^2} + \frac{\Delta\meta^2}{\fetap^2}.
    \end{equation}
    Since this mass term, after fixing $\fs = m\EFmat$, breaks the chiral symmetry explicitly, it introduces contact interactions between the dark pions and $\etap$. The part of the lowest order Lagrangian, describing all the interactions among dark $\pid$ and dark $\etap$ fields, is given by  
    \begin{align} \label{eq:lagrangian_eta_pi_interactions}
        \tilde{\mathcal{L}}_{\pid\etap}^{IR;(2)} &= \mathcal{L}_{4\pid}^{IR;(2)} + \frac{\mpi^2 \fpi^2}{8\nf \fetap^4} \etap^4 + \frac{\mpi^2}{2{\fetap^2} \nf} \delta_{ab}\,\pid^a\pid^b\etap^2 + \frac{\sqrt{32} \, }{\sqrt{9\nf}} \frac{\mpi^2}{\fpi\fetap}  D_{abc}\,\pid^a\pid^b\pid^c\etap
        \end{align}
    with $\mathcal{L}_{4\pid}^{IR;(2)}$ given in \eqref{eq:interactions_pions} and $2\,D_{abc} = \Tr{T_a^\F\acombr{T_b^\F,T_c^\F}}$ the totally symmetric $D$-symbol of the flavour algebra. 
    At this order no other terms enter at the same order in the HLS formalism. Especially it seems that the vector mesons do not obtain any contribution from explicit symmetry breaking at the order $\mathcal{O}(\delta^2)$. It is interesting to note here that the vanishing of the $D$-symbol indicates the absence of $\pid\pid \to \pid\etap$ processes. For the $SO(\nc)$ theories presented here this term is present, while it is absent in the two-flavour $Sp(4)$ theory \cite{Kulkarni:2022bvh}. The presence of such interactions is interesting for $\etap$ phenomenology, as discussed in section \ref{sec:anomalous_decays}.
    
\subsubsection{Anomalous action}
    The anomalous action may be derived along the same lines in section \ref{sec:wzw_action}, but for the enlarged HLS group $\Tilde{G}_{F, \,\mathrm{local}}^\mathrm{HLS} \times \Tilde{H}_{F, \,\mathrm{local}}^\mathrm{HLS} = U(2\nf) \times O(2\nf)$. 
    This amounts to the same result, now interpreting the pNGb fields $\ngb$ as in \eqref{eq:reinterpret_ngb}. 
    Especially, for the ungaged action we obtain $\Tilde{S}_{WZW}[\xi] = S_{WZW}[\pi]$ i.e. the result in \eqref{eq:ungauged_wzw_action} remains valid and is independent of $\etap$. While one might guess this already from the structure of the chiral multiplets, one can obtain this result by a simple calculation. By first using \eqref{eq:mcf_tilde} for $\Tilde{\mcf}$ and the fact that $\etap$ commutes with the dark pions, one may verify $(\tilde{\mcf}_{\tau})^2_k = \left(\mcf_{\tau}\right)^2_k$. Hence, the only terms where $\etap$ can enter have totally anti-symmetric\footnote{The square brackets denote total anti-symmetrization i.e. $C_{[i_1,\dots,i_n]} = \frac{1}{n!}\sum_{\sigma \in S^n} \mathrm{sign}\,(\sigma)C_{i_{\sigma(1)},\dots,i_{\sigma(n)}}$. $S^n$ denotes the group of all permutations of $n$ objects. $\mathrm{sign}(\sigma) = \pm 1$ indicates if a permutation $\sigma$ is even or odd. } coefficients $\propto \Tr{\unity T^\F_{[a}T_b^F T_c^F T_{d]}^F}$. These expressions vanish for $SU(2\nc)$, as can be verified by explicit calculation. 
    For SIMP dark matter this means that $\etap$ can not influence the freeze-out via the $3\rightarrow2$ process, even in the limit of large $\nc$, where it is mass degenerate with the dark pions. However, the situation becomes more delicate when also considering the presence of the portal mediator $\zp$, since the $\etap$, as a flavour singlet, may decay to the SM. 

\subsection{The dark photon}
\label{sec:dark_photon_in_ir}
    The dark photon can be included in the IR description by gauging an appropriate one-parameter subgroup of the chiral symmetry. The expression for the resulting non-anomalous terms in the Lagrangian have already been discussed in section \eqref{sec:HLS_lagrangian}. The dark photon enters also in the WZW action and details will be discussed below. The crucial assumption here is that, with decreasing energy, the $U(1)_D$ coupling runs to a fixed value $\ed$, small compared to the scale of the strong interactions, which becomes large in the IR. This allows to treat the dark photon as a perturbation to the strongly interacting system. 
    In terms of the vector meson coupling $\gv$, this amounts to
    \begin{equation}
        r = \frac{\ed}{\gv} \ll 1.
    \end{equation}
    
\subsubsection{Mixing and mass term}
    In the UV description we provided a mass to the $\zp$ via an abelian BEH mechanism. The scalar field involved can always be considered as sufficiently heavy and thus integrated out in the IR theory. 
    The Lagrangian \eqref{eq:HLS_lagrangian_3} already contains a mass term for the dark photon. However, the origin of this mass term is not the BEH mechanism in the UV, but rather results from another BEH-like phenomena, responsible for the masses of the vector mesons in the HLS description \cite{Bando:1987br}. This becomes evident from two observations. First, we realize the fact that this mass term vanishes if the vector mesons are integrated out \cite{Harada:2003jx}. Second, we observe that the Lagrangian  \eqref{eq:HLS_lagrangian_3} is not diagonal in $\Av_\mu$ and $\Af_\mu$. If one introduces a diagonalizing basis, one obtains a massless field that can be interpreted as the physical dark photon \cite{Bando:1987br}. 
    For the physical dark photon $\zp$ we may put a mass term by hand, implementing the features of the BEH effect.
    However, due to the smallness of $r \ll 1$, this mixing is negligible and for all phenomenological purposes the field $B_\mu = -i \ed \zp_\mu Q$ can be interpreted as the physical dark photon. 
    Hence, the dark photon mass term can be introduced directly as 
    \begin{equation}\label{eq:massterm_dark_photon}
        \mathcal{L}^\mathrm{IR}_{\mzp} = -\frac{\mzp^2}{2} \zp_\mu {\zp}^{\mu}.
    \end{equation}
    The same holds true for the inclusion of the kinetic mixing term \eqref{eq:dark_kinetic_mixing}.
    If the vector mesons are not present, the interpretation of the field $B_\mu = -i \ed \zp_\mu Q$ as the physical dark photon is exact. 
    Moreover, the non-diagonal structure of \eqref{eq:HLS_lagrangian_3} also automatically captures mixing between the neutral hadronic singlet $\Tilde{\rho}^{13}$ and $\zp$. This is a phenomenon analogous to $\rho-\gamma$ mixing in the SM. However, in the dark sector considered, there exists only one such neutral vector mesons singlet. The analog of the $\omega$-meson in real world QCD, in terms of quark content, in this theory is called $\tilde{\omega}$ and is part of a flavour triplet. Thus we do not expect an analog to $\omega-\gamma$ mixing and only the $\Tilde{\rho}^{13}$ mixes with $\zp$.
    
\subsubsection{Anomalous decays}\label{sec:anomalous_decays}
\begin{figure}[h]
    \centering
    \includegraphics[width=\textwidth]{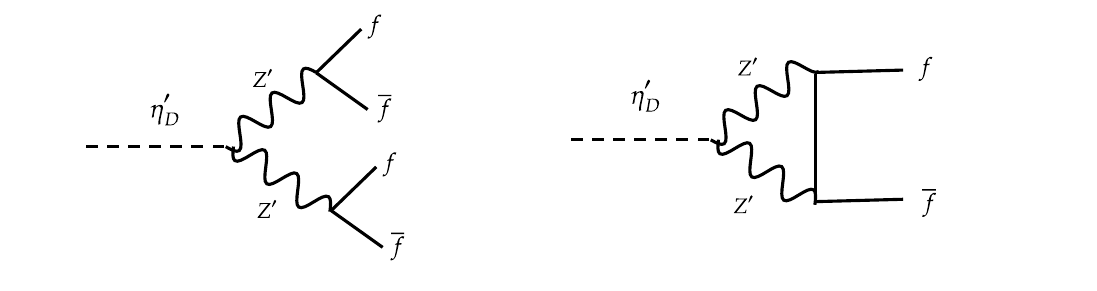}
    \caption{$\etap$ decays to off-shell $\zp$ mediated $4f$ final state (left) and the helicity suppressed $2f$ final state (right).}
    \label{fig:eta_f_decay}
\end{figure}

    In order to include the dark photon into the anomalous terms one simply uses the result \eqref{eq:wzw_action} or \eqref{eq:generalized_wzw_action}, with non vanishing 1-form $B=-i\ed\zp_\mu Q \d x^\mu$. This requires plugging also non zero $F_\tau := F_B[\mcfh_\tau] = \mathrm{d}\mcfh_\tau + \mcfh_\tau^2 \neq 0$ in the formula for the chiral anomaly \eqref{eq:anomaly_expression}. The 1-form $\mcfh_\tau$ was defined in \eqref{eq:schifted_parameterized_maurer_cartan_form}.
    Using appendix \ref{sec:chiral_expansion_appendix}, one may consistently expand the obtained action to lowest order. If we are only interested in scattering processes of five dark pions in the final states, these comprise $5\pi$, $3\pi\zp$ and  $\pi\rightarrow 2\zp$ vertices. All other vertices can only contribute via loops and are thus dropped. The result is
    \begin{align}
        \mathcal{L}_{WZW} 
        \label{eq:wzw_action_expanded_1}
        &\approx \frac{d_\R}{15\fpi^5} \epsilon^{\mu\nu\rho\sigma} \pid^{a} \partial_\mu\pid^{b}\partial_\nu\pid^{c}\partial_\rho\pid^{d}\partial_\sigma\pi^{e} \Tr{T^\F_a T^\F_b T^\F_c T^\F_d T^\F_e} \\
        \label{eq:wzw_action_expanded_2}
        &+i \frac{d_\R \ed}{6} \epsilon^{\mu\nu\rho\sigma} \partial_\mu \ngb^a \partial_\nu \ngb^b \partial_\rho \ngb^c \zp_\sigma \Tr{T_a^F T_b^F T_c^F \Q} \\
        \label{eq:wzw_action_expanded_3}
        & - \frac{d_\R \ed^2}{8} \ngb \epsilon^{\mu\nu\rho\sigma}\partial_\mu Z_{\nu} \partial_\rho Z_{\sigma} \Tr{T_a^\F \Q^2}.
    \end{align}
    At this truncation, the Lagrangian also describes $\ngb \rightarrow \zp\zp$ decay processes and scattering with three dark pions and a dark photon in the final state consistently i.e. it takes into account all relevant effects at same order $\mathcal{O}(\delta^4)$. 
    Note that, due to the discussion in \ref{sec:include_etap}, the $\etap$ will not appear in \eqref{eq:wzw_action_expanded_1}, which we emphasise in our notation. However, in \eqref{eq:wzw_action_expanded_2} and \eqref{eq:wzw_action_expanded_3}, the pNGb fields $\ngb^a$, may be interpreted as either the pions \eqref{eq:interpret_ngb} only or to include also the $\etap$ according to \eqref{eq:reinterpret_ngb}. The results look the same in both cases. In \eqref{eq:wzw_action_expanded_3} we meet condition \eqref{eq:anomaly_cancelation_condition}, guaranteeing the stability of the dark pions, causing the anomalous decay vertex to vanish. Thus, $\pid \rightarrow \zp\zp$ decays are absent and the dark pions are stable. However, for $\etap$ this vertex does not vanish and the singlet may decay into two $\zp$, which may further decay into the standard model. Two processes which make this possible are depicted graphically in figure \ref{fig:eta_f_decay}. In case $\meta \approx \mpi$, this might actually lead to dark pions scattering into $\etap$ via the contact interaction \eqref{eq:lagrangian_eta_pi_interactions}, introduced by the mass term \eqref{eq:explicit_mass_breaking_contribution_eta}, followed by decays of $\etap$ to the SM. Such a reaction may for example lead to additional depletion of DM during freeze out. We comment on phenomenological detail in section \ref{sec:effect_of_eta}. From the discussion in section \ref{sec:dark_photon_in_uv}, it becomes evident that the issue of heaving at least one particle among the $\ngb^a$ that decays via $\zp$ is generic. The charge assignment where $\etap$ is unstable seems to be the best one can do from a stability point of few but for future investigations it might be useful to look into scenarios where meta stability is introduced via a different charge assignment or an explicit mass splitting of the dark quarks. The vertex \eqref{eq:wzw_action_expanded_2} may give resonant dark photon contribution to the thermally averaged five pion scattering cross-section if the mass $\mzp \sim 2\mpi$ is close to the two pion threshold. Such scenarios have already been investigated for $SU(\nc)$  dark sectors \cite{Braat:2023fhn}. Further investigations, using the theory descriptions presented here, may be carried out for the (pseudo-)real case in the future. 
    
\subsubsection{Vector meson dominance} \label{sec:wzw_vmd_limit}
    The particular solution of the anomaly equation does not have any information on the vector mesons, which enter only via the homogeneous part. 
    For the following we again use the language of differential forms to stay concise in notation. Expanding and truncating the homogeneous solutions \eqref{eq:non_anomalous_wzw_1} to \eqref{eq:non_anomalous_wzw_4} gives 
        \begin{align}
        \label{eq:non_anomalous_wzw_expand_1}
        \mathcal{L}^\mathrm{IR, anom.}_{1} &\approx 
        \Tr{\ngb \d \ngb \d \ngb \d \ngb \d \ngb} + \Tr{\d \ngb\d \ngb\d \ngb \Af} - \Tr{\d \ngb\d \ngb\d \ngb \Av}\\
        \label{eq:non_anomalous_wzw_expand_2}
        \mathcal{L}^\mathrm{IR, anom.}_{2} &\approx 0 
        \\
        \label{eq:non_anomalous_wzw_expand_3}
        \mathcal{L}^\mathrm{IR, anom.}_{3} &\approx 
        \Tr{\ngb\d V \d V } - \Tr{\ngb\d V \d B} - \Tr{\d\ngb\d\ngb\d\ngb V}\\
        \label{eq:non_anomalous_wzw_expand_4}
        \mathcal{L}^\mathrm{IR, anom.}_{4} &\approx 
        \Tr{\ngb\d B \d V } - \Tr{\ngb\d B \d B } - \Tr{\d\ngb\d\ngb\d\ngb B}.
    \end{align}
    For the expansion we used explicitly the restriction of $B$ to the unbroken algebra $\liea{h}_F$, applicable for the dark photon. In order to gain an intuition on what values the undetermined parameters $C_i^\mathrm{anom.}$ may adopt, it is useful the see how we can adjust these parameters to incorporate complete vector meson dominance (cVMD) in the anomalous vertices. By cVMD we mean here the suppression of all vertices containing a dark photon in \eqref{eq:wzw_action_expanded_1}-\eqref{eq:wzw_action_expanded_3} and replacing them by vertices with a neutral vector meson. In this case all the interactions in \eqref{eq:wzw_action_expanded_1}-\eqref{eq:wzw_action_expanded_3} are described by interactions of pions and neutral vector-mesons, which then mix with the dark-photon. But all direct anomalous interactions with the dark photon are absent. Expanding the general solution \eqref{eq:generalized_wzw_action} to leading order consistently results in 
    
    \begin{align}
        \label{eq:generalized_wzw_action_expanded_1}
       \mathcal{L}_{WZW} 
        &\approx  \frac{i\, d_\R}{8 \pi^2} \left(\frac{8}{15}+C^\mathrm{anom.}_{1} \right)\Tr{\ngb \d\ngb\d\ngb\d\ngb\d\ngb} \\
        \label{eq:generalized_wzw_action_expanded_2}
        &+  \frac{i\, d_\R}{8 \pi^2}C^\mathrm{anom.}_{3}\Tr{\ngb\, \d\Av\d\Av} \\
        \label{eq:generalized_wzw_action_expanded_3}
        &- \frac{i\, d_\R}{8 \pi^2}\left(C^\mathrm{anom.}_{1}+C^\mathrm{anom.}_{3}\right)\Tr{\d\ngb\d\ngb\d\ngb \,\Av}\\
        \label{eq:generalized_wzw_action_expanded_4}
        &+ \frac{i\, d_\R}{8 \pi^2}\left(C^\mathrm{anom.}_{4}-C^\mathrm{anom.}_{3}\right)\Tr{\ngb \d\Av\d\zp} \\
        \label{eq:generalized_wzw_action_expanded_5}
        &+ \frac{i\, d_\R}{8 \pi^2}\left(1-C^\mathrm{anom.}_{4}\right)\Tr{\ngb\d\zp\d\zp} \\
        \label{eq:generalized_wzw_action_expanded_6}
        &+ \frac{i\, d_\R}{8 \pi^2}\left(C^\mathrm{anom.}_{1}-C^\mathrm{anom.}_{4}+\frac{4}{3}\right)\Tr{\d\ngb\d\ngb\d\ngb \zp}.
    \end{align}
    Here we used the 1-forms $\Av = -i \gv \Av^A_\mu  T^{\F}_A\,\d x^\mu$ and $\zp = -i \ed \zp_\mu \Q\,\d x^\mu$ together with $\tr{\ngb\d \Av \d \zp} = \tr{\ngb\d \zp \d \Av}$.
    Then cVMD would demand the last three vertices \eqref{eq:generalized_wzw_action_expanded_4}, \eqref{eq:generalized_wzw_action_expanded_5} and
    \eqref{eq:generalized_wzw_action_expanded_6} to vanish. This requirement fixes all the relevant LECs to the values $C_1^\mathrm{anom.} = 1/3$ and $C_3^\mathrm{anom.} = C_4^\mathrm{anom.} = 1$. Additionally, to cVMD in the anomalous sector, one may require $\clhs = 2$ to implement vector meson dominance in the pion form factor, as discussed in \ref{sec:HLS_lagrangian}. It should be noted that similar to the modification of the $4\pi$ interaction vertex, discussed in \eqref{eq:explicit_mass_breaking_contribution}, the $5\pi$ interaction is also modified through the coefficient $C^\mathrm{anom.}_{1}$. This modification does not vanish in the cVMD limit. However, when computing the $3\to2$ interactions with $C^\mathrm{anom.}_{i}\neq 0$, all terms should be consistently taken into account. In real world QCD, a limit of complete vector meson dominance (VMD) like this does not seem to be favored by experiment and the physical parameters specify a small deviation of this point \cite{Harada:2003jx}. Although one can not  experimentally verify cVMD properties for dark matter, we may use this as guidance for choosing these parameters. 
    
    For $SU(\nc)$ it has been demonstrated that including the vector mesons in the low energy effective description might help to resolve some problems with perturbativity and related issues of the validity of the EFT including pions only \cite{Choi:2018iit}. An investigation of this issue in the $SO(\nc)$ case is missing and left for future investigations, given the provided framework in this paper.   

\section{First phenomenological applications}
    \label{sec:first_phenomenological_applications}
The chiral Lagrangian developed in the previous section can now be used to study dark matter phenomenology. In particular inclusion of heavier states such as the vector mesons $V^\mu$ and the flavour singlet $\etap$, may have implications on the viable dark matter parameter space. In this section, we therefore analyse the effect of the flavour singlet on dark matter phenomenology while ignoring the vector mesons which may be nearby. By doing so, we can explicitly demonstrate the effect of $\etap$ without worrying about vector meson induced effects. We expect vector meson induced number changing or semi-annihilation processes to be relevant in this theory as well. Their effects will be analysed in a future work. The relevant free parameters of our analysis are $\mpi, \mpi/\fpi$. All other quantities such as the masses of vector mesons and flavour singlet as well as the related decay constants and couplings are a function of the two free variables. Non-perturbative methods e.g. lattice calculations are necessary to establish these functions. Some of this analysis can be found in \cite{Hietanen:2012sz}, however not all relations are yet available. In particular, computing the properties of flavour singlet is a challenging task, e.g. see \cite{Bennett:2023rsl} for an analysis in the context of $Sp(4)$ theories. For our phenomenological analysis, we therefore treat the mass of the $\etap$ and the corresponding decay constant $\fetap$ as free parameters. For sufficiently large $\nc$ we know $\meta\sim\mpi$ and $\fetap\sim\fpi$, which can be used to choose meaningful values. 
\subsection{Boltzmann equations}
For our numerical analysis we solve the following system of Boltzmann equations allowing for the possibility for $\etap$ to decay out of equilibrium. 

\begin{eqnarray}
\dot{n_\pi} + 3\,H\,n_\pi &=& \langle \sigma v\rangle_{\etap\etap\to\pi\pi} \left[n^2_\etap - \frac{n^2_\pi}{n^2_{\pi, eq}}n^2_{\etap,eq} \right]\, + n_\pi\langle \sigma v\rangle_{\pi\etap\to\pi\pi} \left[n_\etap - \frac{n_\pi}{n_{\pi, eq}}n_{\etap,eq} \right]\, \nonumber\\ 
&-& \langle \sigma v^2\rangle_{3\pi\to 2\pi}\,\left( n^3_\pi - n^3_{\pi,eq}\right)\\
\dot{n}_{\etap} + 3\,H\,n_\etap &=& - n_\pi\langle \sigma v\rangle_{\pi\etap\to\pi\pi} \left[n_\etap - \frac{n_\pi}{n_{\pi, eq}}n_{\etap,eq} \right] - \langle \sigma v\rangle_{\eta\etap\to\pi\pi} \left[n^2_\etap - \frac{n^2_\pi}{n^2_{\pi, eq}}n^2_{\etap,eq} \right] \nonumber \\
&-& \langle\Gamma_\etap\rangle \left(n_\etap - n_{\etap,eq} \right),
\label{eq:Boltzmann_system}
\end{eqnarray}
where $n_\pi$, $n_\etap$ denote pion and $\etap$ number densities and $\langle\ldots\rangle$ denote thermal averages, which are detailed below. We define the Hubble constant and the entropy as 
\begin{equation}
    H = \displaystyle\frac{\sqrt{g_*} \pi T^2}{\sqrt{90} M_{pl}} \qquad \qquad
    s = \frac{2\pi^2 g_{*s}}{45} T^3.
\end{equation}
with $g_*, g_{*s}$ being the effective SM degrees of freedom. We use the data for the SM effective degrees of freedom given in \cite{Saikawa:2018rcs}. Finally, we approximate 
\begin{equation}
    \langle\Gamma_\etap\rangle \simeq \displaystyle\frac{K_1(\meta/T)}{K_2(\meta/T)} \, \Gamma(\etap), 
\end{equation}
where $K_1, K_2$ are modified Bessel functions of 1st and 2nd kind. 
It is clear that the system decouples when $ \langle \sigma v\rangle_{\pi\etap\to\pi\pi}  = \langle \sigma v\rangle_{\eta\etap\to\pi\pi}  = 0$ and an analytical approximation for the resulting $3\pi\to2\pi$ Boltzmann equation can be found. We employ the formalism given in \cite{Braat:2023fhn} for such an analytical treatment.
\subsection{Relevant $2\to2$ and $3\to 2$ cross sections}
We compute thermally averaged cross sections using {\tt Mathematica} and by explicitly summing over relevant generators. We use {\tt FeynCalc} to compute Lorentz traces. We compare our results with \cite{Hochberg:2014kqa}. It should be noted that the global flavour symmetry in \cite{Hochberg:2014kqa} is $SU(\nf)$ while in our convention it is $SU(2\nf)$. Therefore when comparing we substitute $\nf = 4$ for results from \cite{Hochberg:2014kqa} and $\nf = 2$ for our results. 
We first present the form of the $2\to 2$ self-scattering cross section as it does not need thermal averaging. 

\subsubsection{$2\pi \to 2\pi$ self-scattering}
The self-scattering cross section among all pions ($N_\pi=9$) of the theory is given by 
\begin{eqnarray}
    \sigma_{2\pi\to 2\pi} &=& \displaystyle\frac{1}{N^2_\pi}\frac{1}{64 S_f\,\pi^2\,s} \displaystyle\int|\mathcal{M}|_{2\to 2}\,d\cos\theta d\phi \nonumber\\
    &\approx& \displaystyle\frac{1}{4608\pi\,m^2_{\pi}}\displaystyle\frac{\left(145\,m^4_{\pi} +384\,m^2_{\pi}\,p^2 +320\, p^4 \right)}{f^4_{\pi}} \nonumber
\end{eqnarray}
where we have used $S_f=2$ and $s\approx 4\,m^2_\pi$. Our result agrees with \cite{Hochberg:2014kqa} in the limit $p\to 0$ where we substitute $\nf=4$ in their calculations to be consistent with their global flavour symmetry and accounting for different definitions of $\fpi$ ($f^{\mathrm{Ref.Ref.[1]}}_\pi = 2\,\fpi$). For our numerical calculations we subsequently use $p\to 0$. In order to match to the upper limit on DM self-interaction cross section we use $2 {\rm cm}^2/g$ \cite{Robertson:2016xjh, Wittman:2017gxn} and obtain
\begin{equation}
    \displaystyle\frac{\sigma_{2\pi\to 2\pi}}{\mpi} = 2.2\times 10^5\displaystyle\frac{\rm{cm}^2}{g}\,\frac{145}{4608\pi}\,\frac{{\rm MeV^{-3}}}{m^3_{\pi}}\displaystyle\frac{m^4_\pi}{f^4_\pi} \lesssim 2 \frac{\rm{cm}^2}{g}.
\end{equation}
This leads to a limit on the pion mass of 
\begin{equation}
    \mpi\gtrsim 10.32\,{\rm MeV} \displaystyle\left(\frac{\mpi}{\fpi}\right)^{4/3}.
\end{equation}

While in complete isolation, all nine pions are expected to be present today in the Universe, in presence of coupling with the external $\zp$, this may not be the case \cite{Kondo:2022lgg}. Coupling  with $\zp$ breaks the flavour symmetry and in turn leads to radiative corrections to the masses of charged pion. These are proportional to $2\,\kappa\,e^2_D/f^2_\pi$, where $\kappa$ is low energy constant, and thus the charged pions are expected to be heavier than the neutral counterparts. Once the $3\pi\to 2\pi$ interactions freeze-out, the residual forward annihilation processes $\pi^+ \pi^- \to \pi^0 \pi^0$ continue depleting the abundance of all charged pions. These forward annihilation processes can be desirable as it eliminates any millicharged dark matter from the present Universe and evades any $\zp$ mediated direct detection constraints. The details of exact charged pion abundance depend on the details of the mediator sector. In order to estimate the effect of such forward annihilation we consider here the two extremes, one where all nine pions remain in the present Universe and second, when only the neutral pions remain. Correspondingly, we also compute the self-interaction cross section among the three neutral species only ($N_\pi = 3$). This results in 

\begin{equation}
    \displaystyle\frac{\sigma_{2\pic^0\to 2\pic^0}}{\mpi} = 2.2\times 10^5\displaystyle\frac{\rm{cm}^2}{g}\,\frac{23}{1536\pi}\,\frac{{\rm MeV^{-3}}}{m^3_{\pi}}\displaystyle\frac{m^4_\pi}{f^4_\pi} \lesssim 2 \frac{\rm{cm}^2}{g},
\end{equation}
and leads to a lower bound on pion mass of $\sim$ 8 MeV at $\mpi/\fpi = 1$.

\subsubsection{$3\pi\to 2\pi$ cannibalisation process }
All nine pions participate in the $3\to2$ process. The corresponding annihilation cross section is given by 
\begin{equation}
\langle\sigma\,v^2\rangle_{3\to 2} = \displaystyle\frac{d^2_\R}{N^3_{\pi}}\displaystyle\frac{25}{2048}\frac{\sqrt{5}}{\pi^5}\frac{m^5_{\pi}}{x^2 f^{10}_\pi}.
\end{equation}
Our results differ by a factor of 1/12 with respect to \cite{Hochberg:2014kqa} after rescaling for $f^{\mathrm{Ref.[1]}}_\pi = 2\,\fpi$. There are two reasons behind this, first, explained in\eqref{eq:WZW_coeff} $d_\R = \nc$ and the factor of 1/3 arises from correcting for Galilean invariant thermally averaged cross section~\cite{Chu:2024rrv, Kamada:2022zwb}.

\subsubsection{$\eta\pi \to \pi\pi, \etap\etap \to \pi\pi$ processes} \label{sec:eta_scattering}
After explicit symmetry breaking by the charge assignment, the remaining $SU(2)_I\times U(1)_B$ symmetry restricts the possible scattering processes. For example in $\etap\pi \to \pi\pi$ scattering, only vertices where all three pion states are charged differently are non-vanishing. $U(1)_B$ conservation further demands that $\etap\pi^0$ or $\etap\etap$ scatter into a pair of anti-particles $\pi^+\pi^-$. In fig. \ref{fig:classification_of_particles} we illustrated that the nine pions of the theory break into three triplets corresponding to neutral and $\pm$ charged states. Hence, all of the pions in the scattering processes $\pi \etap \to \pi \pi$, $\etap \etap \to \pi \pi$ must belong each to a different multiplet. Considering this, the squared amplitudes are

\begin{equation}
    |\mathcal{M}_{\pi \etap \to \pi \pi}|^2 = \displaystyle\frac{9\,m^4_{\pi}}{2 f^2_{\etap}\,f^2_{\pi}} \qquad \qquad\qquad
    |\mathcal{M}_{\etap \etap \to \pi \pi}|^2 = \displaystyle\frac{9\,m^4_{\pi}}{8\,f^4_{\etap}}.
\end{equation}
The corresponding thermally averaged cross-section is given by 
\begin{equation}
    \renewcommand{\arraystretch}{1.5}
     \langle \sigma v_{\rm rel} \rangle  = \left\{
     \begin{array}{ll}
         \displaystyle\frac{9}{512\pi } \displaystyle\frac{m^2_{\pi}}{f^4_{\etap}} \sqrt{1-\displaystyle\frac{m^2_\pi}{m^2_\eta}} &  \quad {\rm for  }\,\, \etap \etap\to\pi\pi  \\
         \displaystyle\frac{9}{128\pi } \displaystyle\frac{m^2_{\pi}}{f^2_{\etap}f^2_{\pi}} \sqrt{1-\displaystyle\frac{4 m^2_\pi}{(\mpi+\meta)^2}} &\quad {\rm for  }\,\, \etap \pi\to\pi\pi
     \end{array}\right.     
\end{equation}
where we have used $s = 4 \meta^2$ or $s = (\meta + \mpi)^2$ for $\etap \etap\to\pi\pi$ and $\etap \pi\to\pi\pi$ processes respectively in the non-relativistic limit. 

\subsection{Numerical results}

\subsubsection{$3\pi\to 2\pi$ WZW processes}
\begin{figure} [h!]
\centering
    \includegraphics[width=0.43\textwidth]{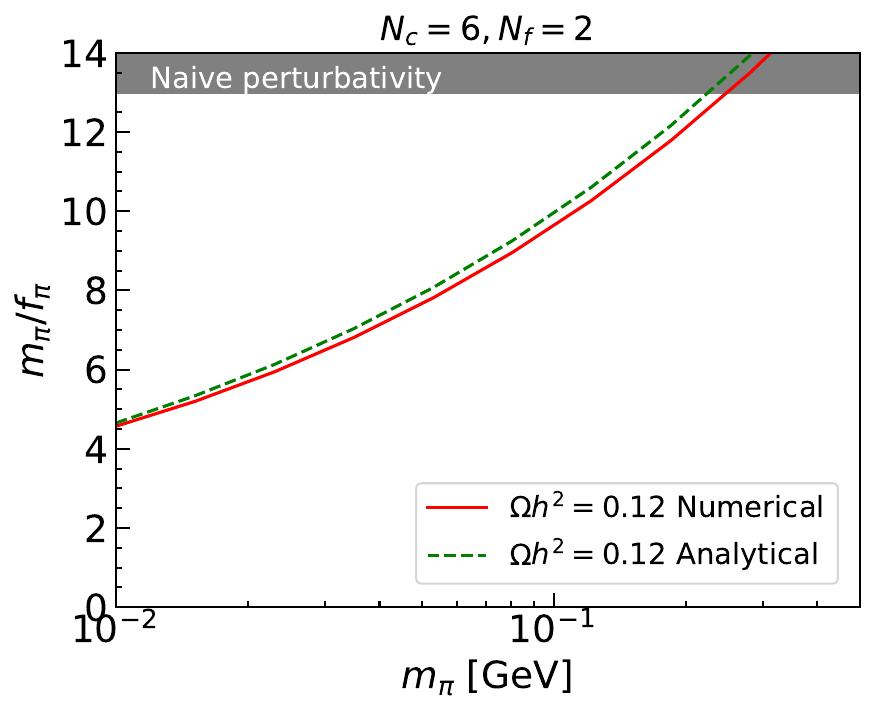}
    \includegraphics[width=0.45\textwidth]{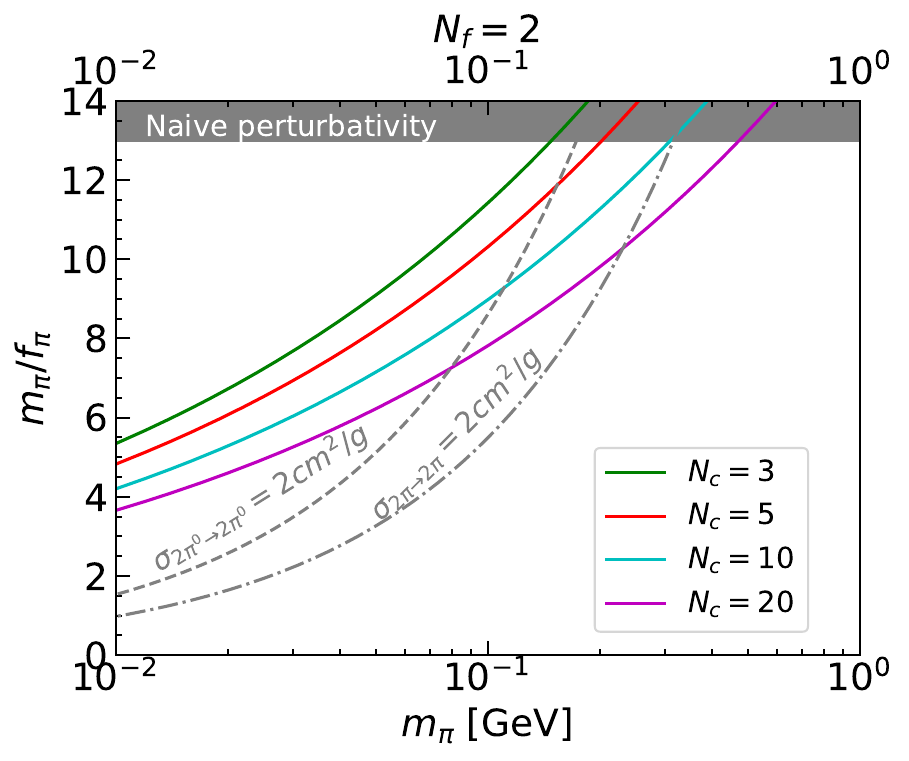}
    \caption{(left): Relic density contour obtained by numerically solving Boltzmann equation (red solid line) and corresponding analytical solution (green dashed line) for fixed $\nc = 6, \nf = 2$. Relic density contours ($\Omega h^2 = 0.12$) in $\mpi/\fpi$ -- $\mpi$ plane (solid lines) for various values of $\nc$ and two Dirac fermions. Contours representing the DM self interaction cross section of $2~\rm{cm}^2/g$ are also shown if all 9 pions take part in the interaction (light green dot dashed line) or only neutral pions ($\pic^1,\pic^2,\pic^3$) are accounted for (dashed dark green line). }
    \label{fig:relic_summary}
\end{figure}
We first demonstrate the region of viable parameter space by requiring correct relic density and obeying the self interaction cross section for pion only processes. Correspondingly, in~\eqref{eq:Boltzmann_system} we set $\langle\sigma v\rangle_{\eta\etap\to\pi\pi} = \langle\sigma v\rangle_{\eta\pi\to\pi\pi} = \langle\Gamma_\etap\rangle = 0$ and solve the resulting Boltzmann equation numerically. We also employ analytical approximation as shown in \cite{Braat:2023fhn}. 
Fig.~\ref{fig:relic_summary} (left) shows a remarkable agreement between the numerical solution of the Boltzmann equation (red solid line) and the analytical approximation (green dashed line) for $\nc = 6, \nf = 2$. Given this agreement, we use the  analytical approximation to compute relic density in the right panel. In fig.~\ref{fig:relic_summary} (right), we show the combined results of self-interaction cross-section and relic density constraints for several values of $\nc$ with two Dirac flavours. We compute the self interaction cross section among all nine pions in the theory as well as using only the charge neutral three pions ($\pic^1,\pic^2,\pic^3$). Given the smaller number of neutral states the self-interactions and relic density can be reconciled for smaller pion mass for a given $\nc$. As the self-interaction cross-section is independent of $\nc$, the self-interaction favoured region does not depend on $\nc$. $\nc > 10$ is required for self-interactions and relic density to be satisfied at the same time for self interactions among all pion states, while $\nc$ can be smaller for neutral only states. The pion mass required for a phenomenologically viable parameter space decreases for larger $\nc$.

\subsubsection{Effect of $\etap$} \label{sec:effect_of_eta}
Above discussion demonstrates that for large $\nc$ one can fulfill relic density requirement for smaller pion masses. For such large values of $\nc$, the purely gluonic contributions to $\meta$ will become suppressed. Therefore, we investigate the importance of $\etap$ for relic density calculations in a regime where the relative deviation $\frac{\meta - \mpi}{\mpi}$ is small. 

We begin with estimating the $\etap$ lifetime as it could strongly affect the relic density estimates. A short lived $\etap$ would decay in equilibrium and act as a semi-annihilation partner much similar to the $\rho$ meson illustrated in \cite{Berlin:2018tvf, Hochberg:2015vrg, Bernreuther:2023kcg}, while a very long-lived state which mixes with $\pi$ and decays out of equilibrium e.g. in mass split theories can also be of phenomenological interest \cite{Katz:2020ywn, Hochberg:2018vdo}. 

The flavour singlet $\etap$ decays via an anomalous vertex given in \eqref{eq:wzw_action_expanded_3}. This leads to two possible $\etap$ decay modes. First decay mode is analogous to the anomalous SM neutral pion ($\pi^{0}_{\rm{SM}}\to\gamma\gamma$ decay), $\etap \to \zp \zp \to 4f$, where $f$ denotes SM fermion and second is the loop induced $\etap \to 2f$ final state analogous to electromagnetic decays of SM neutral pion. 

The exact lifetime of $\etap$ is not relevant for our phenomenological studies, therefore we follow \cite{Hochberg:2018vdo} to estimate the lifetime. Taking into account symmetry factors for our setup the lifetime estimates are 
\begin{eqnarray}
    \Gamma(\etap \rightarrow 4f)  
    &=& \frac{\meta^3}{\pi}\left(\frac{\alpha_D d_\R}{8\pi \fetap} \tr{T_0^F\mathcal{Q}^2} \right)^2\left(\frac{\alpha}{2\pi} \epsilon^2 \left(\frac{\meta}{2 \mzp}\right)^4\right)^2 \nonumber\\
    & = & 1.06\times10^{-11}\displaystyle\frac{\alpha_D^2 d^2_\R \mpi^9 \epsilon^4}{\mzp^8}\displaystyle\left(\frac{\meta}{\mpi}\right)^{11}\displaystyle\left(\frac{\mpi}{\fpi}\right)^2
    \label{eq:etap_4f}
\end{eqnarray}

where we substituted $Q_1 = 1, Q_2= -1$ and assumed $\fetap = \fpi$. Similarly
\begin{eqnarray}
    \Gamma(\etap \rightarrow 2f)  
    &=& \frac{\meta^3}{\pi}\left(\frac{\alpha_D d_\R}{8\pi \fetap} \tr{T_0^F\mathcal{Q}^2} \right)^2\left(\frac{\alpha}{2\pi} \epsilon^2 \left(\frac{m^2_\etap}{\mzp^2}\right)\left(\frac{m_f}{\mpi}\right)\right)^2 \nonumber\\
    &=& 2.72\times 10^{-9} \displaystyle\frac{\alpha_D^2 d^2_\R m^2_f m^3_\pi \epsilon^4}{\mzp^4}\displaystyle\left(\frac{\meta}{\mpi}\right)^{7}\displaystyle\left(\frac{\mpi}{\fpi}\right)^2
    \label{eq:etap_2f}
\end{eqnarray}
where $m_f$ is the mass of heaviest phase space allowed Standard Model fermion. As $\etap \rightarrow 2f$ interaction is helicity suppressed, the $\etap$ decays to the heaviest available SM fermion via this decay mode.

From eq.\eqref{eq:etap_4f}-\eqref{eq:etap_2f} it is clear that $\etap \rightarrow 4f$ decay mode dominates the lifetime due to larger $\mzp$ suppression. As a benchmark scenario, for $\nc = 5, \alpha_D = 1/(4\pi), \epsilon = 10^{-4}, \meta/\mpi = 1.01, \mpi/\fpi = 10, \mpi= 0.1\,\rm{GeV}$ and $\mzp = 3\,\rm{GeV}$ the lifetime is $\sim 10^8\,\rm{sec}$. This shows that the lifetime is generally relatively large $\gg 1\,{\rm sec}$. Owing to this observation, we set $\Gamma_\etap = 0$ in the Boltzmann equations since it is irrelevant for the timescales of interest. It is important to note that Big Bang Nucleosynthesis constraints may play a role for such scenarios \cite{Hufnagel:2018bjp}, these constraints can be evaded by appropriately adjusting $\zp$ mass and couplings \cite{Katz:2020ywn}.

\begin{figure} [htb!]
\centering
    \includegraphics[width=0.45\textwidth]{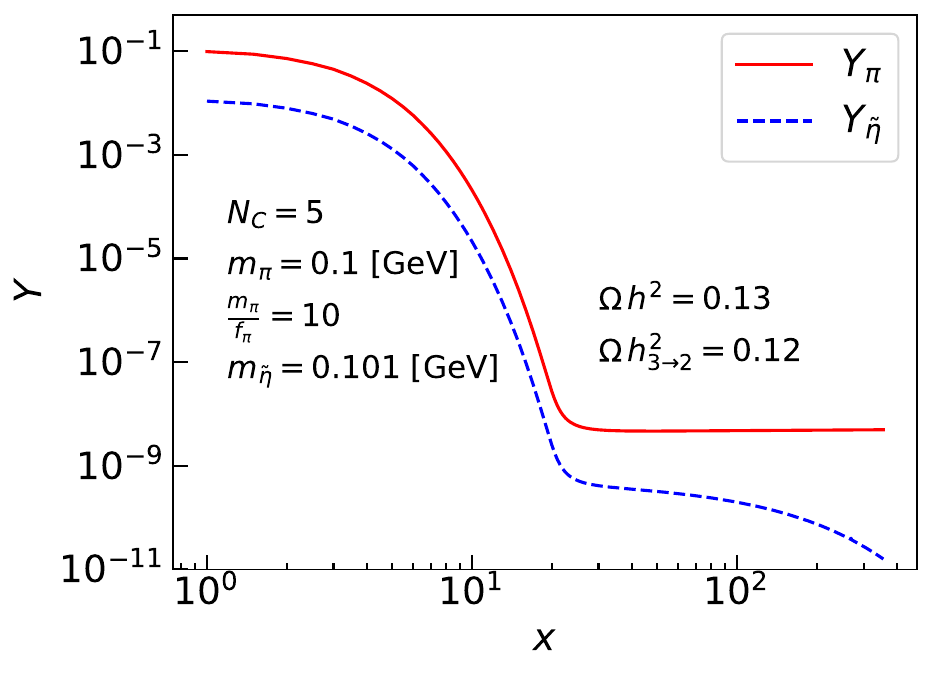}
    \includegraphics[width=0.45\textwidth]{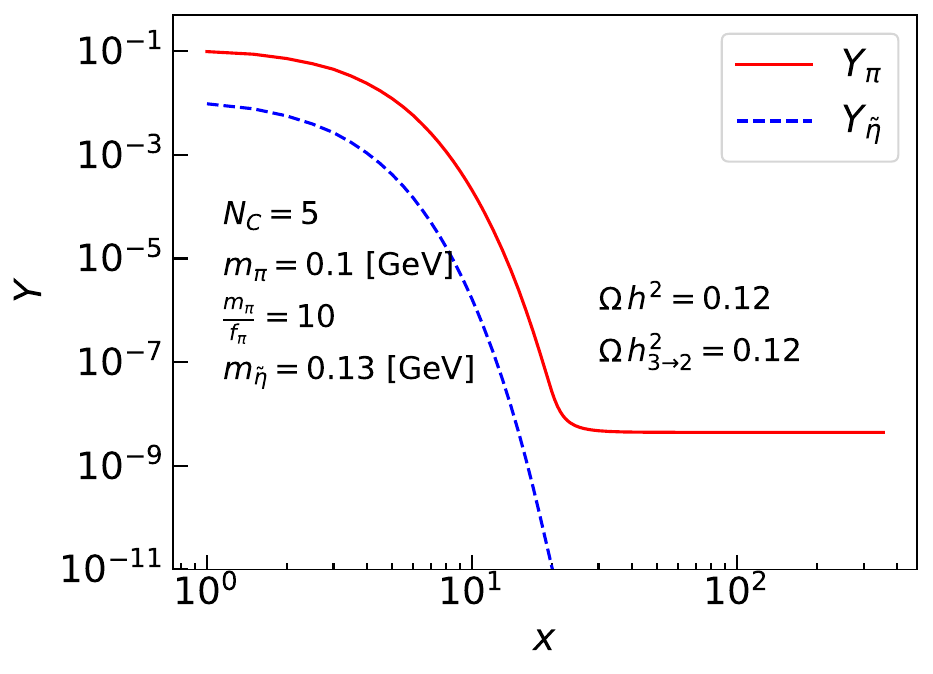}
    \caption{Evolution of pion and $\etap$ abundance as a function of $x$ for two different values of $\meta$. $\Gamma_\etap$ is set to 0 to obtain these results. The $3\to2$ only relic density for this benchmark satisfies the DM relic density, relic density including $\etap$ is 0.13 and 0.126 for left and right panel respectively.}
    \label{fig:etap_relic}
\end{figure}

In fig.~\ref{fig:etap_relic}, we show the effect of inclusion of $\etap$ in relic density. There are three number changing processes of interest here $3\pi\to 2\pi$, $\etap\etap\to\pi\pi$ and $\pi\etap\to\pi\pi$. Including all three processes, leads to a small increase in the overall relic density if the mass difference between $\etap$ and $\pi_D$ is small. We obtain the correction to be up to 8\% for a percent level $\mpi-\meta$ splitting. This relative increase rapidly vanishes as $\etap$--$\pi_D$ mass difference increases and by 30\% mass splitting the $\etap$ makes no difference to the relic density. The increase in the relic density can be understood as a effect of residual forward annihilation processes $\etap\etap\to\pi\pi$ and $\pi\eta\to\pi\pi$ given that $\mpi \lesssim \meta$. It is also interesting to understand the relative importance of $\etap\etap\to\pi\pi$ and $\pi\etap\to\pi\pi$ processes. The processes involving two $\etap$ suffer a stronger Boltzmann suppression compared to processes involving one $\etap$. The $\etap\etap\to\pi\pi$ cross section depends on both $\meta$ and $\fetap$ however given that this processes is more Boltzmann suppressed compared to $\pi\etap\to\pi\pi$, the final relic density does not sensitively depend on the value of $\fetap$. Finally, the increase in the relic density is more pronounced for larger $\mpi$ as the $3\to 2$ cross sections become comparable to semi-annihilation cross sections.

\section{Generalizations of the $SO(N_C)$-vector model}
    \label{sec:generalizations}
    So far the model discussed was based on two Dirac fermions in the vector representation of $SO(\nc)$.  
The essential property used was the reality of the fermions representation. Hence generalisations to arbitrary gauge groups $G_D$ and number of Dirac fermions $\nf$ are possible.
The latter is straightforward and we kept the arbitrary number $\nf$ in the notation where the generalised statement holds. The case $\nf = 2$ was of special interest since it is minimal in the sense that for $\nf = 1$ no WZW term exists\footnote{Although there might be a different portal mechanism \cite{Davighi:2024zip} which makes this scenario interesting.}. The intermediate case of a theory built from 3 Majorana forms \cite{Hochberg:2014kqa}, denoted as $\nf = 1.5$, allows a WZW. However, in this case the neutral pion is always a flavour singlet, once the dark photon is introduced.

Most of the results may be generalised for a general dark gauge group $G_D$ as long as the fermions transform in a real representation $\R$ of $G_D$. 
However, there are two major differences to be taken into account when deviating from the $SO( \nc)$ vector scenario. First, in order to guarantee occurrence of chiral symmetry breaking, the theory must lie below conformal window \cite{Sannino:2009aw, Lee:2020ihn}, which strongly depends on the choice of $G_D$ and $\R$ and in turn constraints theory realisations. 

Secondly, there are additional features related to the anomalously broken axial symmetry. As discussed in \ref{sec:dark_particles}, the criterion \eqref{eq:large_nc_criterion} may tell us if we can expect the $\etap$ to be light in an appropriate t'Hooft large $\nc$ limit. If $\etap$ can not be expected to become massless in the large $\nc$ limit, the methods developed in \ref{sec:include_etap} can  not be applied. This was recently argued also in \cite{Anber:2023yuh} in the context of 2 index chiral gauge theories. 

Furthermore, the Dynkin index $T_\R \neq 1$ may be different from unity, see tabular \ref{tab:properties of representation} in appendix \ref{sec:topological_charge}. According to equation \eqref{eq:set_of_non_anomalous_determinants}, not only charge conjugation but a $\mathbb{Z}_{2T_\R}$ subgroup of $U(2N_F)$  matrices, with complex determinants $ e^{-i k \pi/T_\R}$, can be non-anomalous. Here $k = 0,\dots,2T_\R-1$. To give a representation of this subgroup, it is more useful to work with a basis of Majorana fermions, rather than the Nambu-Gorkov formalism\footnote{In order to see how the representation of the flavour matrices can be related, see appendix \ref{sec:generators_of_su2nf}.}. Then the matrices
\begin{equation}
    C_k := 
    \left(\begin{matrix}
        1&0&0&0\\
        0&\unity_{\nf-1} & 0 & 0 \\
        0& 0 & e^{i k \pi/T_\R} & 0 \\
        0&0 & 0 & \unity_{\nf-1} 
    \end{matrix}\right)
\end{equation}
furnish a representation of $\mathbb{Z}_{2T_\R}$. Note that $C_{T_R}= C_u$ given in \eqref{eq:cu_mat}, when changing back to the Nambu-Gorkov formalism. When multiplying with a suitable flavour transformation $U^\F \in SU(2\nf)$ of unit determinant i.e. $\det{U^\F} = 1$, we may obtain
\begin{equation}
        U^\F C_k := 
    \left(\begin{matrix}
        e^{i k \pi/2T_\R}&0&0&0\\
        0&\unity_{\nf-1} & 0 & 0 \\
        0& 0 & e^{i k \pi/2T_\R} & 0 \\
        0&0 & 0 & \unity_{\nf-1} 
    \end{matrix}\right).
\end{equation}
This is the representation of an axial transformation of the first Dirac fermion $q^{(1)}$ in the Majorana basis adopted. The path integral measure changes as
\begin{equation}
   \mathcal{D}q\mathcal{D}\overline{q} \xmapsto{\quad U^\F C_k \quad} e^{-i 2 k(n_L-n_R) \pi/2T_\R} \mathcal{D}q\mathcal{D}\overline{q}
\end{equation}
with the difference of fermion zero modes $(n_L-n_R) = 2T_\R \Qtopo$ given via the Atiyah-Singer index theorem \cite{Bilal:2008qx}. Since $\Qtopo$ is always an integer, all these transformations leave the path-integral invariant. The occurrence of these symmetries is consistent with predictions made with the effective 't Hooft vertex \cite{Creutz:2007yr, tHooft:1976rip}. 

However, the only transformations satisfying the isotropy condition \eqref{eq:invariance_condition_mass_term} are given by $\mathbb{Z}_2 = \{C_0 = \unity, C_{T_R} = C_u\}$. Thus, the discrete axial symmetry is spontaneously broken by the chiral condensate. For sufficiently low energies, the description of the dark pions is expected to still hold as derived in section \ref{sec:long_range_description}. 

However, one expects that the spontaneous breakdown of discrete global symmetries leads to the formation of domain walls \cite[Chpt. 23]{Weinberg:1996kr}. Due to the explicit symmetry breaking terms in the theory, those domain walls will either not form at all or are unstable and eventually collapse. The latter leads to potential gravitational wave signatures \cite{Saikawa:2017hiv}. These potential signatures are complementary to the ones produced by a first order phase transition as suggested in \cite{Schwaller:2015tja}. Further investigation is beyond the scope of this paper and left for the future.

\section{Summary and conclusion}
    \label{sec:summary_and_conclusion}
    Pseudo-Nambu Goldstone bosons as dark matter candidates emerging from new confining strongly interacting scenarios present an interesting opportunity to reconcile dark matter relic density generated by $3\to2$ WZW interactions together with large self-interaction cross sections generated by $2\to2$ Goldstone scatting processes. Such confining non-Abelian sectors also present new signatures at colliders. Investigations of these scenarios are thus important to establish the viability of dark matter compatible parameter space.

Despite their appeal, construction and analysis of such theories remains a challenging task. It involves identifying local and global symmetries of the theory, their breaking patterns and construction of underlying effective Lagrangian for efficient perturbative calculations. In this context, we concentrated on realisation of non-Abelian gauge group accommodating a real representation with Dirac fermions, which have been studied little so far.  In particular we analysed theories with two Dirac fermions. These theories are interesting due to their topologically non-trivial coset geometry, rendering the standard construction of the WZW terms inconclusive. We therefore used an alternative construction of WZW terms. Using this construction, we not only fixed the form of the WZW terms but also the overall normalisation coefficient which otherwise remains to be fixed via experimental measurements or via anomaly mediated decays. Finally, we included $\etap$, the flavour singlet meson in the effective Lagrangian. 

In order to thermalise the dark sector with the SM bath, we used the well established $\zp$ mediator mechanism. While the stability of the pNGBs can be preserved even after introduction of this mediator, it destabilises the singlet $\etap$ typically resulting in a long lived state. The mediator also introduces a mass splitting between neutral and charged Goldstones due to radiative corrections.The precise value of the mass split remains to be estimated. 

We then used this framework for phenomenological study. The aim of this study was twofold. One was to establish the dark matter relic density and self interaction favoured regions while considering an isolated dark sector. The second aim was to investigate the effect of mediator mechanism on relic density and self interaction cross sections. 

The inherent non-perturbative nature of such dark sector theories presents several interesting challenges in making systematic progress. While usage of chiral effective theories is well established in treating such sectors in the chirally broken phase, several questions remain unanswered. Some of these questions are, at what value of $\nc, \nf$ does the theory enter conformal phase, what is the dependence of LECs in the chiral Lagrangian (e.g. masses and decay constants) on the fundamental parameters of the theory such as $\nc, \nf, \mpi/\fpi$, at what $\nc$ does the $\etap$ becomes mass degenerate with Goldstones? 

While the main part of the article was concerned with an $SO(\nc)$ gauge theory with mass-degenerate vector fermions, we also discussed generalisations for other gauge theories with real fermion representations finding interesting deviations related to the axial anomaly. We also discussed the expected symmetry structure and mass-spectrum in the mass non-degenerate case. 
Our investigations based purely on usage of effective theories were preliminary steps towards answering these questions for real representations. These conclusions can now be taken as inputs for further lattice investigations.

\section*{Acknowledgements} 
    \label{chap:ack}
    We thank B. Lucini, A. Maas, J. H. Giansiracusa and R. Alkofer, F. Zierler and F. Kahlhoefer for useful discussions. We are indebted to H. Kolesova and M. Piai for useful comments on the manuscript. SK is supported by FWF research group grant FG1. JP is supported by FWF standalone project grant P 36947-N for part of the work. SK thanks Aspen center of physics which is supported by National Science Foundation grant PHY-2210452, for hospitality during completion of part of this work.
    \newpage

\appendix
\section*{Appendix}
    \label{sec:appendix}
    \section{Generators of $SU(2\nf)$} 
\label{sec:generators_of_su2nf}

    In figure \ref{fig:generators_of_su2nf} we provide a convenient choice of generators for the $U(2\nf)$ flavour symmetry, that is useful for explicit calculations due to its compatibility with all the symmetry breaking patterns.
        
    \begin{figure}[h!]
        \centering
        \resizebox{1\textwidth}{!}{\tikzset{every picture/.style={line width=0.75pt}} 

\begin{tikzpicture}[x=0.75pt,y=0.75pt,yscale=-1,xscale=1]

\draw  [draw opacity=0][fill={rgb, 255:red, 247; green, 247; blue, 247 }  ,fill opacity=1 ] (0,40) -- (560,40) -- (560,120) -- (0,120) -- cycle ;
\draw  [draw opacity=0][fill={rgb, 255:red, 217; green, 217; blue, 217 }  ,fill opacity=1 ] (0,120) -- (560,120) -- (560,200) -- (0,200) -- cycle ;
\draw  [draw opacity=0][fill={rgb, 255:red, 247; green, 247; blue, 247 }  ,fill opacity=1 ] (0,200) -- (560,200) -- (560,350) -- (0,350) -- cycle ;
\draw  [draw opacity=0][fill={rgb, 255:red, 217; green, 217; blue, 217 }  ,fill opacity=1 ] (0,350) -- (560,350) -- (560,430) -- (0,430) -- cycle ;
\draw  [draw opacity=0][fill={rgb, 255:red, 247; green, 247; blue, 247 }  ,fill opacity=1 ] (0,430) -- (560,430) -- (560,510) -- (0,510) -- cycle ;
\draw  [draw opacity=0][fill={rgb, 255:red, 217; green, 217; blue, 217 }  ,fill opacity=1 ] (0,510) -- (560,510) -- (560,590) -- (0,590) -- cycle ;
\draw  [draw opacity=0][fill={rgb, 255:red, 217; green, 217; blue, 217 }  ,fill opacity=1 ] (0,0) -- (560,0) -- (560,40) -- (0,40) -- cycle ;
\draw    (450,0) -- (450,590) ;
\draw    (0,40) -- (560,40) ;

\draw (161,132.4) node [anchor=north west][inner sep=0.75pt]  [font=\scriptsize]  {$T_{2}^{\mathcal{F}} =\frac{1}{\sqrt{8}}\begin{pmatrix}
0 & 1 & 0 & 0\\
1 & 0 & 0 & 0\\
0 & 0 & 0 & 1\\
0 & 0 & 1 & 0
\end{pmatrix}$};
\draw (11,132.4) node [anchor=north west][inner sep=0.75pt]  [font=\scriptsize]  {$T_{1}^{\mathcal{F}} =\frac{1}{\sqrt{8}}\begin{pmatrix}
1 & 0 & 0 & 0\\
0 & -1 & 0 & 0\\
0 & 0 & 1 & 0\\
0 & 0 & 0 & -1
\end{pmatrix}$};
\draw (310,132.4) node [anchor=north west][inner sep=0.75pt]  [font=\scriptsize]  {$T_{3}^{\mathcal{F}} =\frac{1}{\sqrt{8}}\begin{pmatrix}
0 & -i & 0 & 0\\
i & 0 & 0 & 0\\
0 & 0 & 0 & i\\
0 & 0 & -i & 0
\end{pmatrix}$};
\draw (11,210.4) node [anchor=north west][inner sep=0.75pt]  [font=\scriptsize]  {$T_{4}^{\mathcal{F}} =\frac{1}{2}\begin{pmatrix}
0 & 0 & 1 & 0\\
0 & 0 & 0 & 0\\
1 & 0 & 0 & 0\\
0 & 0 & 0 & 0
\end{pmatrix}$};
\draw (161,212.4) node [anchor=north west][inner sep=0.75pt]  [font=\scriptsize]  {$T_{5}^{\mathcal{F}} =\frac{1}{\sqrt{8}}\begin{pmatrix}
0 & 0 & 0 & 1\\
0 & 0 & 1 & 0\\
0 & 1 & 0 & 0\\
1 & 0 & 0 & 0
\end{pmatrix}$};
\draw (311,212.4) node [anchor=north west][inner sep=0.75pt]  [font=\scriptsize]  {$T_{6}^{\mathcal{F}} =\frac{1}{2}\begin{pmatrix}
0 & 0 & 0 & 0\\
0 & 0 & 0 & 1\\
0 & 0 & 0 & 0\\
0 & 1 & 0 & 0
\end{pmatrix}$};
\draw (11,279.4) node [anchor=north west][inner sep=0.75pt]  [font=\scriptsize]  {$T_{7}^{\mathcal{F}} =\frac{1}{2}\begin{pmatrix}
0 & 0 & i & 0\\
0 & 0 & 0 & 0\\
-i & 0 & 0 & 0\\
0 & 0 & 0 & 0
\end{pmatrix}$};
\draw (161,282.4) node [anchor=north west][inner sep=0.75pt]  [font=\scriptsize]  {$T_{8}^{\mathcal{F}} =\frac{1}{\sqrt{8}}\begin{pmatrix}
0 & 0 & 0 & i\\
0 & 0 & i & 0\\
0 & -i & 0 & 0\\
-i & 0 & 0 & 0
\end{pmatrix}$};
\draw (311,282.4) node [anchor=north west][inner sep=0.75pt]  [font=\scriptsize]  {$T_{9}^{\mathcal{F}} =\frac{1}{2}\begin{pmatrix}
0 & 0 & 0 & 0\\
0 & 0 & 0 & i\\
0 & 0 & 0 & 0\\
0 & -i & 0 & 0
\end{pmatrix}$};
\draw (161,442.4) node [anchor=north west][inner sep=0.75pt]  [font=\scriptsize]  {$T_{13}^{\mathcal{F}} =\frac{1}{\sqrt{8}}\begin{pmatrix}
1 & 0 & 0 & 0\\
0 & 1 & 0 & 0\\
0 & 0 & -1 & 0\\
0 & 0 & 0 & -1
\end{pmatrix}$};
\draw (86,522.4) node [anchor=north west][inner sep=0.75pt]  [font=\scriptsize]  {$T_{14}^{\mathcal{F}} =\frac{1}{\sqrt{8}}\begin{pmatrix}
0 & 0 & 0 & 1\\
0 & 0 & -1 & 0\\
0 & -1 & 0 & 0\\
1 & 0 & 0 & 0
\end{pmatrix}$};
\draw (251,522.4) node [anchor=north west][inner sep=0.75pt]  [font=\scriptsize]  {$T_{15}^{\mathcal{F}} =\frac{1}{\sqrt{8}}\begin{pmatrix}
0 & 0 & 0 & i\\
0 & 0 & -i & 0\\
0 & i & 0 & 0\\
-i & 0 & 0 & 0
\end{pmatrix}$};
\draw (11,362.4) node [anchor=north west][inner sep=0.75pt]  [font=\scriptsize]  {$T_{10}^{\mathcal{F}} =\frac{1}{\sqrt{8}}\begin{pmatrix}
1 & 0 & 0 & 0\\
0 & -1 & 0 & 0\\
0 & 0 & -1 & 0\\
0 & 0 & 0 & 1
\end{pmatrix}$};
\draw (161,362.4) node [anchor=north west][inner sep=0.75pt]  [font=\scriptsize]  {$T_{11}^{\mathcal{F}} =\frac{1}{\sqrt{8}}\begin{pmatrix}
0 & 1 & 0 & 0\\
1 & 0 & 0 & 0\\
0 & 0 & 0 & -1\\
0 & 0 & -1 & 0
\end{pmatrix}$};
\draw (311,362.4) node [anchor=north west][inner sep=0.75pt]  [font=\scriptsize]  {$T_{12}^{\mathcal{F}} =\frac{1}{\sqrt{8}}\begin{pmatrix}
0 & -i & 0 & 0\\
i & 0 & 0 & 0\\
0 & 0 & 0 & i\\
0 & 0 & -i & 0
\end{pmatrix}$};
\draw (459,137.4) node [anchor=north west][inner sep=0.75pt]  [font=\normalsize]  {$\ \begin{pmatrix}
H_{1} & 0\\
0 & H_{1}^{*}
\end{pmatrix}$};
\draw (462,251.4) node [anchor=north west][inner sep=0.75pt]  [font=\normalsize]  {$\ \begin{pmatrix}
0 & S\\
S^{\dagger } & 0
\end{pmatrix}$};
\draw (458,526.4) node [anchor=north west][inner sep=0.75pt]  [font=\normalsize]  {$\ \begin{pmatrix}
0 & A\\
A^{\dagger } & 0
\end{pmatrix}$};
\draw (464,452.4) node [anchor=north west][inner sep=0.75pt]  [font=\normalsize]  {$\ \begin{pmatrix}
\mathbb{1} & 0\\
0 & -\mathbb{1}
\end{pmatrix}$};
\draw (454,367.4) node [anchor=north west][inner sep=0.75pt]  [font=\normalsize]  {$\ \begin{pmatrix}
H_{2} & 0\\
0 & -H_{2}^{*}
\end{pmatrix}$};
\draw (228,12.4) node [anchor=north west][inner sep=0.75pt]    {$U( 4)$};
\draw (468,10.4) node [anchor=north west][inner sep=0.75pt]    {$U( 2N_{F})$};
\draw (161,52.4) node [anchor=north west][inner sep=0.75pt]  [font=\scriptsize]  {$T_{0}^{\mathcal{F}} =\frac{1}{\sqrt{8}}\begin{pmatrix}
1 & 0 & 0 & 0\\
0 & 1 & 0 & 0\\
0 & 0 & 1 & 0\\
0 & 0 & 0 & 1
\end{pmatrix}$};
\draw (464,62.4) node [anchor=north west][inner sep=0.75pt]  [font=\normalsize]  {$\ \begin{pmatrix}
\mathbb{1} & 0\\
0 & \mathbb{1}
\end{pmatrix}$};

\end{tikzpicture}}
        \caption{Generators of $U(2\nf)$. 
        On the left we have explicit, properly normalised generators for $SU(4)$. To the right their general structure is given for arbitrary values of $\nf$. 
        The matrices $H_{1,2}$ denote hermitian matrices. The matrix $S = S_R + i S_I$ denotes a complex, traceful, symmetric matrix. The matrix $A = A_R + i A_I$ denotes a complex, anti-symmetric matrix. All matrices are defined with respect to the Nambu-Gorkov basis \eqref{eq:dark_strong_quark_lagrangian_nambu_gorkov}.} 
        \label{fig:generators_of_su2nf}
    \end{figure}
    
    \noindent  We state one choice of matrices explicitly for the case of $SU(4)$, while also providing the general parametrization of the generators in terms complex (anti-)symmetric and hermitian matrices for general $SU(2\nf)$.   The matrices given explicitly are normalised such that $\Tr{T_N^\F T_M^\F} = \frac{1}{2} \delta_{NM}$.
    This choice of basis makes evident the multiplet structures under the various global symmetries. The dark pions states correspond to the matrices $T^\F_{1-9}$, which split into matrices that furnish the Adjoint and the complex 2-index symmetric representation of $U(\nf) \cong SU(\nf)_I\times U(1)_B$. The Adjoint representation, parametrized by all $\nf\times \nf$ hermitian matrices $H_1$, form the neutral pion multiplet under $U(1)_D$. The generators parametrized by the complex, traceful, symmetric matrix $S$ relate to the two multiplets of all the charged pions and their anti-particles with respect to $U(1)_D$. 
    The generator $T_{13} = \Q$, corresponds to the charge assignment matrix and is the generator of $U(1)_B$. The generators parametrized by the  hermitian $\nf\times \nf$ matrices $H_2$ correspond to the global isospin symmetry $SU(\nf)_I$. Simultaneously, since $\Q$ commutes with the generators of $SU(\nf)_I$, they can be used to parameterize the neutral vector meson flavour multiplet, substituting the generalization to the $\omega$-meson in QCD. 
    These matrices stated in figure \ref{fig:generators_of_su2nf} furnish the fundamental representation of $\liea{su}(2\nf)$, with the matrix components given with respect to the Nambu-Gorkov basis \eqref{eq:dark_strong_quark_lagrangian_nambu_gorkov}. 
    Instead, one could have used the Majorana basis $q_M^{(j)}$ for the fermions, which lead to a different representation $T^\mathcal{M}_N$ of the flavour generators. Both representations are related via a basis transformation $V$ on flavour space i.e. $T^\F_N = V T^\mathcal{M}_N V^\dagger$. The matrix $V$ is given by  
    \begin{equation}\label{eq:majorana_weyl_rotation}
        V =  \Tilde{V} \otimes \unity_{\nf} \quad\quad \text{with} \quad\quad \Tilde{V} := \frac{1}{{\sqrt{2}}} \left(\begin{array}{cc}1&i\\1&-i
        \end{array}\right)
    \end{equation}
    and establishes the connection between the Weyl fermions $\psi^{(k)}$ of the Nambu-Gorkov formulation and the Majorana basis $q_M^{(k)}$ (no summation convention)
    \begin{align*}
        \psi^{(k)}
        &= \frac{1+\gamma_5}{2}\left( V^{k}_{k}~q^{(k)}_M + V^{k}_{k+N}~q^{(k+N)}_M \right) \\
        \psi^{(k+N)}
        &= \frac{1+\gamma_5}{2} \left( V^{k+N}_{k}~q^{(k)}_M + V^{k+N}_{k+N}~q^{(k+N)}_M \right)
    \end{align*}
    In the representation $T^\mathcal{M}_N$ the matrices $T^\mathcal{M}_{10-15}$ become antisymmetric and hence span a $\liea{so}(4)$ Lie-subalgebra of $\liea{su}(4)$. The matrices $T^\mathcal{M}_{1-9}$ are symmetric and span the 2-index symmetric representation of $SO(4)$, which is irreducible and substitutes the pion multiplet in the isolated case. 
    These statements generalize for arbitrary $\nf$.
    It is interesting to note that the invariant tensor $\EFmat$ transforms as a covariant rank two tensor  under the change of basis i.e.  $V^\top \omega V = \unity$. Hence, the Lie-algebra automorphism \ref{eq:lie_algebra_involutive_automorphism} with respect to the Majorana fermions is given by $\Nar(A) = -A^\top$, consistent with the anti-symmetric matrices substituting the unbroken generators.
    All the generators in figure \ref{fig:generators_of_su2nf} are hermitian. This requires for example the associated dark pion fields to be real valued. Thus, they can only describe neutral fields or fields that have no definite charge under the dark photon. Since not all the generators commute with the charge assignment matrix $\Q$, some dark pion states must be charged under $U(1)_D$. This especially holds for the matrices parameterized by a symmetric or anti-symmetric complex matrix. 
    Let us see how we obtain generators that are associated with particles of definite charge. 
    In the following we choose the generators such that for each generator $T^\F[S]$, parameterized by a complex, symmetric matrix $S$, there exists another generator $T^\F[iS]$, parameterized by $i S$. This is always possible and for example satisfied by the matrices $T_{4-9}^\F$. 
    Then the linear combinations $\Tilde{T}^\F = T^\F[S] \pm iT^\F[iS]$ give matrices for which $\combr{\Q,\Tilde{T}^\F} \propto \Tilde{T}^\F$ holds. Note that the new matrices $\Tilde{T}^\F $ are not hermitian anymore. Hence, the associate pion fields associated with these matrices must be complex and from the adjoint action of $\Q$ on the matrices $\Tilde{T}^\F$ one can read of the $U(1)_D$ charge. The associated dark pion state thus has a definite charge under $U(1)_D$. The same procedure is applicable to generators parameterized by anti-symmetric matrices $A$, which relate to the charge eigenstates of the vector mesons.
    If one wants to consider an explicit mass splitting of the fermions, yet another basis of fermions, and thus another representation of the $SU(2\nf)$ generators, is best suited. For the mass split case it is advantageous to organise the degrees of freedom in $\Psi$ not by collecting first all left-handed Weyl-fermions related to left-handed dark quarks and then anti-quarks, but to collect pairwise the degrees of freedom of each Dirac fermion. The Nambu-Gorkov parametrization \eqref{eq:dark_strong_quark_lagrangian_nambu_gorkov} of a Dirac spinor hence becomes
        \begin{equation} \label{eq:dark_strong_quark_lagrangian_nambu_gorkov_reshuffled}
        q^{(j)} = 
        \begin{pmatrix}
            \psi^{(2j-1)} \\
            \ESmat \ECmat \psi^{(2j)*}
        \end{pmatrix}
    \end{equation}
    This means that the new basis $T_N^\mathcal{P}$ is related to the one given in figure \ref{fig:generators_of_su2nf} via $T_N^\mathcal{P} = \mathcal{P}T_N^\F\mathcal{P}^\dagger $ with $\mathcal{P}$ a permutation matrix. 
    All entries of $\mathcal{P}$ are zero, except for
    \begin{align*}
        \mathcal{P}^{j}_{2j-1} = 1 \quad \quad \quad \quad \quad \quad 
        \mathcal{P}^{j+\nf}_{2j} = 1 
    \end{align*}
    with $j = 1,\dots, \nf$. 
    In the representation $T_N^\mathcal{P}$, one explicitly checks that in the case of $\nf = 2$ the only generators, satisfying the invariance condition \eqref{eq:invariance_condition_mass_term}, are given by $T_{10}^\F$ and $T_{13}^\F$. It also becomes obvious, in this basis, that these are the generators of the group $SO(2) \times SO(2)$. The $\mathbb{Z}_2$ extension of negative determinant matrices from $SO(2)$ to $O(2)$ are not anomalous and also satisfy \eqref{eq:invariance_condition_mass_term}. Hence, the breaking pattern discussing in section \ref{fig:comparison_of_symmetries_in_fundamental_two_flavor_theories} is established. 

\section{Forth homotopy group of $SU(4)/SO(4)$} \label{sec:homotpy_groups_of_coset_space}
    Wittens construction of the Wess-Zumino term in QCD \cite{Witten:1983tw}, as well as several generalizations of it \cite{DHoker:1994rdl, DHoker:1995mfi, Brauner:2018zwr} have the preliminary assumption that the forth homotopy group\footnote{A good explanation of what homotopy groups are can be found in \cite{Nakahara2015}. Also, don't confuse the symbol of the homotopy group with the pion field. These are completely different things.} $\hg_4\left(G/H\right)$ of the corresponding coset space is trivial. 
    We will show that in the case of our theory, where $G/H = SU(4)/SO(4)$, this preliminary assumption is violated. Hence, the geometrical construction by Witten is not applicable in this case and the classifications based on it are inconclusive \cite{Davighi:2018inx}. Unfortunately, $SU(4)/SO(4)$ is out of the range of Bott's periodicity theorem \cite{BottPeriodicity}, which can be used to prove the trivially of the fourth homotopy group of $SU(2k)/SO(2k)$ for $k>2$. However, the homotopy groups of $SO(4)$ and $SU(4)$ are known and summarised in table \ref{tab: homotopy_groups}.
    \begin{table}[ht!]
        \centering
        \renewcommand{\arraystretch}{1.5}
        \caption{Homotopy groups of $SO(4)$ and $SU(4)$ \cite[Appendix A, Table 6.VII] {itoEncyclopedicDictionaryMathematics1993}} 
        \begin{tabular}{c|ccc}
             &  $\hg_3 $ & $\hg_4$ & $\hg_5$\\ \hline
            $SO(4)$&$\mathbb{Z}\oplus\mathbb{Z}$ &$\mathbb{Z}_2\oplus\mathbb{Z}_2$& $\mathbb{Z}_2\oplus\mathbb{Z}_2$\\
             $SU(4)$ & $\mathbb{Z}$ & $0$&$\mathbb{Z}$
        \end{tabular}
        \label{tab: homotopy_groups}
    \end{table}
    
    \noindent For the case at hand, $SO(4)$ is an embedded Lie-subgroup of $SU(4)$. Results from differential geometry and Lie-theory \cite{Hamilton:2017gbn} tell us that $SU(4)/SO(4)$ then has a uniquely defined manifold structure and the projection map $\Pi: SU(4)\rightarrow SU(4)/SO(4)$ defines a fiber bundle $SO(4)\rightarrow SU(4)\xrightarrow{\Pi}SU(4)/SO(4)$. In Algebraic Topology \cite{HatcherAlgebraicTopology} such a fibration gives rise to a long exact sequence
    \begin{equation*}
        \renewcommand{\arraystretch}{1.5}
        \begin{array}{ccccccccc}
            \rightarrow & \hg_4\left(SU(4)\right) & \xrightarrow{h_1}  & \hg_4\left(SU(4)/SO(4)\right) & \xrightarrow{h_2}& \hg_3\left(SO(4)\right) & \xrightarrow{h_3}& \hg_3\left(SU(4)\right)  & \rightarrow\\
             \rightarrow & 0 & \xrightarrow{h_1}  & ? & \xrightarrow{h_2}& \mathbb{Z}\oplus\mathbb{Z} & \xrightarrow{h_3}& \mathbb{Z} & \rightarrow
        \end{array}
    \end{equation*}
    of group homomorphism between the homotopy groups. From this sequence we can extract a lot of information. First we observe that $\{0\} = h_1(\{0\})= \img h_1 = \ker h_2$. This lets us conclude that $h_2$ is injective. Henceforth, $\hg_4\left(SU(4)/SO(4)\right) \cong \img h_2 = \ker h_3$. But $h_3$ is a group homomorphism mapping the rank two group $\mathbb{Z}\oplus\mathbb{Z}$ on the smaller rank one group $\mathbb{Z}$. This means that $\ker h_3$ cannot be trivial and conclusively $\hg_4\left(SU(4)/SO(4)\right)$ is non-trivial. This answers a footnote remark in \cite{Brauner:2018zwr}, concerning the applicability of their methods to this coset space: They never seem to be applicable, independent of how $SO(4)$ sits inside $SU(4)$.

\section{Topological charge, Instantons and Dynkin index} \label{sec:topological_charge}
    Instantons, being gauge field configurations $\Ac_\mu$ of finite action that satisfy the classical equation of motion, can be classified by the fact that at the ``boundary`'' of $\partial \mathcal{M} \cong S^3$ they may be characterised by the fact that they approach pure gauge field configuration.
    \begin{equation} \label{eq:a_approach_pure_gauge}
        \Ac  \xrightarrow[r\rightarrow\infty]{} U^{-1}(\hat{x})\mathrm{d} U(\hat{x}) + \mathcal{O}\left(r^{-1}\right)
    \end{equation}       
    Here $\hat{x}$ is the unit vector, specifying points on $S^3$ and $U: S^3 \rightarrow \R(G)$. Again, $\R$ denotes the representation of the gauge-group $G$,  in which we put the matrix valued 1-form $\Ac = \Ac_\mu \d x^\mu$. Using \eqref{eq:a_approach_pure_gauge} and fixing a point on the sphere that must always be mapped to the neutral element of the group, one may establish a one-to-one correspondence between a distinct instanton configuration and the third homotpy group of the gauge group $\hg_3(G)$ \cite{Vandoren:2008xg}. For the classical groups $\hg_3(G) = \mathbb{Z}$, allowing to assign a unique integer $\nu$, called the ``instanton number'' or ``topological charge'', to each configuration.
    Following \cite[Chpt. 23.4]{Weinberg:1996kr}, this number is given by 
    \begin{equation} \label{eq:deiniton_topological_charge}
        \Qtopo[\Ac] = -\frac{1}{64\pi^2\mathcal{N}} \int_{\mathcal{M}}\mathrm{Tr}_R\left\{F^\R \wedge F^\R\right\}
    \end{equation}
    where $F^\R = \d \Ac + \Ac^2$ is the matrix valued field-strength 2-form in some representation $\R$. The trace contracts all indices related to the representation space  of $\R$.
    The quantity $\Qtopo[\Ac]$ assigns a unique real number to each instanton. However, the normalisation $\mathcal{N}$ must still be chosen such that $\Qtopo[\Ac]$ gives an integer and that the absolute value of the smallest possible integer is unity. This choice depends on the representation $\R$ and the chosen basis of the Lie-algebra $\liea{g}$.
   In order to obtain the correct normalisation for the classical groups, one may use a result, first obtained by Bott \cite{bottApplicationMorseTheory1956}, that  any map $U: S^3 \rightarrow G$ may be continuously deformed to a map $\tilde{U}: S^3 \rightarrow \mathrm{Std}(SU(2)) \subset G$, where $\mathrm{Std}: SU(2) \rightarrow G$ denotes a standard embedding of $SU(2)$ into $G$. Since the integral \eqref{eq:deiniton_topological_charge} depends only on the equivalence class $[U]_{\hg_3(G)}$ all the information on the normalisation $\mathcal{N}$ is given by the standard embedding and the normalisation of the correct $\liea{su}(2)$ generators within $\liea{g}$. 
   The standard embedding $\mathrm{Std}: SU(2) \rightarrow G$ may be defined for the classical groups \cite{bernardPseudoparticleParametersArbitrary1977} as follows:
    \begin{table}[h!]
        \centering
        \renewcommand{\arraystretch}{1}
        \begin{tabular}{ll}
            $SU(N)\;$ : & $\mathrm{Std}(SU(2))$ is corresponds to the $SU(2)$ subgroup of $SU(N)$ acting only\\
            {\footnotesize $N\geq3 $} & on the  first two components in the defining representation. \vspace{6pt}\\ 
            $Sp(2N)$  : & $\mathrm{Std}(SU(2))$ corresponds to the $Sp(2) \cong SU(2)$ subgroup of $Sp(2N)$,\\
            {\footnotesize $N\geq2$}    &  acting only on the first two components in the defining representation. \vspace{6pt}\\ 
            $SO(N)\;$ : & Using that $SO(4) \cong SU(2)\times SU(2)$, $\mathrm{Std}(SU(2))$ corresponds to either \\ 
            {\footnotesize $N\geq5$} & $SU(2)$ subgroup in the $SO(4)$ subgroup of $SO(N)$, acting on the first \\
            & four components in the defining vector representation.
        \end{tabular}
    \end{table}\\
    \noindent Following \cite[Chpt. 23.4]{Weinberg:1996kr}, the normalisation is given determined by the following relations
    \begin{align} 
        \label{eq:fix_standard_su2_generators_lambda}
        \left[T^\R_\alpha,T^\R_\beta\right] 
        &= \sqrt{\lambda}~\epsilon_{\alpha\beta\gamma}\,\delta^{\gamma \gamma'}\, T^\R_{\gamma'} \\
        \label{eq: fix_standard_su2_generators_T}
        tr_\R\left\{T^\R_\alpha T^\R_\beta\right\} 
        &= \lambda\, \mathcal{N} ~\delta_{\alpha \beta}
    \end{align}
    where $\lambda > 0$ is some free parameter, determining the normalisation of the generator basis and $\epsilon_{\alpha\beta\gamma}$ is the Levi-Civita symbol. 
    The first relation \eqref{eq:fix_standard_su2_generators_lambda} is independent of the representation and metric on the Lie-algebra. Hence, for explicit calculations, we may choose a basis such that $\lambda =1$. In doing so, we obtain the normalisation $\mathcal{N} = T_\R$ to be the Dynkin index of the generators in the representation $\R$. Under the assumptions of always adopting such a properly normalised basis, we obtain a formula for the topological charge, agnostic to the (matter) representation $\R$.
    \begin{equation}
        \Qtopo = \gd^2\frac{\epsilon^{\mu\nu\rho\sigma} \delta_{\alpha\beta}}{64\pi^2} \int \d x^4 \Fc_{\mu\nu}^\alpha \Fc_{\rho\sigma}^\beta
    \end{equation}
    Note that for $F^\R$ we used a convention such that the coupling constant $\gd$ is absorb within the gauge-connection 1-form i.e. that the Yang-Mills Lagrangian is normalised as in \eqref{eq:dark_strong_yang_mills_lagrangian}. 
    \begin{equation}
        F^\R = -i \gd \frac{\Fc_{\mu\nu}^\alpha}{2} \,\left(\d x^\mu\wedge \d x^\nu\right) \otimes T_\alpha^R
    \end{equation}
    In order to calculate the Dynkin index $T_\R$ in an arbitrary representation, one has to fix a metric $\kappa$ on $\liea{g}$.  
    The common choice adopted in physics is given by the trace of the generators represented in the adjoined representation normalised by a constant $c_\mathrm{adj}$.
    \begin{equation}
        \kappa_{\alpha\beta} = \frac{1}{c_\mathrm{adj.}} \Tr{T^\mathrm{adj}_\alpha T^\mathrm{adj}_\beta}
    \end{equation}
    By definition, $c_\mathrm{adj}$ determines the quadratic Casmir number of the adjoint representation. 
    All the relevant group invariants for this work, calculated within the conventions described above, are summarised in table \ref{tab:properties of representation}.
    \begin{table}
    \renewcommand{\arraystretch}{1.5}
    \centering
    \caption{Dimension, Casimir number and Dynkin index of various representations $\R$ of the classical matrix groups \cite{Ryttov:2007cx, Sannino:2009aw}. $T_\R$ denotes the Dynkin index, $c_\R$ denotes the quadratic Casimir number. In the description of the representation ``(anti-)symmetric'' always refers to the 2-index representations.}
    \begin{tabular}{c|c|c|c|c|c}
        G        & $\R$      &  $\dim \R$          & $T_\R$          & $c_\R$             & complex or real \\
        \hline
        $SU(N)$  & fundamental  &  $N$                &  $\frac{1}{2}$  & $\frac{N^2-1}{2N}$ & complex\\
        $SU(N)$  & adjoint   &  $N^2-1$            &  $N$            & $N$                & real\\
        \hline
        $SO(N)$  & vector  &  $N$                &  $1$            & $\frac{N-1}{2}$    & real\\
        $SO(N)$  & adjoint   &  $\frac{N(N-1)}{2}$ &  $N-2$          & $N-2$    & real\\
        $SO(N)$  & symmetric   &  $\frac{N(N+1)}{2}-1 $ &  $N+2 $          & $ \frac{N(N-1)(N+2)}{N(N+1)-2}$    & real\\
        \hline
        $Sp(2N)$ & fundamental  &  $2N$               &  $\frac{1}{2}$  & $\frac{2N-1}{4}$   & pseudo real\\
        $Sp(2N)$ & adjoint   &  $2N^2+N $          &  $N+1$          & $N+1$              & real \\
        $Sp(2N)$ & anti-symmetric     &  $N(2N-1)-1$        &  $N-1$          & $N $               & real\\
    \end{tabular}
    \label{tab:properties of representation}
\end{table}

\section{Technical details on the kinematic perturbative expansion} \label{sec:chiral_expansion_appendix}
    In what follows we replace the pNGb fields $\ngb \rightarrow \tau \ngb$, to also make contact with the expressions used in section \ref{sec:wzw_action}. The parameter can be used to count the number of pNGbs $\ngb$ in the vertex and may be set to $1$ at the end of the calculation. We use the quantity $\ed$ to count how many fields $\Af_\mu$ are in the vertices and adopt a language of differential forms, which might be translated back easily. For example the exterior derivative is given by $\d := \partial_\mu \d x^\mu$ and the connection 1-form $\Af = \Af_\mu \mathrm{d}x^\mu$. 
    The perturbative kinematic expansion can be performed  most conveniently by taking into account the commutator properties \eqref{eq:symmetric_lie_algebra_splitting} for the symmetric splitting of $\liea{g}_F$ and the formulas
    \begin{align}
        \exp{X} Y \exp{-X}            &= \sum_{k=0}^{\infty} \frac{\combr{X,Y}_k}{k!} \\
        \exp{X} \d \,\exp{-X} &= -\sum_{k=0}^{\infty} \frac{\combr{X,\d X}_k}{(k+1)!}\, .
    \end{align}
    Here $\combr{X,Y}_k := \combr{X, \combr{X,Y}_{k-1}}$ is recursively defined and $\combr{X,Y}_0 := Y$. This allows to obtain compact expansions of the projected quantities like $\mcfh_{\tau,h} = \mcf_{\tau,h} + \Afh_{\tau,h}$ and $\mcfh_{\tau,k} = \mcf_{\tau,k} + \Afh_{\tau,k}$. 
    \begin{align}
        \mcf_{\tau,h} &= -\frac{\tau^2}{2}\combr{\ngb,\d\ngb} - \frac{\tau^4}{24}\combr{\ngb,\d\ngb}_3 + \mathcal{O}(\tau^6) \\
        \mcf_{\tau,k} &= -\tau \d \ngb - \frac{\tau^3}{3}\combr{\ngb,\d \ngb}_2 + \mathcal{O}(\tau^5)\\
        \Afh_{\tau,h} &= B + \frac{\tau^2}{2}\combr{\ngb, B}_2 + \frac{\tau^4}{24}\combr{\ngb, B}_4 + \mathcal{O}(\tau^6 \ed)\\
        \Afh_{\tau,k} &= \tau\combr{\ngb ,B} + \frac{\tau^3}{6}\combr{\ngb, B}_3 + \mathcal{O}(\tau^5\ed)
    \end{align}
    The convenience lies in the fact that the perturbative expansion is formulated solely in terms of commutators and all coefficients can be expressed easily in terms of the structure constants of the Lie-algebra $\liea{g}_F$. Relations like
    \begin{align}
        \Tr{X\combr{Y,Z}} &= -\Tr{Z\combr{Y,X}}\\
        \Tr{\combr{X,Y}\combr{X,Z}} &= - \Tr{Y\combr{X,Z}_2} \\
        \Tr{\combr{X,Y}_{n+1}\combr{X,Z}_{m+1}} &= - \Tr{\combr{X,Y}_{n}\combr{X,Z}_{m+2}} 
    \end{align}
    become handy in explicit calculations.  
    Another important quantity is $F_{\tau} = \d \mcfh_{\tau} + \mcfh_{\tau}^2$. Using $\d\mcf = -\mcf^2$ we can proof \eqref{eq:relation_fmc_fb}, relating $F_\tau$ with $F_B = \d B + B^2$ and  use it to perform the perturbative expansion. 
    \begin{align} \label{eq:relation_fmc_fb}
        F_{\tau} &= \gamma^\dagger F_B \gamma =\exp{\tau\ngb}F_B\,\exp{-\tau\ngb} \\
        F_{\tau, h} &= F_B + \frac{\tau^2}{2}\combr{\ngb, F_B}_2 + \mathcal{O}(\tau^4 \ed)\\
        F_{\tau, k} &= \tau\combr{\ngb ,F_B} + \frac{\tau^3}{6}\combr{\ngb, F_B}_3 + \mathcal{O}(\tau^5\ed)
                    \end{align}
                    
\section{Interpolating operators for composite states} \label{sec:interpolating_linear_operators}
    We discuss how to construct all the interpolating operators build from two quark fields for scalar and vector states. Alternative constructions of such operators for (pseudo-)real theories can be found in \cite{Bennett:2019cxd}. We list explicit expressions for $\nf = 2$ in table \ref{tab:interpolating_operators}, which we could not find in the literature. 
    
\subsubsection*{Scalar operators}
    There are $2\times(2\nf)^2$ bilinear operator 
    \begin{equation}
        \psi^{(k)\top}\ESmat^*\ECmat^*\psi^{(l)}
        \quad\quad\quad\text{and}\quad\quad\quad
        \psi^{(k)\dagger}\ESmat\ECmat\psi^{(l)*}
    \end{equation}
    of which only $2\nf(2\nf+1)$ are linearly independent due to the fact that the pairings are symmetric.
    \begin{equation}
        \psi^{(k)\top}\ESmat^*\ECmat^*\psi^{(l)} = \psi^{(l)\top}\ESmat^*\ECmat^*\psi^{(k)}
    \end{equation}
    Under spatial parity \eqref{eq:spatial_parity_nambu_gorkov_basis}, these operators transform according to 
    \begin{equation}
        \left(\psi^{(k)\top}\ESmat^*\ECmat^*\psi^{(l)}\right) (t,\vec{x})
        \,\,\xmapsto{\Par}\,\,
        -\delta^{kn^\prime}\EFmat_{n^\prime n}
        \,\delta^{lm^\prime}\EFmat_{m^\prime m}
        \,\left(\psi^{(n)\dagger}\ESmat\ECmat\psi^{(m)*}\right)(t,-\vec{x})
    \end{equation}
    In order to construct $2\times \nf(2\nf+1)$ linearly independent operators, that transform either as proper scalar or pseudo-scalar under parity, one may take linear combinations with the coefficients $O_{kl} = \EFmat_{kn}\left\{T^\F_a\right\}_{\;\,\,l}^n$, where $T^\F_a$ are the broken generators of $\liea{u}(2\nf)$. This gives
    \begin{align}
        \mathcal{O}^\mathrm{S}_a &= \Psi^\top \ESmat^*\ECmat^*\EFmat T^\F_a \Psi - \Psi^\dagger \ESmat\ECmat\EFmat^* T^{\F*}_a \Psi^{*}  \\
        \mathcal{O}^\mathrm{PS}_a &=\Psi^\top \ESmat^*\ECmat^*\EFmat T^\F_a \Psi + \Psi^\dagger \ESmat\ECmat\EFmat^* T^{\F*}_a \Psi^{*} 
    \end{align}
    For practical applications, like lattice calculations, it is useful to express these operators via Dirac fermions. For this one might either follow the strategy in appendix F of \cite{Bennett:2019cxd} or use the relations  
    \begin{align}
        -\overline{q}^{(i)}\Gamma_{(\mp)}q^{(j)} &= 
        \psi^{(i+\nf)\top}\ESmat^*\ECmat^*\psi^{(j)} \mp \psi^{(i)\dagger}\ESmat\ECmat\psi^{(j+\nf)*} \\
        -\overline{q}^{(i)}\Gamma_{(\mp)}q^{(j)}_{\CC} &= 
        \psi^{(i+\nf)\top}\ESmat^*\ECmat^*\psi^{(j+\nf)} \mp \psi^{(i)\dagger}\ESmat\ECmat\psi^{(j)*}\\
        -\overline{q}^{(i)}_\CC\Gamma_{(\mp)}q^{(j)} &= 
        \psi^{(i)\top}\ESmat^*\ECmat^*\psi^{(j)} \mp \psi^{(i+\nf)\dagger}\ESmat\ECmat\psi^{(j+\nf)*}
    \end{align}
    with $\Gamma_{(+)} = \gamma_5$ and $\Gamma_{(-)} = \unity$. For these one can easily see that for example $\overline{q}^{(i)}\Gamma_{(\mp)}q^{(j)}_{\CC} = \overline{q}^{(j)}\Gamma_{(\mp)}q^{(i)}_{\CC}$. 
    
\subsubsection*{Vector operators}
    There are $(2\nf)^2$ linearly independent  operators
    \begin{equation}
        \psi^{(k)\dagger} \overline{\sigma}^\mu \psi^{(l)} = -\left(\psi^{(l)\dagger} \overline{\sigma}^\mu \psi^{(k)}\right)^*
    \end{equation}
    which transform as vectors under Lorentz transformations. Due to \eqref{eq:parity_and_currents}, we know that the linear combinations
    \begin{align}
        \mathcal{O}^\mathrm{V}_A &= \Psi^\dagger \overline{\sigma}^\mu T_A^\F\Psi\\
        \mathcal{O}^\mathrm{AV}_a &=\Psi^\dagger \overline{\sigma}^\mu T_a^\F\Psi
    \end{align}
    transform as proper vectors or as axial vectors, depending on whether $T_a^\F$ is a broken or unbroken generator of $\liea{u}(2\nf)$. 
    In order to relate them to a basis expressed in terms of Dirac fermions, one may use
    \begin{align}
        \overline{q}^{(i)}\Gamma_{(\mp)}^\mu q^{(j)} &= 
        \psi^{(i)\dagger} \overline{\sigma}^\mu \psi^{(j)} \mp  
        \psi^{(j+\nf)\dagger} \overline{\sigma}^\mu \psi^{(i+\nf)} \\
        \overline{q}^{(i)}\Gamma_{(\mp)}^\mu q^{(j)}_{\CC} &= 
        \psi^{(i)\dagger} \overline{\sigma}^\mu \psi^{(j+\nf)} \mp  
        \psi^{(j)\dagger} \overline{\sigma}^\mu \psi^{(i+\nf)}\\
        \overline{q}^{(i)}_{\CC}\Gamma_{(\mp)}^\mu q^{(j)} &= 
        \psi^{(i+\nf)\dagger} \overline{\sigma}^\mu \psi^{(j)} \mp  
        \psi^{(j+\nf)\dagger} \overline{\sigma}^\mu \psi^{(i)}\, .
    \end{align}
    Here $\Gamma_{(-)}^\mu = \gamma^\mu$ and $\Gamma_{(+)}^\mu = \gamma^\mu\gamma^5$. Alternatively, one may use the strategy presented in appendix F of \cite{Bennett:2019cxd}. 

    \begin{table}[h!]
        \centering
        \renewcommand{\arraystretch}{1.41}
        \begin{tabular}{c|c|c||c|c|c}
            & $\quad\Psi^\top \ESmat^*\ECmat^*\EFmat T^\F_a \Psi + \text{h.c.}\quad$ & $J^{\Par}$ & & $\quad\Psi^\top \ESmat^*\ECmat^*\EFmat \Tilde{T}^\F_a \Psi + \text{h.c.}\quad$ & $B$  \\ \hline
            $\pid_1$& $\frac{1}{\sqrt{2}}\left(\overline{d}\gamma_5d - \overline{u}\gamma_5u\right)$ &$0^-$ 
                & $\tilde{\pid}_{1}$&$\frac{1}{\sqrt{2}}\left(\overline{d}\gamma_5d - \overline{u}\gamma_5u\right)$  &0 \\ 
            $\pid_2$& $\frac{-1}{\sqrt{2}}\left(\overline{u}\gamma_5d + \overline{d}\gamma_5u\right)$ &$0^-$ 
                & $\tilde{\pid}_{2}$&$\frac{-1}{\sqrt{2}}\left(\overline{u}\gamma_5d + \overline{d}\gamma_5u\right)$ &0 \\ 
            $\pid_3$& $\frac{i}{\sqrt{2}}\left(\overline{u}\gamma_5d - \overline{d}\gamma_5u\right)$ &$0^-$ 
                & $\tilde{\pid}_{3}$&$\frac{i}{\sqrt{2}}\left(\overline{u}\gamma_5d - \overline{d}\gamma_5u\right)$  &0 \\ \hdashline
            $\pid_4$& $\frac{-1}{2}\left(\overline{u}\gamma_5u_{\CC}+\overline{u}_{\CC}\gamma_5u\right)$ &$0^-$ 
                & $\tilde{\pid}_{4}$& $\frac{-1}{\sqrt{2}}\, \overline{u}\gamma_5 u_ {\CC}$ &-1 \\ 
            $\pid_5$& $\frac{-1}{\sqrt{2}}\left(\overline{u}\gamma_5d_{\CC}+\overline{u}_{\CC}\gamma_5d\right)$ &$0^-$ 
                & $\tilde{\pid}_{5}$& $- \overline{u}\gamma_5 d_ {\CC}$&-1 \\ 
            $\pid_6$& $\frac{-1}{2}\left(\overline{d}\gamma_5d_{\CC}+\overline{d}_{\CC}\gamma_5d\right)$ &$0^-$ 
                & $\tilde{\pid}_{6}$& $\frac{-1}{\sqrt{2}}\, \overline{d}\gamma_5 d_{\CC}$&-1 \\ \hdashline
            $\pid_7$& $\frac{i}{2}\left(\overline{u}_{\CC}\gamma_5 u - \overline{u}\gamma_5 u_{\CC}\right)$ &$0^-$ 
                & $\tilde{\pid}_{7}$& $\frac{-1}{\sqrt{2}}\, \overline{u}_{\CC}\gamma_5 u$ &1 \\ 
            $\pid_8$& $\frac{i}{\sqrt{2}}\left(\overline{u}_{\CC}\gamma_5 d - \overline{u}\gamma_5 d_{\CC}\right)$ &$0^-$ 
                & $\tilde{\pid}_{8}$&$ - \overline{u}_{\CC}\gamma_5 d$ &1 \\ 
            $\pid_9$& $\frac{i}{2}\left(\overline{d}_{\CC}\gamma_5 d - \overline{d}\gamma_5 d_{\CC}\right)$ &$0^-$ 
                & $\tilde{\pid}_{9}$& $\frac{-1}{\sqrt{2}}\, \overline{d}_{\CC}\gamma_5 d$ &1 \\ \hline
            $\etap$ & $\frac{1}{\sqrt{2}}\left(\overline{d}\gamma_5d + \overline{u}\gamma_5u\right)$ &$0^-$ 
                & $\etap$           & $\frac{1}{\sqrt{2}}\left(\overline{d}\gamma_5d + \overline{u}\gamma_5u\right)$ &0 \\ \hline\hline
            & $\Psi^\dagger \overline{\sigma}^\mu T_a^\F\Psi$ & $J^{\Par}$ & & $\Psi^\dagger \overline{\sigma}^\mu \Tilde{T}_a^\F\Psi$ & $B$ \\ \hline
            $\omega_{10}$&$\frac{1}{\sqrt{8}} \left(\overline{u}\gamma^\mu u - \overline{d}\gamma^\mu d\right)$ &$1^-$ 
                & $\tilde{\omega}_{10}$& $\frac{1}{\sqrt{8}} \left(\overline{u}\gamma^\mu u - \overline{d}\gamma^\mu d\right)$ &0 \\ 
            $\omega_{11}$&$\frac{1}{\sqrt{8}} \left(\overline{u}\gamma^\mu d + \overline{d}\gamma^\mu u\right)$ &$1^-$ 
                & $\tilde{\omega}_{11}$& $\frac{1}{\sqrt{8}} \left(\overline{u}\gamma^\mu d + \overline{d}\gamma^\mu u\right)$ &0 \\ 
            $\omega_{12}$&$\frac{i}{\sqrt{8}} \left(\overline{u}\gamma^\mu d - \overline{d}\gamma^\mu u\right)$ &$1^-$ 
                & $\tilde{\omega}_{12}$& $\frac{i}{\sqrt{8}} \left(\overline{u}\gamma^\mu d - \overline{d}\gamma^\mu u\right)$&0 \\ \hdashline
            $\rho_{13}$&$\frac{1}{\sqrt{8}} \left(\overline{u}\gamma^\mu u + \overline{d}\gamma^\mu d\right)$ &$1^-$ 
                & $\tilde{\rho}_{13}$& $\frac{1}{\sqrt{8}} \left(\overline{u}\gamma^\mu u + \overline{d}\gamma^\mu d\right)$ &0 \\ \hdashline
            $\rho_{14}$&$\frac{1}{\sqrt{8}} \left(\overline{u}\gamma^\mu d_{\CC} + \overline{u}_{\CC}\gamma^\mu d\right)$ &$1^-$ 
                & $\tilde{\rho}_{14}$& $\frac{1}{2} \overline{u}_{\CC} \gamma^\mu d$ &1 \\ \hdashline
            $\rho_{15}$&$\frac{i}{\sqrt{8}} \left(\overline{u}\gamma^\mu d_{\CC} -\overline{u}_{\CC}\gamma^\mu d\right)$ &$1^-$ 
                & $\tilde{\rho}_{15}$&$\frac{1}{2} \overline{u} \gamma^\mu d_{\CC}$ &-1 \\ 
        \end{tabular}
        \caption{Summary of all interpolating operators for the composite states relevant for DM in the IR. $J^{\Par}$ denotes their angular momentum quantum number, while $B$ denotes their dark Baryon number charge or equivalently their charge under the dark photon. The latter can only be assigned for the charge eigenbasis. Solid vertical lines separate multiplets under $SO(2\nf)$, dashed line multiplets under $SU(\nf) \times U(1)_B$. We borrowed the notation from QCD e.g. here $u := q^{(1)}$ and $d := q^{(2)}$.
        The construction of the matrices $\Tilde{T}_a^\F$ has been explained in appendix \ref{sec:generators_of_su2nf}.}
        \label{tab:interpolating_operators}
    \end{table}
    
\section{Connection to $SU(\nc)$-QCD}
    The standard literature on $SU(\nc)$ gauge theories typically uses a different but confusingly similar formalism for the description of the low energy effective description. Since we used the general language of Bando et al. \cite{Bando:1987br}, the special syntax of $SU(\nc)$ QCD must be contained. We would like to explicitly demonstrate how this comes about, since this is useful to compare results with the existing literature. 
In the $SU(\nc)$ case, the generators of the chiral $SU(\nf)_L \times SU(\nf)_R$ have the following structure 
\begin{equation}
    T_n^\F = \left(\begin{array}{cc}
        h_1 + h_2 & 0\\
        0 & -(h_1 - h_2)^\top
    \end{array}\right)
    \quad\quad
    \text{where}
    \quad\quad
    \left\{\begin{array}{cc}
        h_1 = 0 & \Rightarrow  T_N^F \in \ks \\
        h_2 = 0 & \Rightarrow  T_N^F \in \liea{h}
    \end{array}\right.
\end{equation}  
if we work in a basis of only left-handed fermions and anti-fermions, analog to the Nambu-Gorkov formalism. The invariant tensor $\EFmat$ in this basis is given by \eqref{eq:invaraint_flavor_tensor}. 
Then one may decompose all the building blocks of the HLS approach as follows. 
\begin{equation*}
    \begin{array}{rlcrlcrlcrl}
        \gamma &= 
        \left(\begin{array}{cc}
            \gamma_L & 0 \\
            0 & \gamma_R^*  
        \end{array}\right)
        &\quad&
        \mcf &= \left(\begin{array}{cc}
            L & 0 \\
            0 & -R^\top  
        \end{array}\right)
        &\quad&
        \Sigma &= \left(\begin{array}{cc}
            0 & U \\
            U^\top & 0  
        \end{array}\right)
        &\quad&
        \Af &= \left(\begin{array}{cc}
            \mathcal{L} & 0 \\
            0 & -\mathcal{R}^\top \\
        \end{array}\right)\\
        &&&&&&&\\
        \Afh &= 
        \left(\begin{array}{cc}
            \hat{\mathcal{L}} & 0 \\
            0 & -\hat{\mathcal{R}}^\top  
        \end{array}\right)
        &\quad&
        \Av &= \left(\begin{array}{cc}
            \mathcal{V} & 0 \\
            0 & -\mathcal{V}^\top  
        \end{array}\right)
        &\quad&
        \hat{F}_\Af &= \left(\begin{array}{cc}
            \hat{F}_\mathcal{L} & 0 \\
            0 & -\hat{F}_\mathcal{R}^\top  
        \end{array}\right)
        &\quad&
        F_\Av &= \left(\begin{array}{cc}
            F_\mathcal{V} & 0 \\
            0 & -F_\mathcal{V}^\top  
        \end{array}\right)
    \end{array}
\end{equation*} 
where $\gamma \in SU(\nf)_L \times SU(\nf)_R $ is the coset representative and $\mathcal{L},\mathcal{R}$ are the gauge connection 1-forms of $SU(\nf)_L$ and $SU(\nf)_R$ respectively. The HLS gauge fields are collected in $\mathcal{V}$.   
\begin{equation*}
    \begin{array}{rlcrlcrlcrl}
        L&= \gamma^\dagger_L \d \gamma_L
        &\quad&
        R&= \gamma^\dagger_R \d \gamma_R
        &\quad&
        U&= \gamma_L\gamma_R^\dagger
        &\quad&
        F_\mathcal{V}&= \d\mathcal{V} + \mathcal{V}^2
        \\
        \hat{\mathcal{L}}&= \gamma_L^\dagger \mathcal{L} \gamma_L  
        &\quad&
        \hat{\mathcal{R}}&= \gamma_R^\dagger \mathcal{R} \gamma_R
        &\quad&
        \hat{F}_{\mathcal{L}}&= \gamma_L^\dagger (\d\mathcal{L} + \mathcal{L}^2) \gamma_L
        &\quad&
        \hat{F}_{\mathcal{R}}&= \gamma_R^\dagger (\d\mathcal{R} + \mathcal{R}^2) \gamma_R
    \end{array}
\end{equation*}
Note that $\gamma$ transforms according to table \ref{tab:transformation_properties_of_HLS_fields}, while $U$ transforms as $U \mapsto U_L U U_R^\dagger$, with $U_L \in SU(\nf)_L$ and $U_R \in SU(\nf)_R$. The quantity $U$ is the one used to construct the chiral Lagrangian in the standard literature like \cite{Scherer:2012xha, Harada:2003jx}. The HLS gauge fixing condition \eqref{eq:HLS_gauge_fixing_condition}, translates into 
\begin{equation}
    \gamma_L = \gamma_R^{\dagger} = e^{i\pi/\fpi}
\end{equation}
Some simple manipulations on the Lagrangian \eqref{eq:HLS_lagrangian} allow to obtain back the results from $SU(\nc)$-QCD as for example given in \cite{Harada:2003jx}. We demonstrate this explicitly for the homogeneous part \eqref{eq:non_anomalous_wzw_1}-\eqref{eq:non_anomalous_wzw_4} of the WZW action.
By application of the automorphism $\Nar$, we may decompose $\mcf + \Afh - V = (\mcfh_h - V) + \mcfh_k$, where 
\begin{equation*}
    \mcfh_h - V = \left(\begin{array}{cc}
        \frac{1}{2}(\hat{\alpha}_L + \hat{\alpha}_R ) &0  \\
        0 & -\frac{1}{2}(\hat{\alpha}_L + \hat{\alpha}_R )^\top
    \end{array}\right)
    \quad\quad
    \mcfh_k = \left(\begin{array}{cc}
        \frac{1}{2}(\hat{\alpha}_L - \hat{\alpha}_R ) &0  \\
        0 & \frac{1}{2}(\hat{\alpha}_L - \hat{\alpha}_R )^\top
    \end{array}\right)
\end{equation*}
and
\begin{align}
    \hat{\alpha}_L &= L + \mathcal{L} -\mathcal{V}\\
    \hat{\alpha}_R &= R + \mathcal{R} -\mathcal{V}
\end{align}
For the homogeneous part of the WZW one may verify that
\begin{align}
    \Tr{(\mcfh_h-V)\mcfh_k^3 + (\mcfh_h-V)^3\mcfh_k} &= \frac{1}{2}\Tr{\hat{\alpha}_R\hat{\alpha}_L^3 - \hat{\alpha}_L\hat{\alpha}_R^3} \\
    \Tr{(\mcfh_h-V)\mcfh_k^3 - (\mcfh_h-V)^3\mcfh_k} &=
    \frac{1}{2}\Tr{\hat{\alpha}_L\hat{\alpha}_R\hat{\alpha}_L\hat{\alpha}_R} \\
    \Tr{F_\Av(\mcfh_h-V)\mcfh_k} &= \frac{1}{2}\Tr{F_\mathcal{V}(\hat{\alpha}_R\hat{\alpha}_L-\hat{\alpha}_L\hat{\alpha}_R)}\\ 
     \Tr{\hat{F}_\Af(\mcfh_h-V)\mcfh_k} &= \frac{1}{4}\Tr{(\hat{F}_\mathcal{L}+\hat{F}_\mathcal{R})(\hat{\alpha}_R\hat{\alpha}_L-\hat{\alpha}_L\hat{\alpha}_R)} 
\end{align}
These results are in agreement with \cite{Harada:2003jx}.

\section{Conventions on spacetime signature, indices and $\gamma$-matrices}
\label{sec:conventions}
    We use the spacetime metric $g_{\mu\nu}$ of signature $(+---)$. Otherwise then explicitly noted, the position of the indices matter and the transformation behaviour of upper and lower indices in general differs\footnote{Typically, they are inverse to each other so that contracted quantities are invariant.}. The Einstein summation convention is only assumed for pairs of an upper and a lower index. 
    The Pauli matrices we define
    \begin{equation} \label{eq:pauli_matrices}
        \overline{\sigma}^0 
        = \left(\begin{array}{cc c cc c cc}
            1 & 0 \\
            0 & 1 
        \end{array}\right)
        \quad 
        \overline{\sigma}^1 
        = \left(\begin{array}{cc c cc c cc}
            0 & 1 \\
            1 & 0 
        \end{array}\right) 
        \quad
        \overline{\sigma}^2 
        = \left(\begin{array}{cc c cc c cc}
            0 & -\mathrm{i} \\
            \mathrm{i} & 0 
        \end{array}\right)
        \quad
        \overline{\sigma}^3 
        = \left(\begin{array}{cc c cc c cc}
            1 & 0 \\
            0 & -1 
        \end{array}\right) 
    \end{equation}
    Further we define $ \overline{\sigma}_\mu = g_{\mu\nu} \overline{\sigma}^\nu $ and spacetime indices are pulled with the metric $\sigma^\mu = g^{\mu\mu} \overline{\sigma}^\mu$. For the $\gamma$-matrices we can choose a special basis as the chiral basis given by 
    \begin{equation} \label{eq:gamma_matrices}
        \gamma^\mu = 
        \left(\begin{array}{cc}
            0                      & \sigma^\mu \\
            \overline{\sigma}^\mu  & 0 
        \end{array}\right)
    \end{equation}
    where again $\gamma_\mu = g_{\mu\nu}\gamma^\nu $. The charge conjugation of a Dirac fermion $q$ is defined as $C\,q^*$ with the charge conjugation matrix $C = i\gamma^2 = -i\gamma_2$. The chiral element $\gamma_5 = \gamma^5 = i\gamma^0\gamma^1\gamma^2\gamma^3$ is defined in the standard way. In the chiral basis both are represented as 
    \begin{equation} \label{eq:charge_conjugation_and_gamma_five_matrix}
        C =
        \left(\begin{array}{cc}
            0&-\i\,\overline{\sigma}^2\\
            \i\,\overline{\sigma}^2&0  
        \end{array}\right) 
        \quad\quad\quad
        \gamma_5 =  
        \left(\begin{array}{cc}
            \unity  & 0\\
            0       & -\unity 
        \end{array}\right)
    \end{equation}
    Typically $\mu,\nu,\rho,\sigma$ denote spacetime indices. The indices $\alpha,\beta,\gamma$ are related to the basis of the colour algebra $\liea{g}_C$ i.e. colour-gauge-indices. The indices $N,M,K$ relate to the flavour algebra $\liea{g}_F = \liea{h}_F + \ks$, while $a,b,c$ indicate broken generators in $\ks$ and $A,B,C $ correspond to unbroken generators in $\liea{h}_F$. The indices $k,l,m,n$ count the basis elements of the representation space of the fundamental representation of $\liea{g}_F$ i.e. flavour indices. 

\label{sec:connection_to_qcd}

\bibliographystyle{JHEP}
\bibliography{main.bib}
\end{document}